\documentclass[aps,prd, twocolumn, superscriptaddress,amsmath,amssymb,floatfix]{revtex4-2}

\usepackage[dvipdfmx]{graphicx}
\usepackage{dcolumn}
\usepackage{bm}
\usepackage{braket}
\usepackage{braket}
\usepackage{breqn}
\usepackage{color}
\usepackage{mathtools} 
\usepackage{hyperref}

\renewcommand{\Re}[1]{\mathrm{Re}\left[ #1 \right]}
\renewcommand{\Im}[1]{\mathrm{Im}\left[ #1 \right]}

\begin{document}

\title{Deci-Hz gravitational waves from the self-interacting axion cloud around the rotating stellar mass black hole}

\author{Hidetoshi Omiya}
\email{omiya@tap.scphys.kyoto-u.ac.jp}
\affiliation{Department of Physics$,$ Kyoto University$,$ Kyoto 606-8502$,$ Japan}
\author{Takuya Takahashi}\email{ttakahashi1@rikkyo.ac.jp}
\affiliation{Department of Physics$,$ Rikkyo University$,$ Toshima$,$ Tokyo 171-8501$,$ Japan}
\author{Takahiro Tanaka}
\email{t.tanaka@tap.scphys.kyoto-u.ac.jp}
\affiliation{Department of Physics$,$ Kyoto University$,$ Kyoto 606-8502$,$ Japan}
\affiliation{Center for Gravitational Physics and Quantum Information$,$ Yukawa Institute for Theoretical Physics$,$ Kyoto University$,$ Kyoto 606-8502$,$ Japan}
\author{Hirotaka Yoshino}
\email{hyoshino@omu.ac.jp}
\affiliation{Department of Physics$,$ Osaka Metropolitan University$,$ Osaka 558-8585$,$ Japan}
\affiliation{Nambu Yoichiro Institute of Theoretical and Experimental Physics (NITEP)$,$ Osaka Metropolitan University$,$ Osaka 558-8585$,$ Japan}

\date{\today}

\begin{abstract}
Gravitational waves from condensates of ultra-light particles, such as axion, around rotating black holes are a promising probe to search for unknown physics. For this purpose, we need to characterize the signal to detect the gravitational waves, which requires tracking the evolution of the condensates, including various effects. 
The axion self-interaction causes the non-linear coupling between the superradiant modes, resulting in complicated branching of evolution. Most studies so far have considered evolution under the non-relativistic approximation or the two-mode approximation. In this paper, we numerically investigate the evolution of the axion condensate without these approximations, taking higher multipole modes into account. We also investigate the possible signature in gravitational waves from the condensate. We show that the higher multipole modes are excited, leading to the gravitational wave signal by the transition of the axion between different levels. The most prominent signal of gravitational waves arises from the transition between modes with their angular quantum numbers different by two. 
The gravitational wave signal is emitted in the deci-Hz band for stellar mass black holes, which might be observable with the future gravitational wave detectors.
\end{abstract}

\maketitle

\section{Introduction}

Black holes and gravitational waves offer a promising avenue to test physics beyond the standard model of particle physics, such as an axion. The interesting possibility suggested by string theory is that our universe may contain many species of axions, the so-called axiverse scenario~\cite{Arvanitaki:2009fg, Cicoli:2012sz, Demirtas:2018akl, Mehta:2021pwf}. In particular, axions in string theory typically have mass with their Compton wavelength around the astrophysical scale and the so-called axion decay constant, which determines the strength of the interaction involving the axion around the grand unification scale ($\sim 10^{16}\mathrm{GeV}$)~\cite{Svrcek:2006yi}. Besides the axiverse scenario, there are other motivations to assume the existence of axions. The QCD axion is needed to solve the strong CP problem~\cite{Peccei:1977hh, Weinberg:1977ma, Wilczek:1977pj, Kim:1979if, Shifman:1979if, Zhitnitsky:1980tq, Dine:1981rt}, while also axion is a dark matter candidate~\cite{Abbott:1982af, Preskill:1982cy, Dine:1982ah, Hui:2016ltb}.

Axions with the Compton wavelength about the size of astrophysical black holes (mass around $10^{-20} - 10^{-10}\mathrm{eV}$) can be probed by observing the condensate of the axion around black holes~\cite{Arvanitaki:2010sy}. The condensate of the axion is spontaneously formed by the superradiant instability~\cite{Zouros:1979iw, Detweiler:1980uk, Brito:2015oca}. Superradiant instability is a phenomenon such that the axion bounded by the gravitational potential exponentially grows by extracting the energy and angular momentum of the black hole. The rate of instability is fastest when the Compton wavelength of the axion is comparable to the size of the black hole. For the solar mass black hole, the rate is around one minute. As a result, the macroscopic number of axions will occupy the superradiant mode within the age of the universe.
In this paper, we call the condensate made solely by a single superradiant mode as an axion cloud and the superposition of axion clouds as an axion condensate.

One way to observe the axion condensate is to observe the gravitational waves emitted from it~\cite{Arvanitaki:2010sy, Arvanitaki:2014wva, Yoshino:2014wwa, Brito:2017zvb, LIGOScientific:2021jlr, Zhu:2020tht, Tsukada:2018mbp, Isi:2018pzk, Siemonsen:2022yyf} (see for example~\cite{Brito:2015oca, Baumann:2018vus, Chen:2019fsq} for other methods to observe axion condensates). The gravitational waves are emitted through two distinct processes: the pair annihilation of axions and the level transition between different superradiant modes. The characteristic feature of the emitted gravitational waves is that they are monochromatic. If a rotating black hole exists in our neighborhood, we have a chance to observe the gravitational waves from the associated axion condensate.

To better characterize the gravitational wave signal, a better understanding of the evolution of axion condensates under various effects, such as the axion self-interaction, is necessary (see ~\cite{Ikeda:2018nhb, Ding:2020bnl, Tong:2021whq, Choudhary:2020pxy, Yoo:2021kyv, Tong:2022bbl, Takahashi:2021eso, 
Takahashi:2021yhy, Takahashi:2023flk, Sakurai:2023hkg, Spieksma:2023vwl, Cannizzaro:2023jle, Sarmah:2024nst} for investigations on other effects such as tidal effects from the companion and the coupling to the photon). The self-interaction causes several effects ~\cite{Yoshino:2015nsa, Fukuda:2019ewf, Omiya:2020vji, Omiya:2022mwv}, but the most significant one is the dissipation of the cloud by transferring the energy to another cloud ~\cite{Gruzinov:2016hcq, Baryakhtar:2020gao, Omiya:2022gwu, Calza:2023rjt}. At a certain amplitude, the dissipation by the self-interaction is balanced with the growth due to the superradiant instability, resulting in a quasi-stationary configuration with various modes excited simultaneously.

So far, the evolution of the self-interacting condensate has been extensively investigated under the two-mode approximation with the non-relativistic approximation~\cite{Baryakhtar:2020gao} or the three-mode approximation neglecting the spin down of the black hole~\cite{Omiya:2022gwu}. These are not satisfactory to fully characterize the gravitational wave signals. The observationally attractive regime where the growth rate is the fastest, and the amplitude of the gravitational wave becomes large is in the relativistic regime. Furthermore, higher multipole modes are excited in the relativistic regime~\cite{Baryakhtar:2020gao, Omiya:2022gwu}. In addition, the inclusion of the spin down of the black hole is necessary since the spin of the black hole determines the lifetime of the condensate~\cite{Omiya:2022gwu}.

In this paper, we numerically examine the co-evolution of the self-interacting axion condensate and the black hole, as well as the resulting gravitational wave radiation in the relativistic regime, including higher multipole modes. We use the adiabatic approximation to track the evolution, which is well-motivated in the current situation~\cite{Brito:2014wla, Omiya:2022mwv}. We find that the higher multipole modes are excited, but representative excited modes depend on the axion mass and the axion decay constant. In addition, we show that the excitation of the higher multipole modes will result in a characteristic gravitational wave signal different from the one predicted in Ref.~\cite{Baryakhtar:2020gao}. More specifically, the transition, including the higher multipole modes, leads to a louder signal than the pair annihilation signal or the transition between the fastest and the second fastest growing modes for the GUT scale decay constant ($10^{17}--10^{16}{\rm GeV}$).

This paper is organized as follows.
In Sec.~\ref{sec:2}, we briefly review the axion cloud formed by the superradiant instability.
In Sec.~\ref{sec:3}, we formulate the evolution equations of the self-interacting axion condensate, including the multiple modes up to $l=m=4$ and the spin down of the central black hole.
In Sec.~\ref{sec:4}, we numerically solve the evolution equations formulated in Sec.~\ref{sec:3} and see how the excitation of the higher multipole modes proceeds.
In Sec.~\ref{sec:5}, we compute the gravitational and axion waves from the axion condensate.
Section~\ref{sec:6} summarizes our findings with some comments on the detectability of the gravitational waves.
In the rest of this paper, we take the unit with $c=G=\hbar = 1$.

\section{Axion cloud}\label{sec:2}

We first review the formation of a condensate of an axion around a rotating black hole by the superradiant instability. For a detailed review of this topic, see~\cite{Brito:2015oca}. In this paper, we employ the action
\begin{align}\label{eq:action}
	S =& \frac{1}{16 \pi} \int d^4 x\, \sqrt{-g} R \cr
 &\quad +  F_a^2 \int d^4 x\,  \sqrt{-g} \left\{ - \frac{1}{2}g^{\mu\nu}\partial_\mu \phi \partial_\nu \phi - V(\phi)\right\}~,
\end{align}
where we normalize the axion field  $\phi$ with the decay constant $F_a$. Note that $F_a$ is measured in the unit of the Planck mass $M_{\rm pl} = G^{-1/2}$. Thus, we have $F_a \sim 10^{-3}$ for the GUT scale decay constant. In this paper, we consider the leading order piece of the cosine type potential,
\begin{align}\label{eq:cospotential}
	V(\phi) = \mu^2 \left(1 - \cos\phi\right) \sim \frac{\mu^2}{2}\phi^2 - \frac{\mu^2}{4!}\phi^4 ~.
\end{align}
The background metric is fixed to the Kerr metric given by
\begin{align}
	\label{kerrmetric}
	ds^2 &=  - \left(1 - \frac{2 M r}{\rho^2}\right)dt^2 - \frac{4 a M r \sin^2\theta}{\rho^2} dt d\varphi \cr
	 &\qquad + \left[(r^2 + a^2)+ \frac{2 M r}{\rho^2} a^2 \sin^2 \theta \right]\sin^2\theta d\varphi^2 \cr
    &\qquad\qquad+ \frac{\rho^2}{\Delta} dr^2 + \rho^2 d\theta^2~,
\end{align}
with
\begin{align}
	\Delta &=(r-r_+)(r-r_-)~, & \rho^2 &= r^2 + a^2\cos^2\theta~,
\end{align}
and
\begin{align}
	\frac{r_\pm}{M} = 1  \pm \sqrt{1 - \chi^2}~.
\end{align}
Here, $M$ is the mass, and $\chi = a/M$ is the dimensionless spin parameter of the black hole. The radii $r_+$ and $r_-$ correspond to the positions of the event horizon and the Cauchy horizon, respectively.

In the initial phase of the evolution, the amplitude of the axion is small, and we can neglect the quartic term in the potential~\eqref{eq:cospotential}. In this case, the equation of motion for the axion is given by
\begin{align}\label{eq:massivescalar}
	(\square_g - \mu^2)\phi = 0~.
\end{align}
As shown in the literature~\cite{Brill:1972xj}, we can solve the equation of motion by the separation of the variables,
\begin{align}
	\phi = e^{-i(\omega t - m \varphi)}R_{lm\omega}(r) S_{lm\omega}(\theta) + {\rm c.c.}~.
\end{align}
Here, ${\rm c.c.}$ denotes the complex conjugate.
The radial part and the angular part of the equation become
\begin{align}
		\frac{d}{dr}\left(\Delta\frac{dR_{lm\omega}}{dr}\right)  + \left[\frac{K^2(\omega)}{\Delta} - \mu^2 r^2 -\lambda_{lm}(\omega) \right]R_{lm\omega} = 0~,\label{eq:EOMrad}
\end{align}
\begin{align}
		\frac{1}{\sin\theta}\frac{d}{d\theta}\left(\sin\theta \frac{d{S}_{lm\omega}}{d\theta}\right) &+ \left[c^2(\omega) \cos^2\!\theta - \frac{m^2}{\sin^2\theta}\right]S_{lm\omega} \cr
  &\qquad \quad = - \Lambda_{lm}(\omega) S_{lm\omega}~,\label{eq:EOMang}
\end{align}
with
\begin{align}
		 c^2 (\omega) &= a^2 (\omega^2 - \mu^2)~, \qquad K(\omega) = (r^2+a^2)\omega - am~,\cr
	 \lambda_{lm}(\omega) & = -2am \omega +a^2\omega^2 +\Lambda_{lm}(\omega)~,
\end{align}
and $\Lambda_{lm}(\omega)$ is the separation constant that is reduced to $l(l+1)$ in the limit $a\to 0$.  The normalization of the spheroidal harmonics $S_{lm\omega}$ is taken to satisfy
\begin{align}
	\int d\cos\theta\ S_{lm\omega}(\theta)S_{l'm\omega}(\theta) = \delta_{ll'}~.
\end{align}

Since we are interested in the bound states of the axion, we solve the radial equation~\eqref{eq:EOMrad} with the boundary conditions
\begin{align}\label{eq:bclinear}
	R_{lm\omega}(r) \to \begin{dcases}
	\left(\frac{r-r_+}{M}\right)^{-i \frac{2 M r_+}{r_+ - r_-}(\omega - m \Omega_H)}~, & (r\to r_+)~,\\
	\left(\frac{r}{M}\right)^{-1 -M\frac{\mu^2 - 2 \omega^2}{\sqrt{\mu^2 - \omega^2}}} e^{-\sqrt{\mu^2 - \omega^2}r}~, & (r\to \infty)~,
	\end{dcases}
\end{align}
where  $\Omega_H = a/(r_+^2 + a^2)$ is the angular velocity of the event horizon. By the analytical~\cite{Detweiler:1980uk} and numerical methods~\cite{Dolan:2007mj}, the solutions are found, and they are labeled by the three integers $n,l,$ and $m$ with $n \ge l+1$, where $n, l,$ and $m$ is called the principal, azimuthal, and magnetic quantum numbers, respectively. We denote a solution as
\begin{align}
	\phi_{nlm} &= e^{-i(\omega_{nlm} t - m \varphi)}S_{lm\omega_{nlm}}(\theta)R_{lm\omega_{nlm}}(r) + {\rm c.c.}~.
\end{align}
Here, $\omega_{nlm}$ is the angular frequency of the solution. We call the solutions with the smallest $n~ ( = l +1)$ as the fundamental modes and those with larger $n$ as the overtone modes. Note that we normalize the radial mode function in such a way that the energy of the configuration calculated within the linearized approximation is equal to $F_a^2 M$.

\begin{figure*}[t]
        \centering
        \includegraphics[keepaspectratio,scale=0.55]{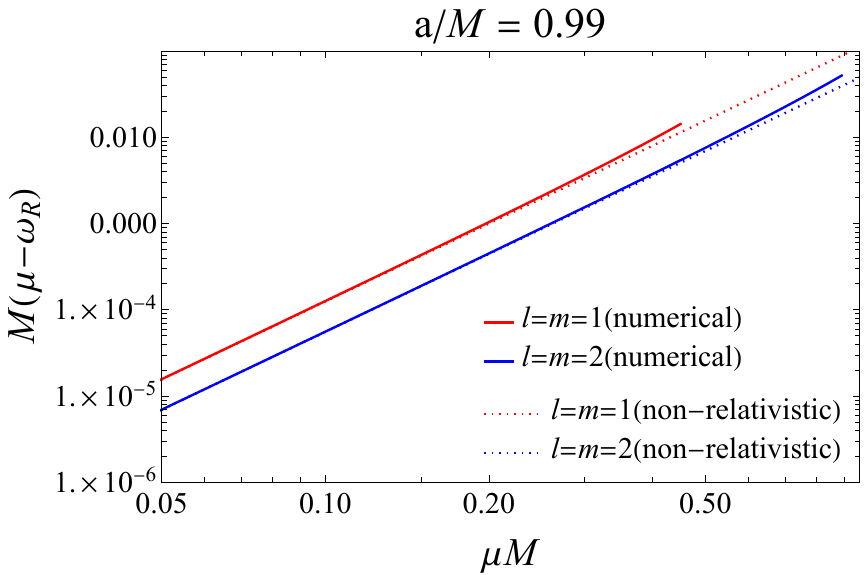}
	\includegraphics[keepaspectratio,scale=0.55]{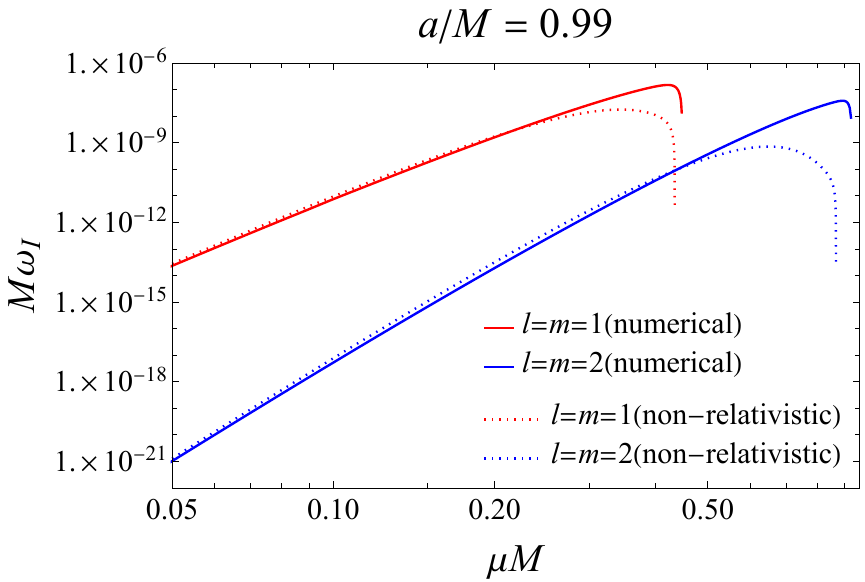}
	\caption{(Left) The behavior of $M (\mu - \omega_{nlm,R})$ as the function of $\mu M$, where $\omega_{nlm,R}$ denotes the real part of the frequency $\omega_{nlm}$. The red and the blue solid curve correspond to the $l=m=1$ and $l=m=2$ fundamental superradiant modes, respectively. The corresponding dotted curves are the ones with the non-relativistic approximation Eq.~\eqref{eq:nonrelwR}. The spin of the central black hole is set to $a/M = 0.99$. (Right) The similar plot for the imaginary part of the frequencies $\omega_{nlm}$.}
	\label{fig:omega}
\end{figure*}

The real and the imaginary parts of the frequency of the $l=m=1$ and $l=m=2$ fundamental modes are shown in Fig.~\ref{fig:omega}. We observe that the frequencies possess the positive imaginary part, indicating the presence of instability. This instability is called {\it superradiant instability}. The superradiant instability occurs for~\cite{Zouros:1979iw}
\begin{align}\label{eq:SRcond}
	0<\Re{\omega_{nlm}} < m \Omega_{H}~.
\end{align}
 Because of the superradiant instability, the axion will occupy the superradiant states. As a result, the condensate of the axion will be spontaneously formed. In this paper, we denote the condensate made purely from a single superradiant mode as an {\it axion cloud} and the superposition of such clouds as an {\it axion condensate}.  

\begin{figure}[t]
\centering\includegraphics[keepaspectratio,scale=0.5]{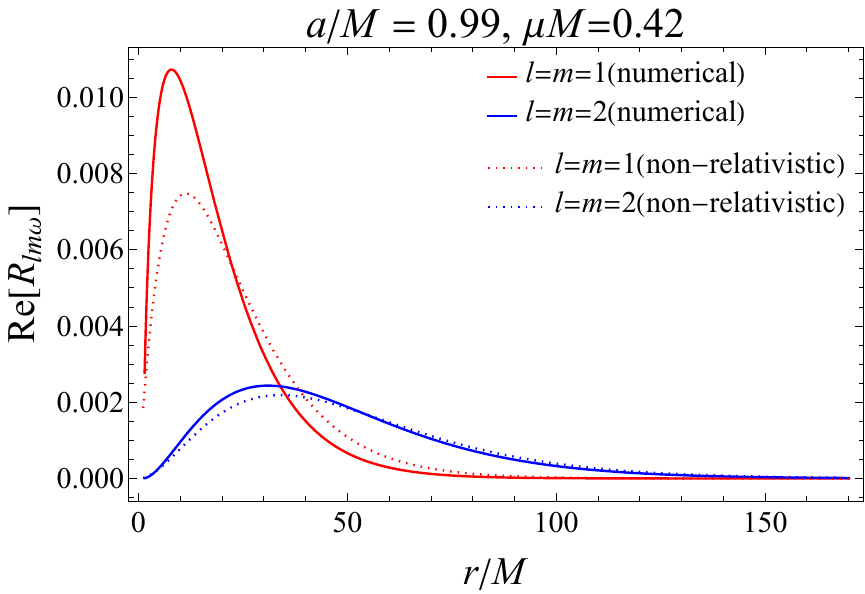}
\caption{The real part of the radial mode function, $R_{lm\omega}$ as a function of $r/M$. The red and blue solid curves, respectively, correspond to the $l=m=1$  and $l=m=2$ superradiant modes without any approximation, while the dashed curves are the counterparts in the non-relativistic approximation. 
The spin of the black hole and the mass of the axion are set to $a/M = 0.99$ and $\mu M = 0.42$, respectively. The phase of the mode functions is chosen so that there is no imaginary part for large $r$.}
\label{fig:radialmode}
\end{figure}

The radial configuration of the axion cloud is shown in Fig.~\ref{fig:radialmode}. We observe that the configuration is quite similar to the radial wave function of the Hydrogen atom. In fact, in the small gravitational coupling regime, $\mu M \ll 1$,  the frequencies, and the mode functions coincide with those of the Hydrogen atom at a distant place~\cite{Detweiler:1980uk}
\begin{align}\label{eq:nonrelwR}
	\omega_{nlm,R} \equiv \Re{\omega_{nlm}} \sim \mu \left(1 - \frac{(\mu M)^2}{2n^2}\right)~,
\end{align}
\begin{align}
	S_{lm\omega}(\theta) &\sim Y_{lm}(\theta) \equiv \sqrt{\frac{(l - m)! (2l+1)}{2(l+m)!}} P^m_l(\cos\theta)~,\\
	\label{eq:radnonrel}
	R_{lm\omega}(r) &\sim R_{nl}(r) \equiv \frac{1}{2\sqrt{\pi} \mu M}\left(\frac{2 (\mu M)^2}{n}\right)^{3/2} \sqrt{\frac{ (n - l - 1)!}{ n^4
   (n + l)!}} \cr
   &\qquad  \times \left(\frac{2 \mu ^2 M  r}{n}\right)^l e^{-\frac{\mu ^2 M r}{n}} L_{n -l -1}^{2
   l+1}\left(\frac{2\mu ^2 M r}{n}\right)~.
\end{align}
Here, ${L^k}_{n}$ is the associated Laguerre polynomials.
By the asymptotic matching method, the approximate formula for the imaginary part of the frequency is found as
\begin{widetext}
    \begin{align}\label{eq:nonrelwI}
	\omega_{nlm,I} \equiv \Im{\omega_{nlm}}  \sim& \frac{2 r_+}{M}(\mu M)^{4l+5}(-\mu+m\Omega_H)\frac{2^{4l+1}(l+n)!}{(n-l-1)! n^{2l+4}}\cr
		&\times\left(\frac{l!}{(2l)!(2l+1)!}\right)^{2}\prod_{j=1}^{l}\left(j^2\left(1-\frac{a^2}{M^2}\right) + 4 r_+^2(\mu-m\Omega_H)^2\right)~.
\end{align}
\end{widetext}
We refer to the formulas using the leading order approximations in $\mu M$ (Eqs.~\eqref{eq:nonrelwR} --~\eqref{eq:radnonrel}) as the ``non-relativistic approximation". The more accurate approximation can be found in Refs.~\cite{Baumann:2019eav, Bao:2022hew}. As the figures show, the non-relativistic approximation breaks down in the $\mu M \sim 1$ regime. For example, the radial mode function becomes more compact than the one predicted by the non-relativistic approximation (see Fig.~\ref{fig:radialmode}).

\section{Evolution equations of the axion-black hole system}\label{sec:3}

As the condensate grows, various effects would alter its evolution from the naive exponential growth predicted by the linear approximation in the previous section. The self-interaction~\cite{Arvanitaki:2010sy, Gruzinov:2016hcq, Fukuda:2019ewf, Baryakhtar:2020gao, Mocanu:2012fd, Omiya:2022gwu, Omiya:2022mwv, Omiya:2020vji, Yoshino:2015nsa, Yoshino:2012kn} and the change of the spin of the central black hole~\cite{Arvanitaki:2010sy, Brito:2014wla} are particularly important. The self-interaction would change the evolution by causing the nonlinear mode coupling, which results in the saturation of the evolution~\cite{Gruzinov:2016hcq, Baryakhtar:2020gao, Omiya:2022gwu}. 
In addition, the spin of the central black hole gradually slows down. 
Once the spin reaches a certain value, the growth of the condensate terminates because the superradiant condition~\eqref{eq:SRcond} is no longer met.

 Since the growth time scale of the cloud is much longer than its dynamical time scale and the light crossing time of the black hole, $\omega_{nlm,I} \ll \omega_{nlm,R}, M^{-1}_{\rm BH}$, we can assume that the evolution is adiabatic. Under the adiabatic evolution, at each moment, the metric is approximated by the Kerr metric with a given black hole mass $M_{\rm BH}(t)$ and an angular momentum $J_{\rm BH}(t)$. The evolution of each axion cloud is captured by that of the normalized mass, $M_{{{\rm cl}},i}(t)$, where $i$ specifies the mode. Since we have normalized the axion by the decay constant $F_a$, the actual mass of the cloud is expressed by the normalized mass as $F_a^2 M_{{{\rm cl}},i}(t)$.

\begin{figure*}[t]
 \centering
 \includegraphics[keepaspectratio, scale=0.4]{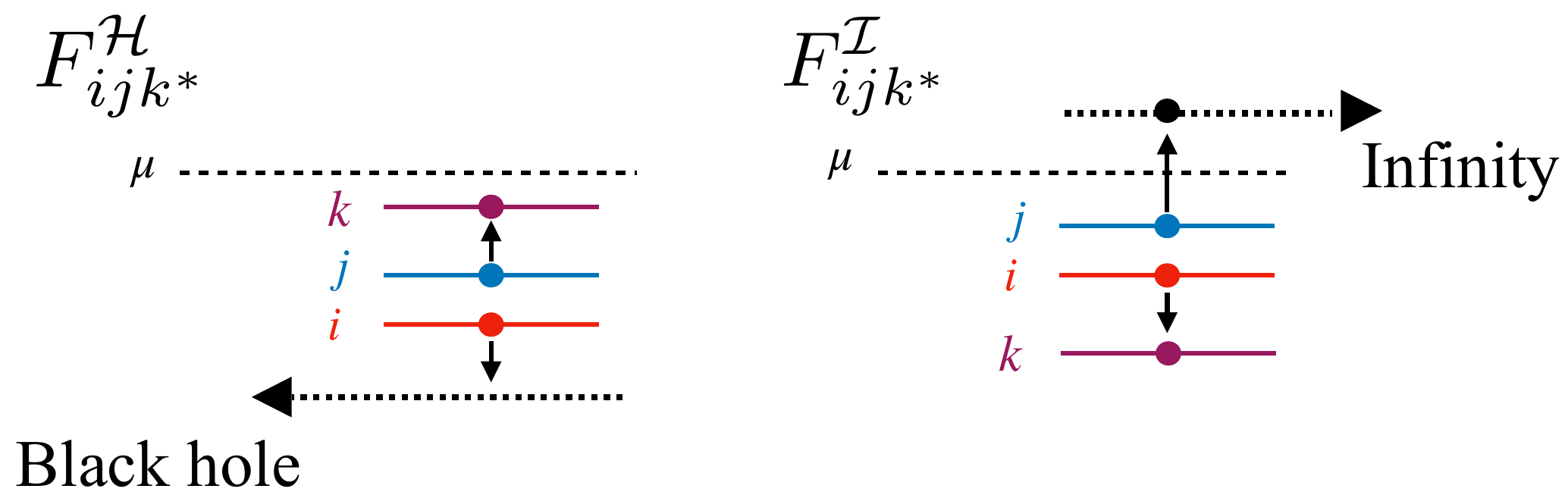}
 \caption{Dissipative processes induced by the self-interaction. Each line represents an energy level. The left panel corresponds to the dissipation of the axion due to the absorption by the black hole, and the right panel corresponds to the dissipation due to the radiation to infinity. The particles in the cloud $i$ and $j$ make a transition to the one in the cloud $k$ and the dissipative mode specified by the superscript ${\mathcal H}$ and ${\mathcal I}$.}
 \label{fig:process}
\end{figure*}

Evolution equations for the clouds are derived from the local conservation laws of the energy and the angular momentum~\cite{Omiya:2022gwu}. The results are summarized in the following form
\begin{widetext}
 \begin{align}
 	\label{eq:Mclevol}
	\frac{d M_{{{\rm cl}},i}}{dt} =&
      + 2 \omega_{i,I}M_{{{\rm cl}},i }\cr
	&  - \sum_{j,k} \frac{(1 + \delta_{ij})\omega_{i,R}}{\omega_{i,R} + \omega_{j,R} -\omega_{k,R}} F^{\mathcal{H}}_{ijk^*}M_{{\rm cl},i} M_{{\rm cl},j}M_{{\rm cl},k} + \sum_{j,k} \frac{\omega_{i,R}}{\omega_{j,R} + \omega_{k,R} -\omega_{i,R}} F^{\mathcal{H}}_{jki^*}M_{{\rm cl},i} M_{{\rm cl},j}M_{{\rm cl},k}\cr
	&  - \sum_{j,k} \frac{(1 + \delta_{ij})\omega_{i,R}}{\omega_{i,R} + \omega_{j,R} -\omega_{k,R}} F^{\mathcal{I}}_{ijk^*}M_{{\rm cl},i} M_{{\rm cl},j}M_{{\rm cl},k} + \sum_{j,k} \frac{\omega_{i,R}}{\omega_{j,R} + \omega_{k,R} -\omega_{i,R}} F^{\mathcal{I}}_{jki^*}M_{{\rm cl},i} M_{{\rm cl},j}M_{{\rm cl},k}~.
\end{align}
\end{widetext}
The first term represents the increase in the energy due to the superradiance.
The coefficients $F^{\mathcal{H}}_{ijk^*}$ and $F^{\mathcal{I}}_{ijk^*}$ are defined by 
\begin{align}
\left.\frac{dE_a}{dt}\right|_{r=r_+} = F_a^2 F^{\mathcal{H}}_{ijk^*} M_{{\rm cl},i}M_{{\rm cl},j}M_{{\rm cl},k}~,\\
	\left.\frac{dE_a}{dt}\right|_{r\to \infty}  = F_a^2 F^{\mathcal{I}}_{ijk^*} M_{{\rm cl},i}M_{{\rm cl},j}M_{{\rm cl},k}~,
\end{align}
which represent the magnitude of the energy flux of the axion to the horizon and that to infinity, respectively, induced by the process involving modes $i,j,$ and $k$. Note that $F_{ijk^*}^{\mathcal{H},\mathcal{I}} = F_{jik^*}^{\mathcal{H},\mathcal{I}}$, since the modes $i$ and $j$ are interchangeable in this process.
These processes are caused by the self-interaction, 
and see Fig.~\ref{fig:process} for the meaning of their subscript.
The terms with the Kronecker delta $\delta_{ij}$ are necessary to take into account that two axions are lost by a single transition when $i = j$.
See appendix~A of~\cite{Omiya:2022gwu} for details of the derivation. 
The factor $F_a^2$ comes from the normalization of the axion field.

The evolution of the black hole mass and the angular momentum is governed by the 
balance equations ~\cite{Brito:2014wla},
\begin{widetext}
\begin{align}\label{eq:evolMBH}
	\frac{d M_{\rm BH}}{d t} &= F_a^2\left(-\sum_i 2 \omega_{i,I}M_{{{\rm cl}},i} +\sum_{(i,j,k)} F^{\mathcal{H}}_{ijk^*}M_{{\rm cl},i} M_{{\rm cl},j}M_{{\rm cl},k}\right) ~,\\
	\label{eq:evolJBH}
	\frac{d J_{\rm BH}}{d t} &= F_a^2\left( -\sum_i 2 \frac{m_i \omega_{i,I}}{\omega_{i,R}} M_{{{\rm cl}},i}+ \sum_{(i,j,k)} \frac{m_i + m_j - m_k}{\omega_{i,R} + \omega_{j,R} - \omega_{k,R}}F^{\mathcal{H}}_{ijk^*}M_{{\rm cl},i} M_{{\rm cl},j}M_{{\rm cl},k} \right)~,\cr
\end{align}
\end{widetext}
where the right hand sides are the energy and angular momentum fluxes through the horizon, respectively. 
The first term in the parenthesis corresponds to the mass and the angular momentum loss (gain) due to the superradiance or absorption of axions. The second term corresponds to the energy and the angular momentum flux induced by the self-interaction (the left panel of Fig.~\ref{fig:process}). 

\subsection{Subleading effects}

There are subleading effects that we do not take into account in the evolution. One is the other effect due to the self-interaction. As summarized in Ref.~\cite{Omiya:2022gwu}, the effect of the self-interaction can be classified into three parts. One is the dissipation due to the low-frequency modes $\omega \sim \mu$, which is considered in this paper. The second one is the dissipation due to the high-frequency modes $\omega \gtrsim 2\mu$. These contributions are small when considering the evolution in the perturbative regime~\cite{Baryakhtar:2020gao, Omiya:2020vji}. The third one is the deformation of the cloud. When the amplitude of the cloud is large, the condensate is deformed significantly~\cite{Omiya:2022mwv}, and for a large amplitude, it leads to the collapse of the condensate~\cite{Yoshino:2015nsa}. However, it has been shown that the dissipative effects due to the self-interaction are large enough to prevent the condensate from entering the collapsing regime in the case the condensate starts with a small amplitude, {\it e.g.}, that produced by the quantum fluctuation~\cite{Baryakhtar:2020gao, Omiya:2022gwu}. For these reasons, we only consider the dissipative effect due to the low-frequency radiation. 

 According to the cloud deformation, the eigenfrequencies of the cloud would change. It becomes essential when we consider the long-term observation of the continuous gravitational waves from the axion cloud~\cite{PhysRevD.99.084042}. When the amplitude of the cloud is not so large, the frequency shift can be calculated by the perturbative method both in the non-relativistic approximation~\cite{Baryakhtar:2020gao} and in the relativistic approach~\cite{Omiya:2020vji, Cannizzaro:2023jle}. When the amplitude of the cloud is large, it should be calculated numerically as in Ref.~\cite{Omiya:2022mwv}. In this paper, we do not compute the frequency shift for ease of calculation, 
since the frequency shift does not change the evolution significantly. 

In addition, we neglect the gravitational wave radiation and self-gravity of the cloud. The energy loss rate of clouds due to the gravitational wave radiation is much smaller than the energy transfer rate due to the superradiant instability~\cite{Yoshino:2013ofa} and the scalar wave radiation induced by the self-interaction. For this reason, we neglect the effect of gravitational waves on the evolution.
The correction to the axion condensate due to the self-gravity is also small~\cite{Chia:2022udn}.

\section{Time evolution with four modes}\label{sec:4}

Now, we numerically solve the evolution equations~\eqref{eq:Mclevol},~\eqref{eq:evolMBH}, and~\eqref{eq:evolJBH}. We analyze the situation in which the four fundamental modes $l=m=1,l=m=2,l=m=3,$ and $l=m=4$ are excited simultaneously. 
In our previous paper~\cite{Omiya:2022gwu}, it has been shown that the amplitudes of overtone modes are always subdominant. In the following we use $i =1,2,3,4$ to label the fundamental modes with $(l,m) = (1,1),(2,2),(3,3),(4,4)$.  

Our four mode approximation would give a correct evolution unless the $l=m=3$ or the $l=m=4$ mode grows to a significantly large amplitude. In the case of the $l=m=3$ or $4$ modes growing to a large amplitude, we need to include much higher multipoles. As we shall see below, the substantial growth of the $l=m=3$ or $4$ mode only happens for a much later time when the $l=m=1$ and $2$ modes have decayed. Therefore, when considering the gravitational wave signal involving the $l=m=1$ or $l=m=2$ mode, our four-mode approximation should give an accurate result.

When considering these modes, the fluxes of the processes that we need to take into account are the following eighteen (ten from the processes dissipating to the black hole and eight from the ones dissipating into infinity):
\begin{description}
\item[Flux to the horizon]

\begin{align}\label{eq:processH}
	 &F^{\mathcal{H}}_{112^*}~,  F^{\mathcal{H}}_{113^*}~,  F^{\mathcal{H}}_{223^*}~,   F^{\mathcal{H}}_{114^*}~, F^{\mathcal{H}}_{224^*}~, \cr
  &F^{\mathcal{H}}_{334^*}~, F^{\mathcal{H}}_{123^*}~, F^{\mathcal{H}}_{124^*}~, F^{\mathcal{H}}_{134^*}~,  F^{\mathcal{H}}_{234^*} ~.
\end{align}

\item[Flux to infinity]

\begin{align}\label{eq:processI}
	F^{\mathcal{I}}_{221^*}~,  F^{\mathcal{I}}_{331^*}~,  F^{\mathcal{I}}_{441^*}~, F^{\mathcal{I}}_{442^*}~,  F^{\mathcal{I}}_{231^*}~,  F^{\mathcal{I}}_{241^*}~,  F^{\mathcal{I}}_{341^*}~, &F^{\mathcal{I}}_{342^*} ~.
\end{align}
\end{description}
Other processes become subdominant because of the following reasons: 1) the frequencies of the dissipative modes are much larger than the axion mass, or 2) for the modes dissipating to the black hole, the angular momentum barrier of the dissipative mode is large enough ($l > 2$)
~\cite{Baryakhtar:2020gao, Omiya:2022gwu}. Note that when considering the flux to infinity, its frequency must be larger than the axion mass.

When solving the evolution equations, we need the numerical values of $\omega_i,  F^{\mathcal{H},{\cal I}}_{ijk^*}$ and so on for given $(\mu M,\chi)$. We numerically calculate these quantities on the grid points in the parameter space of $(\mu M, \chi)$ in the range between $0.02 \leq \mu M \leq 0.45$ and $0.15 <\chi < 0.998$ with interval $\Delta (\mu M) =0.01$ and $\Delta \chi = 0.025$. For the calculation of the time evolution, we use the value of $\omega_i$ and $F^{\mathcal{H},{\cal I}}_{ijk^*}$ obtained by interpolating the values on the grid points. The whole calculation is done using \textsc{Mathematica}.

\subsection{Example of the time evolution}\label{sec:4A}

To demonstrate the behavior of solutions of Eqs.~\eqref{eq:Mclevol},~\eqref{eq:evolMBH}, and~\eqref{eq:evolJBH}, we fix the initial mass of the black hole to $M_{\rm BH, ini} = 10 M_{\odot}$ in this section. Then, the superradiant instability of the $l=m=1$ mode occurs for the axion mass in the range $\mu \lesssim 6\times 10^{-12}{\rm eV}$. We take the initial spin of the black hole to be relatively large $\chi = 0.99$. The initial mass of the cloud is taken to be $M_{{\rm cl}, i} = 10^{-10} M_{\rm BH,ini}$, which is just set to a sufficiently small value. In reality, the cloud mass will start with a much smaller value, but in such a small mass ($M_{{\rm cl}, i} \sim 10^{-80} M_{\odot}$ if the cloud starts with the quantum fluctuation with GUT scale decay constant), self-interaction is totally negligible, and hence our result should give the qualitatively correct results, even if we do not start with such an extremely small number. For the detail of the dependence of the evolution on the initial condition, see Ref.~\cite{Omiya:2022gwu}.

\begin{figure}[tbp]
 \centering
 \includegraphics[keepaspectratio, scale=0.55]{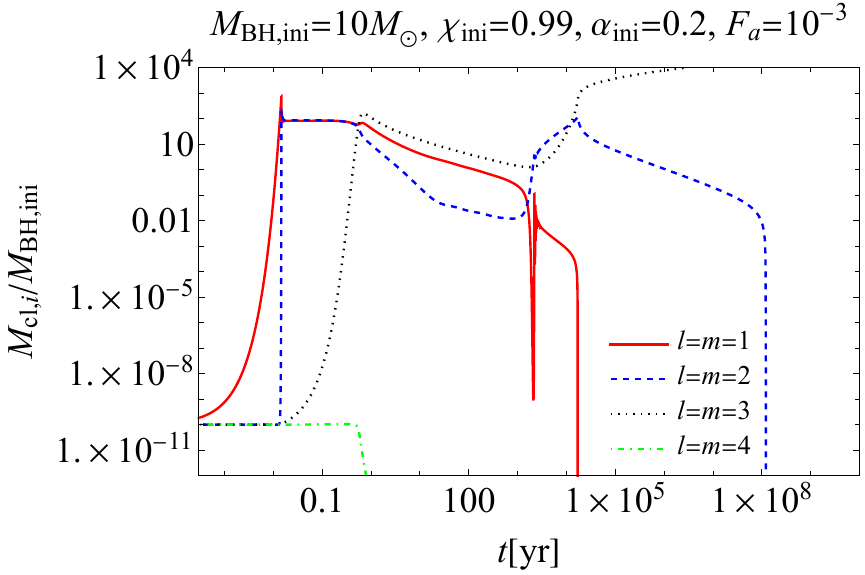}
 \caption{An example of the time evolution of the axion cloud mass $M_{{\rm cl}, i}$. The red solid, blue dashed, black dotted, and green dash-dotted curves correspond to the $l=m=1, 2, 3,$ and $4$ modes, respectively. 
 The initial black hole mass and black hole spin are $M_{\rm BH,ini} = 10M_{\odot}$ and $\chi_{\rm ini} = 0.99$, respectively. We take the axion mass such that $\alpha_{\rm ini} \equiv \mu M_{\rm BH,ini} = 0.2$ and the axion decay constant to be $F_a = 10^{-3}$. }
 \label{fig:examplecloud}
\end{figure}

We first show the evolution of the cloud for the axion mass with $\mu M_{\rm BH, ini} = 0.2,$ and $F_a = 10^{-3}$ in Fig.~\ref{fig:examplecloud}. This axion mass corresponds to $\mu \sim 2.6 \times 10^{-12}{\rm eV}$. The evolutionary track of the cloud is summarized as follows:
\begin{enumerate}
\item The $l=m=1$ cloud grows exponentially by the superradiant instability.
\item The energy of the $l=m=1$ cloud is transferred to the $l=m=2$ cloud. Then, the $l=m=2$ cloud grows exponentially. At a certain amplitude, the energy gained by the superradiance and the dissipation by the self-interaction balance result in a quasi-stationary configuration. The mass of each mode is roughly determined by solving the stationary conditions
\begin{align}
    \frac{dM_{{\rm cl},i}}{dt} &= 0~, & (i&=1,2,3,4)~.
\end{align}

Note that, around the maximum of the cloud mass, we cannot neglect the deformation of the mode function due to the self-interaction~\cite{Omiya:2022gwu}. The clouds become compact at a large amplitude, and the frequencies of the clouds would be modified. However, the numerical analysis shows that the qualitative evolution does not change from the perturbative calculation.  The change in the cloud configuration might enhance the gravitational wave amplitudes, but we neglect this effect for simplicity. The modification to the frequencies is important for the actual observation, but we also neglect it.
\item Higher multipole modes are excited by the same mechanism. In the present example, the $l=m=3$ cloud is excited. Which modes are excited depends on the parameters of the axion, $(\mu, F_a)$, as we briefly discuss below.
\item The $l=m=1$ cloud is depleted because of the spin down of the black hole by the superradiance of the $l=m=2$ mode. As the obstruction due to the presence of the $l=m=1$ cloud disappears, the $l=m=2$ mode starts to grow by the superradiant instability. 
\item Once the energy of the $l=m=2$ cloud dominates, the energy transfer from  the $l=m=2$ cloud to the $l=m=3$ cloud becomes efficient. 
Since there is no process to transfer the energy of the $l=m=3$ mode to the $l=m=2$ mode, the $l=m=3$ mode continues to grow. 
For this case, we need to include much higher multipoles, such as $l=m=6$ mode.
Then, in the end, the energy is transferred to much higher multipoles and settles into the quasi-stationary state.
\end{enumerate} 

\begin{figure}[tbp]
 \centering
 \includegraphics[keepaspectratio, scale=0.55]{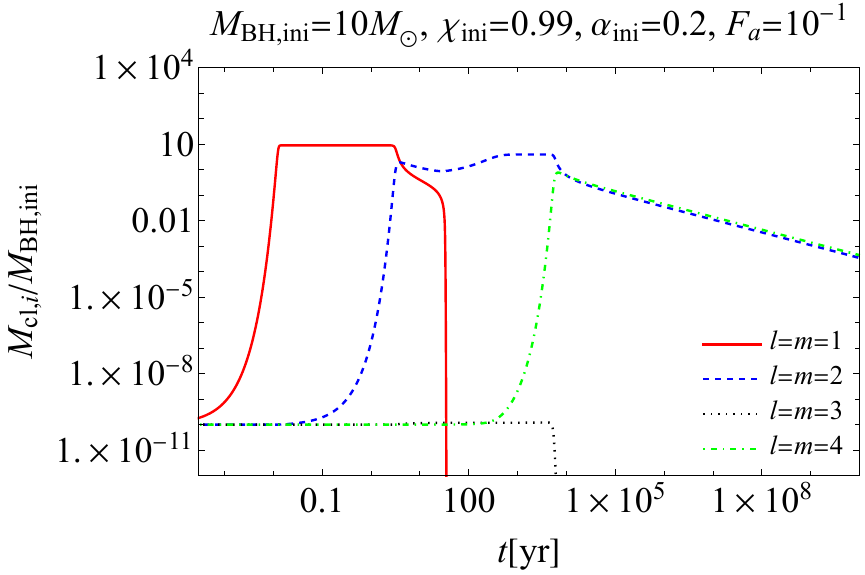}
  \includegraphics[keepaspectratio, scale=0.55]{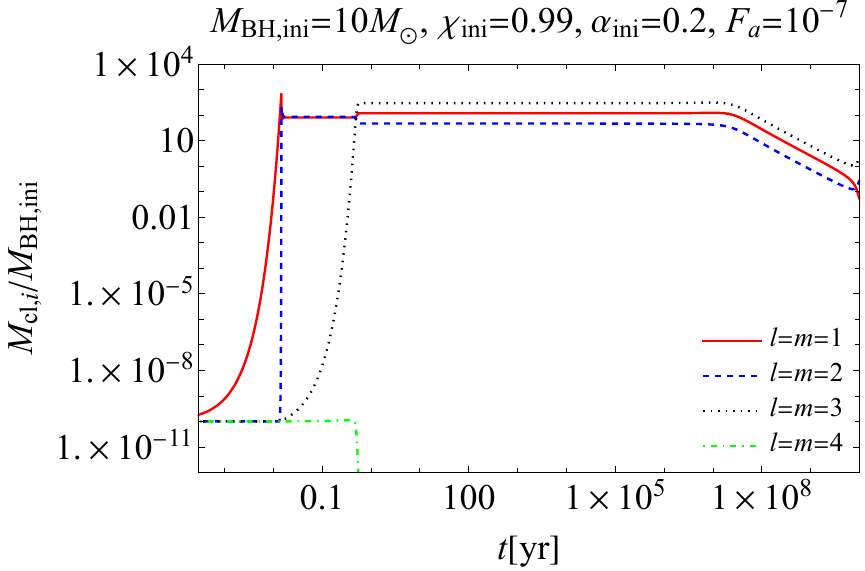}
 \caption{The same figure as Fig.~\ref{fig:examplecloud} but with $F_a = 10^{-1}$(Top) and $10^{-7}$(Bottom), respectively. }
 \label{fig:examplecloudFa}
\end{figure}

Now, let us vary the parameters of the axion, $F_a$ and $\mu$. In Fig.~\ref{fig:examplecloudFa}, we show the case with only the value of $F_a$ changed from that of Fig.~\ref{fig:examplecloud}. Changing the value of $F_a$ is identical to changing the relative size of the black hole spin down rate and the energy transfer rate of the cloud due to the self-interaction. Increasing $F_a$ weakens the self-interaction and increases the relative spin down rate while decreasing $F_a$ does the opposite. The top panel of Fig.~\ref{fig:examplecloudFa} is the case when $F_a$ is increased. We observe that the saturation of the $l=m=1$ cloud occurs much before the excitation of the $l=m=2$ cloud. This saturation occurs because the black hole has spun down to the superradiance threshold of Eq.~\eqref{eq:SRcond}. Although the saturation due to the spin down occurs, the energy transfer from the $l=m=1$ cloud to the $l=m=2$ cloud still occurs, leading to much earlier excitation of the $l=m=2$ cloud than the excitation by the superradiant instability alone. This regime corresponds to the {\it``small self-coupling regime"} in Ref.~\cite{Baryakhtar:2020gao}. On the other hand, we observe a delay in the decay of the $l=m=1$ cloud when we decrease $F_a$ (bottom panel of Fig.~\ref{fig:examplecloudFa}). This regime corresponds to the {\it``large self-coupling regime"} in Ref.~\cite{Baryakhtar:2020gao}.

\begin{figure}[tbp]
 \centering
 \includegraphics[keepaspectratio, scale=0.55]{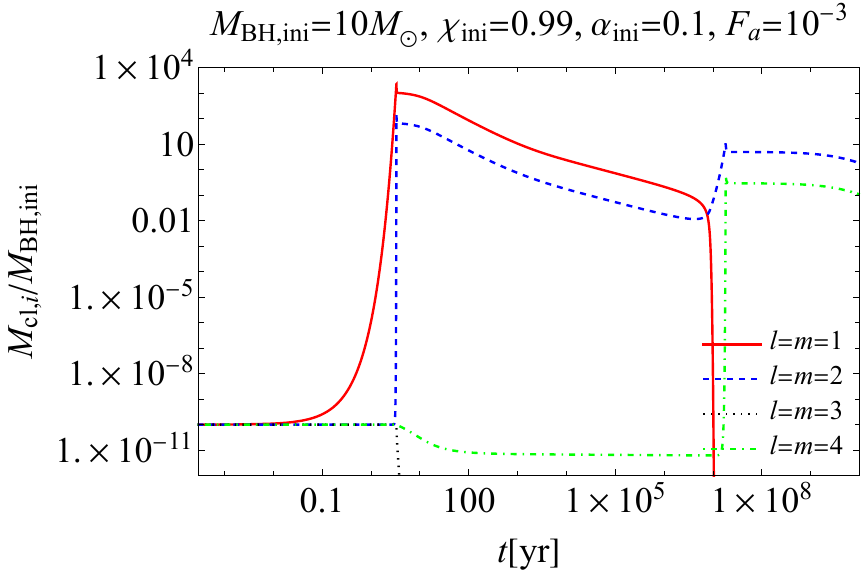}
  \includegraphics[keepaspectratio, scale=0.55]{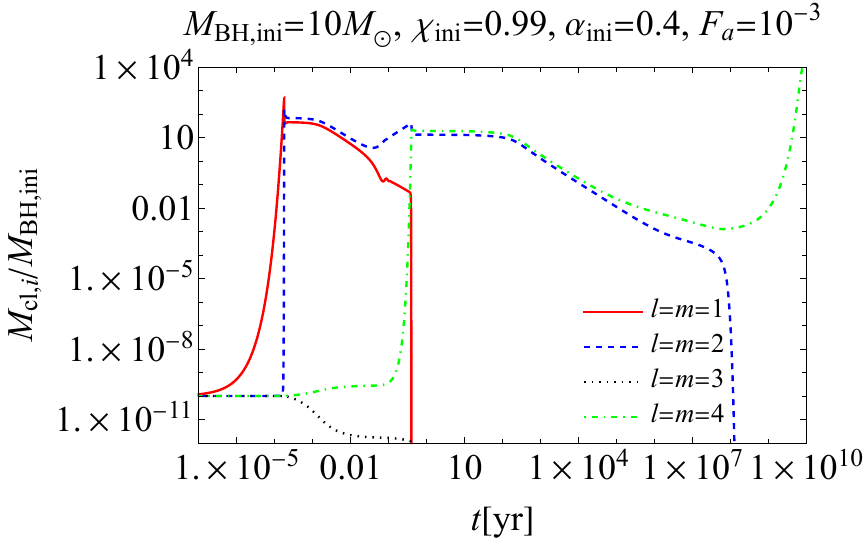}
 \caption{The same figure as Fig.~\ref{fig:examplecloud} but with $\alpha_{\rm ini} = 0.1$(top) and $0.4$(bottom), respectively.}
 \label{fig:examplecloudmu}
\end{figure}

\begin{figure}[tbp]
 \centering
 \includegraphics[keepaspectratio, scale=0.7]{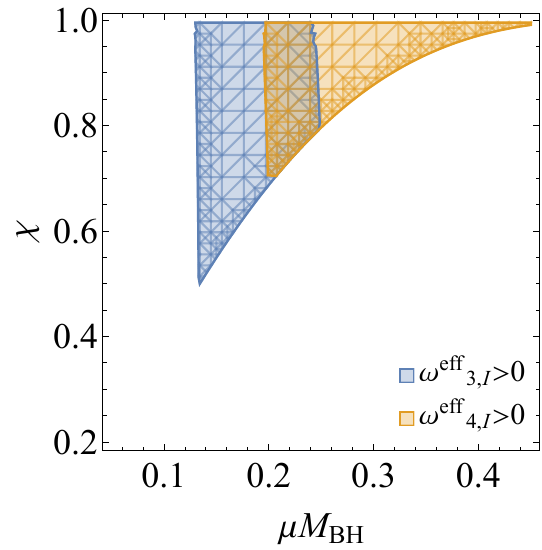}
 \caption{The parameter region where the effective growth rate for the $l=m=3$ ($\omega^{\rm eff}_{3,I}$) and the $l=m=4$ ($\omega^{\rm eff}_{4,I}$) are positive in the $(\mu M_{\rm BH}, \chi)$-plane. The blue and orange region corresponds to $\omega^{\rm eff}_{3,I} > 0$ and $\omega^{\rm eff}_{4,I} > 0$, respectively. In the most of the overlapping region, $\omega^{\rm eff}_{3,I} > \omega^{\rm eff}_{4,I}$ holds.}
 \label{fig:wefflm34}
\end{figure}

We now change the axion mass $\mu$ while the value of $F_a$ is fixed as $10^{-3}$. Figure~\ref{fig:examplecloudmu} shows the cases with $\mu M_{\rm BH,ini} = 0.1$ and $0.4$, which correspond to $\mu \sim 1.3 \times 10^{-12}$ and $5.2 \times 10^{-12}${\rm eV}, respectively. The behavior is similar to the case with $\mu M_{\rm BH,ini} = 0.2$, except that the $l=m=3$ cloud is no longer excited before the depletion of the $l=m=1$ cloud. Since the simultaneous excitation of higher multipole modes will be important when we consider gravitational wave signals (Sec.~\ref{sec:5b}), we discuss the parameter range in which the simultaneous excitation of higher multipole modes is possible. We define the effective growth rate of the $l=m=3$ mode, when the $l=m=1$ and the $l=m=2$ modes are in the quasi-stationary state as
\begin{align}
	\omega_{3,I}^{\rm eff} &\equiv 2\omega_{3,I} + F^{\mathcal{H}}_{113^*}M_{\rm sat,1}^{2} +  F^{\mathcal{H}}_{223^*}M_{\rm sat,2}^{2}  + \cr
 &\left(F^{\mathcal{H}}_{123^*} - F^{\mathcal{I}}_{231^*} \right)M_{\rm sat,1} M_{\rm sat,2}~.
\end{align}
Here, $M_{\rm sat,1}$ and $M_{\rm sat,2}$ are the masses of the $l=m=1$ and $l=m=2$ clouds when the growth is saturated by the self-interaction between the two clouds. We define $\omega_{4,I}^{\rm eff}$ in the same way. The region where $\omega_{3,I}^{\rm eff}$ and $\omega_{4,I}^{\rm eff}$ are positive is shown in Fig.~\ref{fig:wefflm34}.
The excitation of the $l=m=3$ mode is possible when $0.12 \lesssim \mu M_{\rm BH} \lesssim 0.25$, and the excitation of the $l=m=4$ mode is possible when $0.20 \lesssim \mu M_{\rm BH} \lesssim 0.45$. In the overlapping regime, the one with the larger effective growth rate will be excited, and the other mode is depleted by the processes $F^{\mathcal{I}}_{341^*}$ and $F^{\mathcal{I}}_{342^*}$. In the most of the overlapping region, the effective growth rate for $l=m=3$ is larger than that for $l=m=4$.

\begin{figure}[tbp]
 \centering
  \includegraphics[keepaspectratio, scale=0.47]{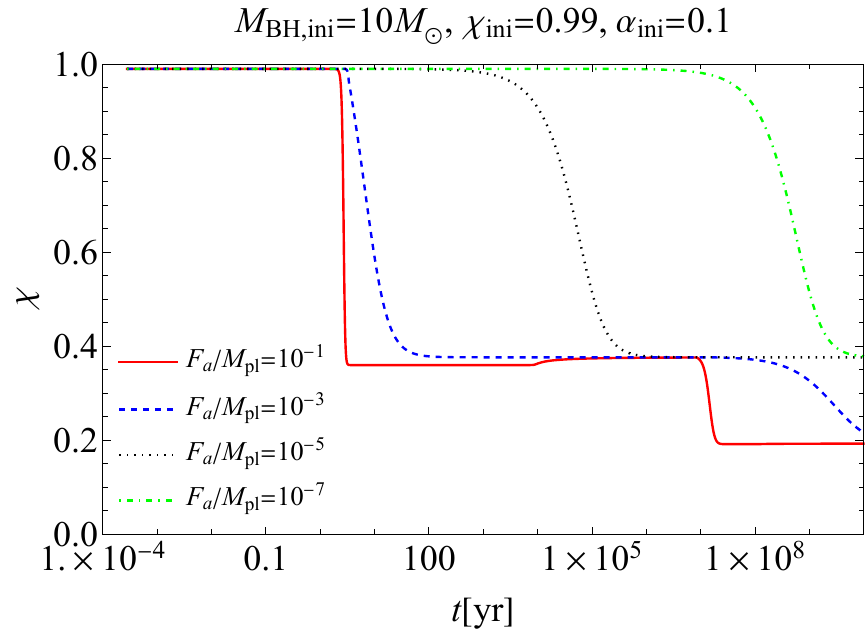}
 \includegraphics[keepaspectratio, scale=0.47]{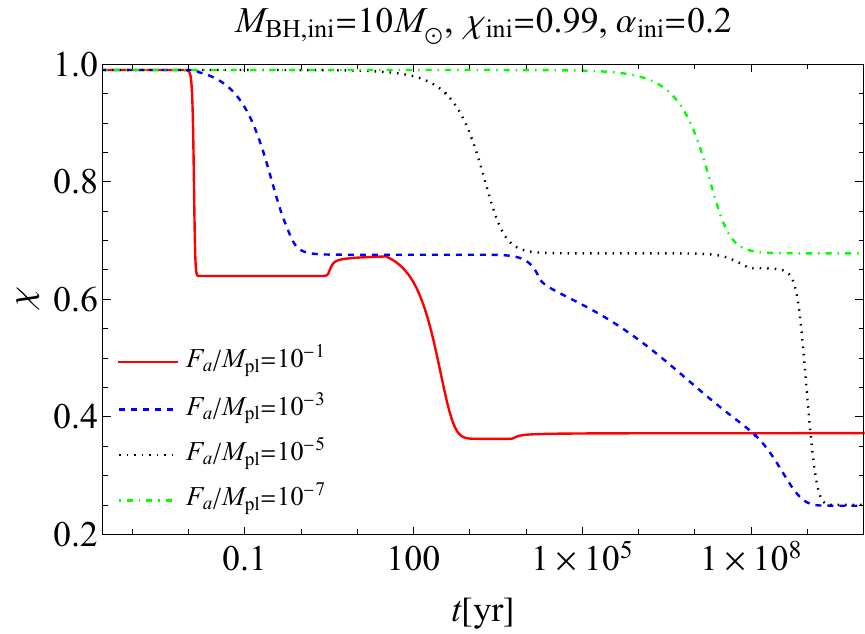}
  \includegraphics[keepaspectratio, scale=0.47]{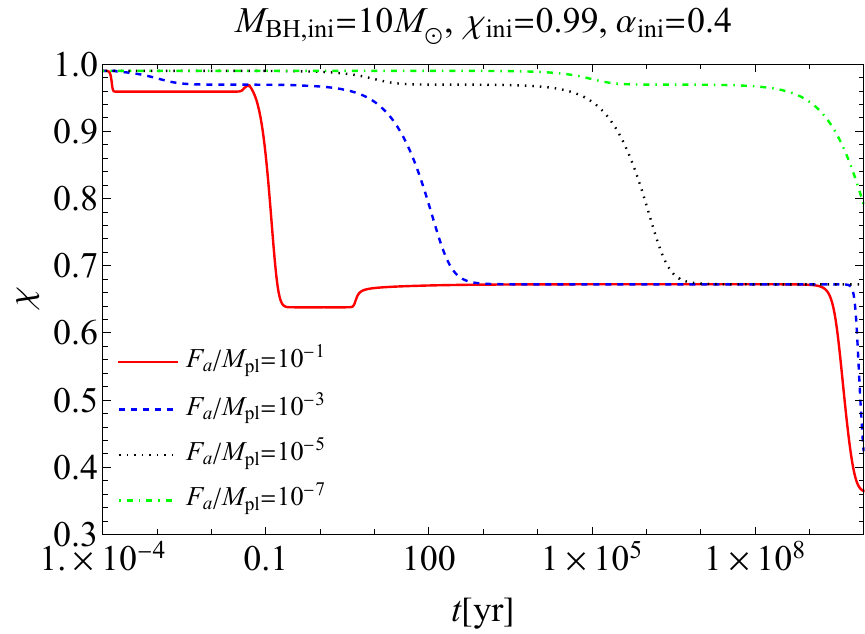}
 \caption{The time evolution of the black hole spin parameters $\chi$ as functions of time. (Top) The red solid, blue dashed, black dotted, and green dash-dotted curves correspond to the decay constant $F_a=10^{-1}, 10^{-3}, 10^{-5}$ and $10^{-7}$, respectively.  The axion mass is chosen to satisfy $\alpha_{\rm ini} = \mu M_{\rm BH, ini} = 0.1$. (Middle) The same figure as the top one but with $\alpha_{\rm ini} = 0.2$. (Bottom) The same figure as the top one but with $\alpha_{\rm ini} = 0.4$.}
 \label{fig:examplespinevol}
\end{figure}

Finally, let us comment on the evolution of the black hole spin. In Fig.~\ref{fig:examplespinevol}, we show the evolution of the black hole spin with various $\mu$ and $F_a$. Since the superradiant rate depends on the axion mass $\mu$, the time when the significant spin down occurs depends on $\mu$. In addition, the spin down time scale becomes longer when one decreases $F_a$. This is evident from the equation governing the angular momentum evolution of the black hole, Eq.~\eqref{eq:evolJBH}. The spin down is initially driven by the $l=m=1$ mode and later by higher multipole modes because of the difference of the threshold spin and the hierarchy in the rate of the superradiant instability (see right panel of Fig.~\ref{fig:omega}). However, we note that the spin down due to the higher multipole modes starts to function much earlier compared with the non-interacting case as long as the decay constant satisfies $F_a \lesssim 10^{-1}$, since the growth of higher multipole modes is accelerated by the mode coupling due to the self-interaction. The spin up observed in the figure is due to the fallback of the lower multipole modes to the black hole.

\section{Observable signals}\label{sec:5}

\begin{figure}[t]
\centering
\includegraphics[keepaspectratio, scale=0.55]{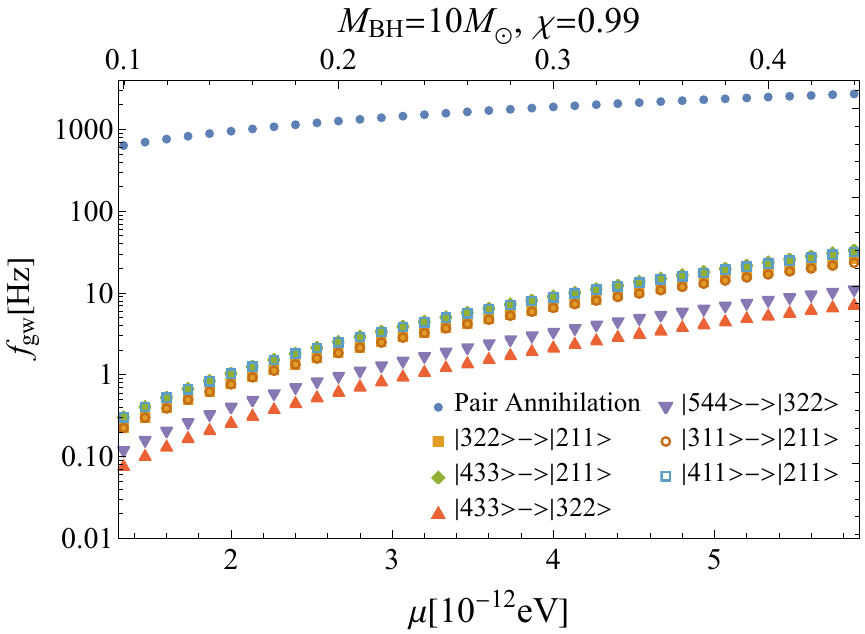}
 \caption{Frequencies of gravitational waves $f_{\rm gw}$ emitted from the axion condensates as functions of the axion mass $\mu$. The blue circle corresponds to the pair annihilation of the $l=m=1$ fundamental mode. The yellow square, green diamond, red triangle, purple reversed triangle, brown open circle, and light blue open square correspond to the level transition signal of $\ket{322}\to\ket{211}$, $\ket{433}\to\ket{211}$,  $\ket{433}\to\ket{322}$,  $\ket{544}\to\ket{322}$,  $\ket{311}\to\ket{211}$, $\ket{411}\to\ket{211}$, respectively. Here, $\ket{nlm}$ corresponds to the mode labeled by $n,l,$ and $m$. Although overtones do not appear in our calculation, we show them for reference. The mass and spin of the black hole are fixed at $M_{\rm BH}=10M_{\odot}$ and $\chi = 0.99$.}
 \label{fig:gwa099freq}
\end{figure}

\subsection{Gravitational waves}

\subsubsection{Calculation of the gravitational waves}

Here, we consider gravitational waves emitted by the axion condensate.  
Gravitational waves are generated from the time-dependent part of the axion energy-momentum tensor $T_{\mu\nu}(\phi)$. When the multiple modes are excited simultaneously, the axion field is given in the form~\footnote{Note that we normalized the axion field by $F_a$, so $\phi$ is non-dimension. In addition, $M_{{\rm cl},i}$ is normalized by $F_a^2 M_{\rm BH}$, so also nondimension.}
\begin{align}
	\phi = \sum_{i}\sqrt{\frac{M_{{\rm cl},i}}{M_{\rm BH}}} e^{-i(\omega_i t - m_i\varphi)}R_{l_im_i\omega_i}S_{l_im_i\omega_i} + {\rm c.c.}~.
\end{align}
Since $T_{\mu\nu}$ is quadratic in $\phi$, the time dependent pieces behaves as $\propto e^{-i(\omega_i + \omega_j) t}$ or $e^{-i(\omega_i - \omega_j) t}$. 
Namely, the frequencies of gravitational waves are either the sum or the difference of the frequencies of two clouds. The former is called the ``pair annihilation signal," while the latter is the ``level transition signal."  The actual frequencies of the gravitational waves for the $10M_{\odot}$ black hole mass with spin $\chi = 0.99$ are shown in Fig.~\ref{fig:gwa099freq}. The pair annihilation signal has a much higher frequency ($\sim 1 {\rm kHz}$) than the level transition signal ($\sim 1 {\rm Hz}$).

The energy flux and the amplitude of gravitational waves can be calculated by solving the Teukolsky equation. We briefly review the calculation of gravitational waves from the axion condensate. The details of the calculations can be found in Refs.~\cite{Yoshino:2013ofa, Guo:2022mpr, Siemonsen:2022yyf}. 

The energy flux $F_{\rm gw}$ and the amplitude $h_{\rm gw}$ of a monochromatic gravitational wave 
with the angular frequency $\omega$ in a distant region from the black hole are described in terms of the Newmann-Penrose variable $\psi_4$ as~\cite{Teukolsky:1973ha}
\begin{align}
	F_{\rm gw} &= \lim_{r\to \infty}\frac{r^2}{4\pi \omega^2} \int d\Omega\ |\psi_4|^2~,\\
	h_{\rm gw} &= \frac{2}{\omega^2}|\psi_4|~.
\end{align}
When the source of the gravitational wave has the form $e^{-i (\omega t - m\varphi)}$, the asymptotic form of the $\psi_4$ is shown to be
	\begin{align}
		\psi_4 &\to \frac{1}{r}e^{-i\omega t + i m\varphi} e^{i \omega r}\sum_{l=2}^{\infty} Z_l^{\rm out}{}_{-2}S_{lm\omega}~, & (r \to& \infty)~,
	\end{align}
	where the amplitude is given by
	\begin{align}
		Z^{\rm out}_{l} &= \frac{1}{{}_{-2}W_{lm\omega}}\int_{r_+}^{\infty} dr' \ \Delta^{-2}{}_{-2}R^{\rm in}_{lm\omega}(r') {}_{-2}T_{lm\omega}(r')~.
	\end{align}
	Here, ${}_{-2}S_{lm\omega}$ is the spin-weighted spheroidal harmonics with the spin weight $s = -2$, ${}_{-2}R_{lm\omega}^{\rm in}$ is the solution to the radial Teukolsky equation with the ingoing boundary condition at the event horizon, and ${}_{-2}W_{lm\omega}$ is the Wronskian of the radial Teukolsky equation. The function ${}_{-2}T_{lm\omega}$ is the source term of the radial Teukolsky equation, related to the energy-momentum tensor of the axion. The precise form of ${}_{-2}T_{lm\omega}$ can be found in Ref.~\cite{Teukolsky:1973ha}. Note that ${}_{-2}T_{lm\omega}$ depends on the mass of the cloud and the decay constant as $\propto F_a^2 \sqrt{M_{{\rm cl}, i}M_{{\rm cl},j}}$, since the energy-momentum tensor is quadratic in $\phi$. Therefore, the energy flux and the amplitude scale as
\begin{align}
	F_{\rm gw} \propto & F_a^4 M_{{\rm cl},i}M_{{\rm cl},j}~,\\
	\label{eq:scaleamp}
	h_{\rm gw} \propto & F_a^2 \sqrt{M_{{\rm cl},i}M_{{\rm cl},j}}~.
\end{align}

\begin{figure*}[t]
 \centering
 \includegraphics[keepaspectratio, scale=0.55]{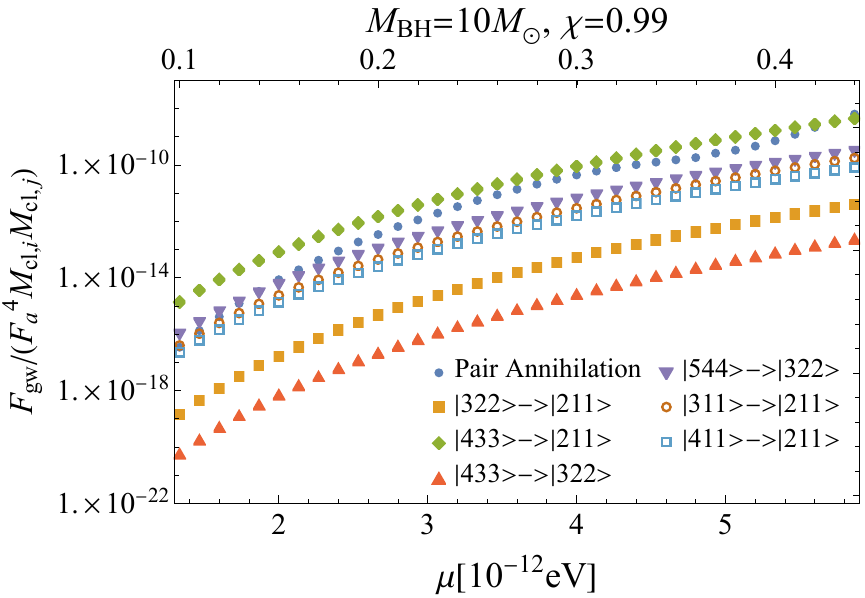}
  \includegraphics[keepaspectratio, scale=0.55]{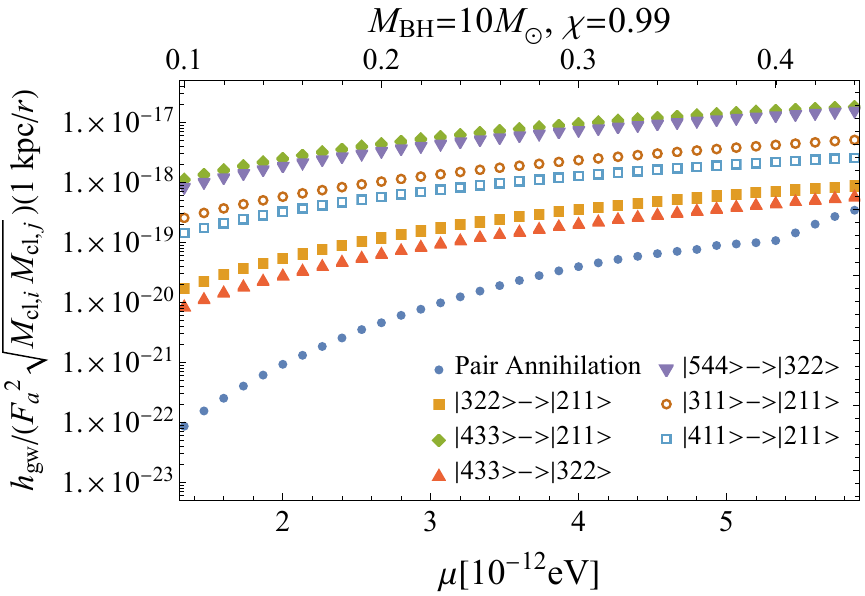}
 \caption{(Left) 
 The energy flux of the gravitational waves from various processes. Each point corresponds to the same process in Fig.~\ref{fig:gwa099freq}.
 (Right) The amplitude of the gravitational waves from various processes. Again, each point corresponds to the same process in Fig.~\ref{fig:gwa099freq}. The black hole is placed at $r =1{\rm kpc}$.}
 \label{fig:gwa099}
\end{figure*}
	
In the left panel of Fig.~\ref{fig:gwa099}, we show the energy flux of gravitational waves from a black hole with the mass $10M_{\odot}$ and the spin $\chi = 0.99$. In terms of the energy flux, the pair annihilation signal of the $l=m=1$ cloud and the level transition signal of the transition from the $l=m=3$ mode to the $l=m=1$ mode dominate if the masses of all clouds are assumed to be the same. The energy dissipation of the cloud due to the pair annihilation is much smaller than the energy gain by the superradiance~\cite{Yoshino:2013ofa}, which justifies neglecting the gravitational wave emission when we consider the evolution of the condensate.

In the right panel of Fig.~\ref{fig:gwa099}, we show the amplitudes of respective gravitational waves. We observe the hierarchy among the amplitudes generated by different processes. First of all, the amplitude of the pair annihilation signal is the smallest. This is the case because the wavelength of the gravitational wave ($\sim 1/\mu$) is much shorter than the size of the condensate ($\sim 1/\mu/(\mu M)$) in the situation we are considering. In contrast, the level transition signal is not suppressed since the wavelength is much longer ($\sim 1/\mu/(\mu M)^2$). The size of the level transition signal is the largest for the transition with $|m_i - m_j| = 2$, followed by $|m_i - m_j| = 0$, and then by $|m_i - m_j| = 1$. The transition with $|m_i - m_j| = 1$ is suppressed because the transition cannot occur through the mass quadrupole radiation because of the parity argument. The relative suppression of the signal with $|m_i - m_j| = 0$  to the one with $|m_i - m_j| = 2$ is due to the axisymmetry of the source for the case of $|m_i - m_j| = 0$. Note that the amplitude of the gravitational waves does not change significantly as one changes the black hole spin. This is because the configuration of the cloud does not depend on the black hole spin prominently.

 Other than the signals presented in Fig.~\ref{fig:gwa099}, it is possible to consider pair annihilation between different modes. However, one can confirm that this ``cross annihilation" process produces much smaller gravitational waves since it starts with the octopole radiation rather than the quadrupole one.

\begin{figure}[t]
 \centering
 \includegraphics[keepaspectratio, scale=0.55]{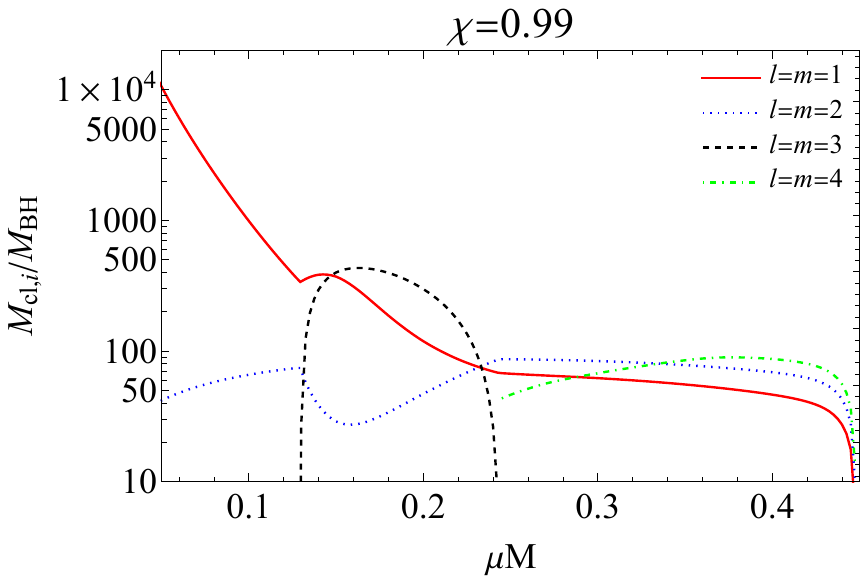}
  \includegraphics[keepaspectratio, scale=0.55]{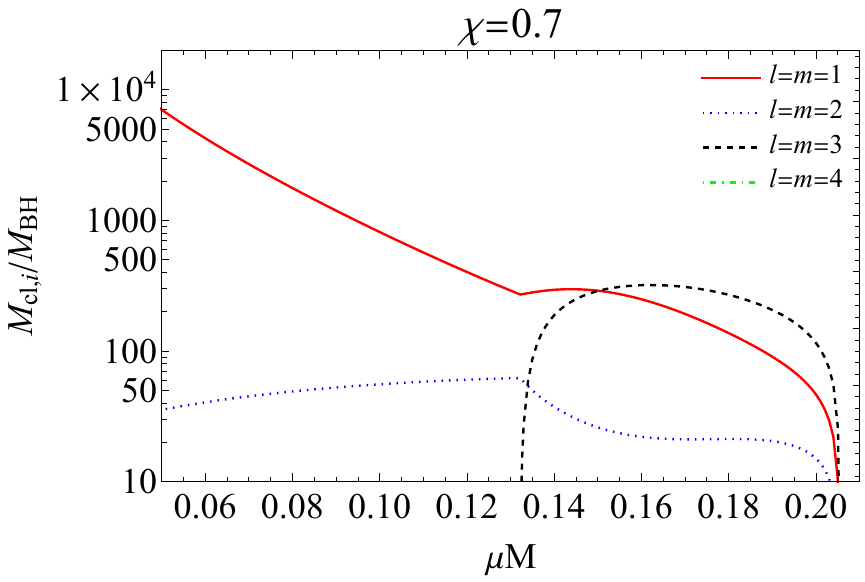}
 \caption{Normalized masses of the axion clouds in the quasi-stationary configuration. The red solid, blue dotted, black dashed, and green dot-dashed curves correspond to the mass of the $l=m=1, 2, 3$ and $4$ cloud when the condensate is in the quasi-stationary configuration. The spin of the black hole is fixed at $\chi = 0.99$(top) and $0.7$(bottom), respectively.}
 \label{fig:satcomp}
\end{figure}

To summarize, we expect a much larger signal than the one previously considered, when
there simultaneously exist two modes whose magnetic quantum numbers are different by two, {\it e.g.}, the $l=m=1$ and $l=m=3$ modes or the $l=m=2$ and $l=m=4$ modes. Let us estimate the amplitude of the gravitational wave emitted during the quasi-stationary state that is found in the previous section. In Fig.~\ref{fig:satcomp}, we show the normalized masses of the clouds $M_{{\rm cl}, i}$ when they settle into the quasi-stationary state. We observe that when the $l=m=3$ cloud is excited, the relation $M_{{\rm cl},1} \sim M_{{\rm cl},3} \sim \mathcal{O}(10^2)M_{\rm BH}$ holds regardless of the spin. The maximum of $\sqrt{M_{{\rm cl},1} M_{{\rm cl},3}}$ is $\sim 380M_{\rm BH}$ for $\chi = 0.99$ and $\sim 290 M_{\rm BH}$ for $\chi = 0.7$, around $\mu M_{\rm BH} \sim 0.15$. From the numerical calculation, the value of  $h_{\rm gw}$ around $\mu M_{\rm BH} \sim 0.15$ normalized as in Fig~\ref{fig:gwa099} is determined to be $\sim 3 \times 10^{-18}$ for both  $\chi = 0.99$ and $0.7$. Thus, the maximum amplitude of the level transition signal from the $l=m=3$ mode to the $l=m=1$ mode is given by 
\begin{widetext}
\begin{align}\label{eq:estimateamp31}
	h_{\rm gw} \sim 
	\begin{cases}
	1.0 \times 10^{-21} \left(\frac{1{\rm kpc}}{r}\right)\left(\frac{M_{\rm BH}}{10M_{\odot}}\right) \left(\frac{F_a}{10^{-3}}\right)^2 \left(\frac{\sqrt{M_{{\rm cl},1}M_{{\rm cl},3}}}{380M_{\rm BH}}\right)~, & (\chi = 0.99,~\mu M_{\rm BH} \sim 0.15)\,,\\
	7.9 \times 10^{-22} \left(\frac{1{\rm kpc}}{r}\right)\left(\frac{M_{\rm BH}}{10M_{\odot}}\right) \left(\frac{F_a}{10^{-3}}\right)^2 \left(\frac{\sqrt{M_{{\rm cl},1}M_{{\rm cl},3}}}{290M_{\rm BH}}\right)~, & (\chi = 0.7,~\mu M_{\rm BH} \sim 0.15)\,.\\
	\end{cases}
\end{align}
\end{widetext}
For a black hole with mass $10M_{\odot}$, the frequency of this gravitational wave is around $1{\rm Hz}$, which is in the target frequencies of the future gravitational wave detectors such as DECIGO~\cite{Kawamura:2011zz}, atomic interferometers~\cite{Proceedings:2023mkp}, TOBA~\cite{Shimoda:2018uiv}, TianGo~\cite{2020PhRvD.102d3001K}, DO~\cite{2020CQGra..37u5011A}, AMIGO~\cite{2020IJMPD..2940007N}, and LGWA~\cite{Ajith:2024mie}. The amplitude does not decrease significantly even if the spin of the black hole is reduced.

\begin{figure}[t]
 \centering
 \includegraphics[keepaspectratio, scale=0.55]{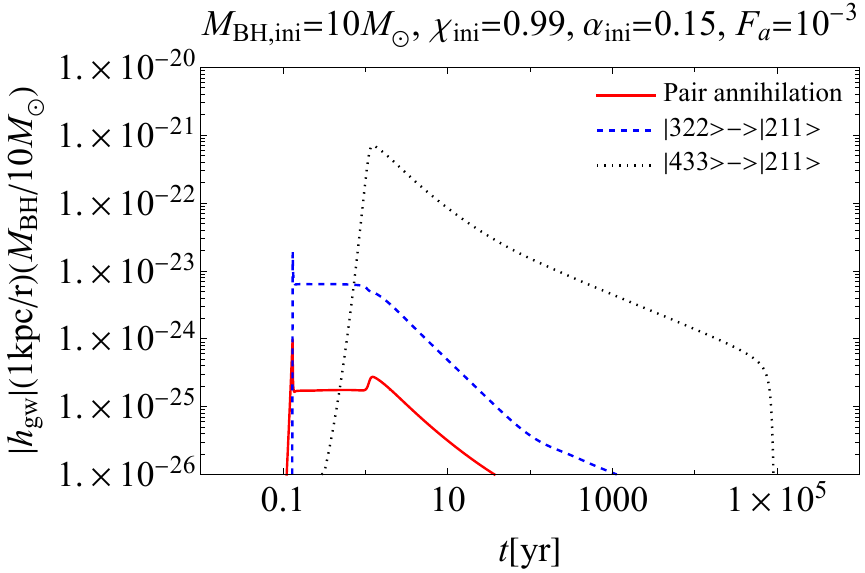}
 \caption{An example of the time evolution of the gravitational wave amplitude from an axion condensate. The red solid, blue dashed, and black dotted curves correspond to the time dependence of the gravitational wave amplitudes for the pair annihilation signal from the $l=m=1$ cloud, the level transition signals from the $l=m=2$ cloud to the $l=m=1$ cloud, and from the $l=m=3$ cloud to the $l=m=1$ cloud,  respectively. We fix the initial black hole mass and spin at $10 M_\odot$ and $0.99$. The axion mass and decay constant are $\mu M_{\rm BH,ini} = 0.15$ and $F_a = 10^{-3}$. The gravitational wave amplitude is estimated from the calculation setting the black hole spin to $\chi = 0.99$.}
 \label{fig:gwampevol}
\end{figure}

The amplitudes of gravitational waves from other processes can be estimated similarly. The maximum amplitude is given by
\begin{widetext}
\begin{align}\label{eq:estimateampPA}
	h_{\rm gw} \sim 
	\begin{cases}
	1.0\times 10^{-24} \left(\frac{1{\rm kpc}}{r}\right)\left(\frac{M_{\rm BH}}{10M_{\odot}}\right) \left(\frac{F_a}{10^{-3}}\right)^2 \left(\frac{M_{{\rm cl},1}}{36 M_{\rm BH}}\right)~, & (\chi = 0.99,~\mu M_{\rm BH} \sim 0.42)~,\\
	2.1 \times 10^{-25} \left(\frac{1{\rm kpc}}{r}\right)\left(\frac{M_{\rm BH}}{10M_{\odot}}\right) \left(\frac{F_a}{10^{-3}}\right)^2 \left(\frac{M_{{\rm cl},1}}{248 M_{\rm BH}}\right)~, & (\chi = 0.7,~\mu M_{\rm BH} \sim 0.2)~,\\
	\end{cases}
\end{align}
 for the $l=m=1$ pair annihilation, and 
\begin{align}\label{eq:estimateamp21}
	h_{\rm gw} \sim 
	\begin{cases}
	4.0 \times 10^{-23} \left(\frac{1{\rm kpc}}{r}\right)\left(\frac{M_{\rm BH}}{10M_{\odot}}\right) \left(\frac{F_a}{10^{-3}}\right)^2 \left(\frac{\sqrt{M_{{\rm cl},1}M_{{\rm cl},2}}}{54M_{\rm BH}}\right)~, & (\chi = 0.99,~\mu M_{\rm BH} \sim 0.42)~,\\
	4.8 \times 10^{-24} \left(\frac{1{\rm kpc}}{r}\right)\left(\frac{M_{\rm BH}}{10M_{\odot}}\right) \left(\frac{F_a}{10^{-3}}\right)^2 \left(\frac{\sqrt{M_{{\rm cl},1}M_{{\rm cl},2}}}{64 M_{\rm BH}}\right)~, & (\chi = 0.7,~\mu M_{\rm BH} \sim 0.2)~,\\
	\end{cases}
\end{align}
\end{widetext}
 for the transition from the $l=m=2$ cloud to the $l=m=1$ cloud.
Thus, we find that the transition from the $l=m=3$ cloud to the $l=m=1$ cloud can provide the strongest signal for the GUT scale decay constant.

We show the typical time evolution of the gravitational wave amplitude for these processes in Fig.~\ref{fig:gwampevol}. After the excitation of the $l=m=1$ cloud, the gravitational waves from the pair annihilation in the $l=m=1$ cloud and the level transition between the $l=m=1$ and the $l=m=2$ clouds are radiated. This phase lasts for the time scale of superradiant instability, which is $\lesssim 1{\rm yr}$ in the current choice of the parameters. After a few years, the larger gravitational wave from the level transition between the $l=m=3$ and the $l=m=1$ clouds appears. The amplitude of this gravitational wave decays as the $l=m=1$ cloud depletes due to the spin down. 

\begin{figure*}[tbh]
 \centering
\includegraphics[keepaspectratio, scale=0.42]{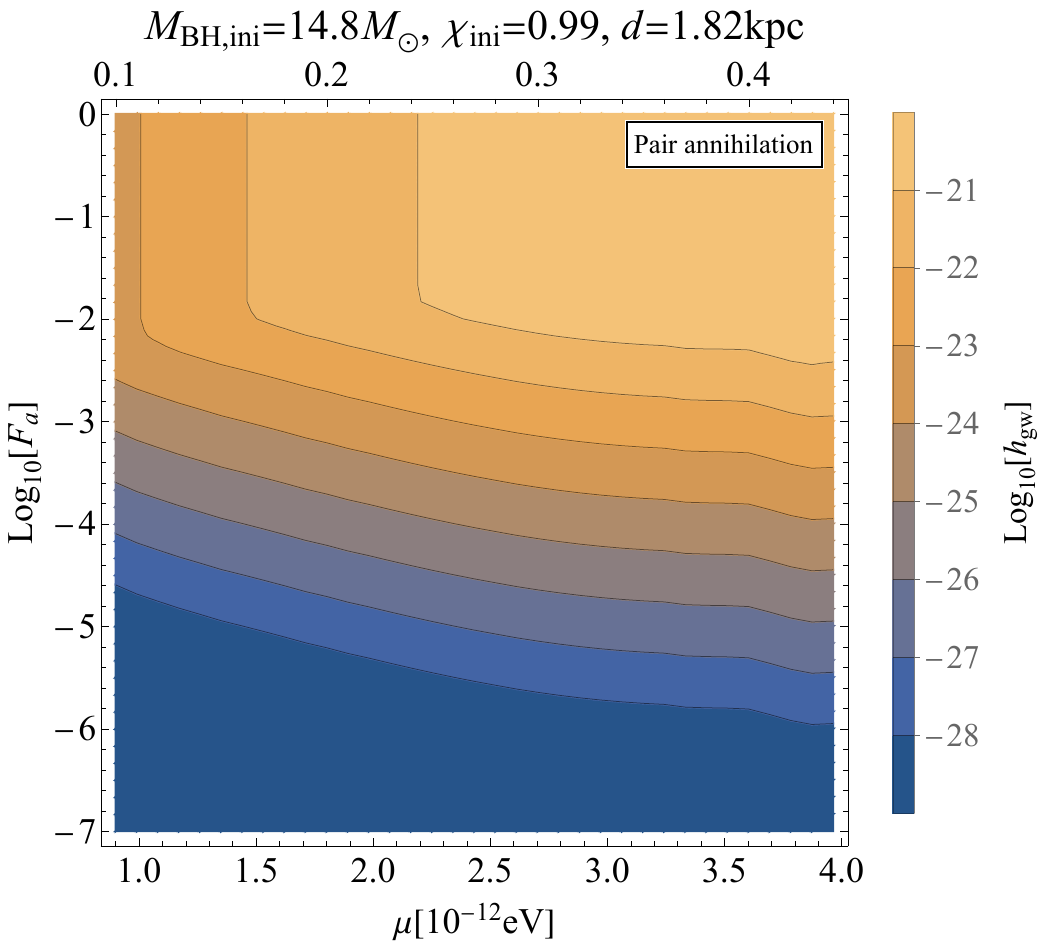}
\includegraphics[keepaspectratio, scale=0.42]{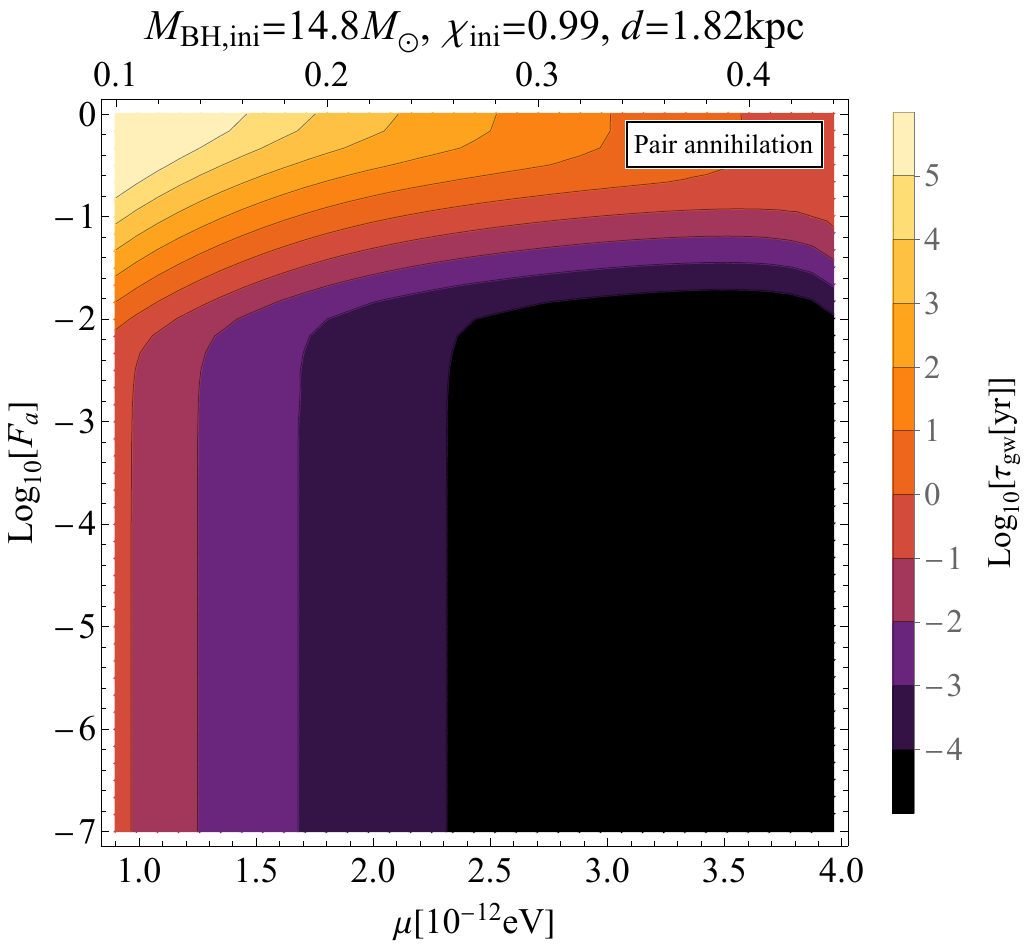}
\includegraphics[keepaspectratio, scale=0.42]{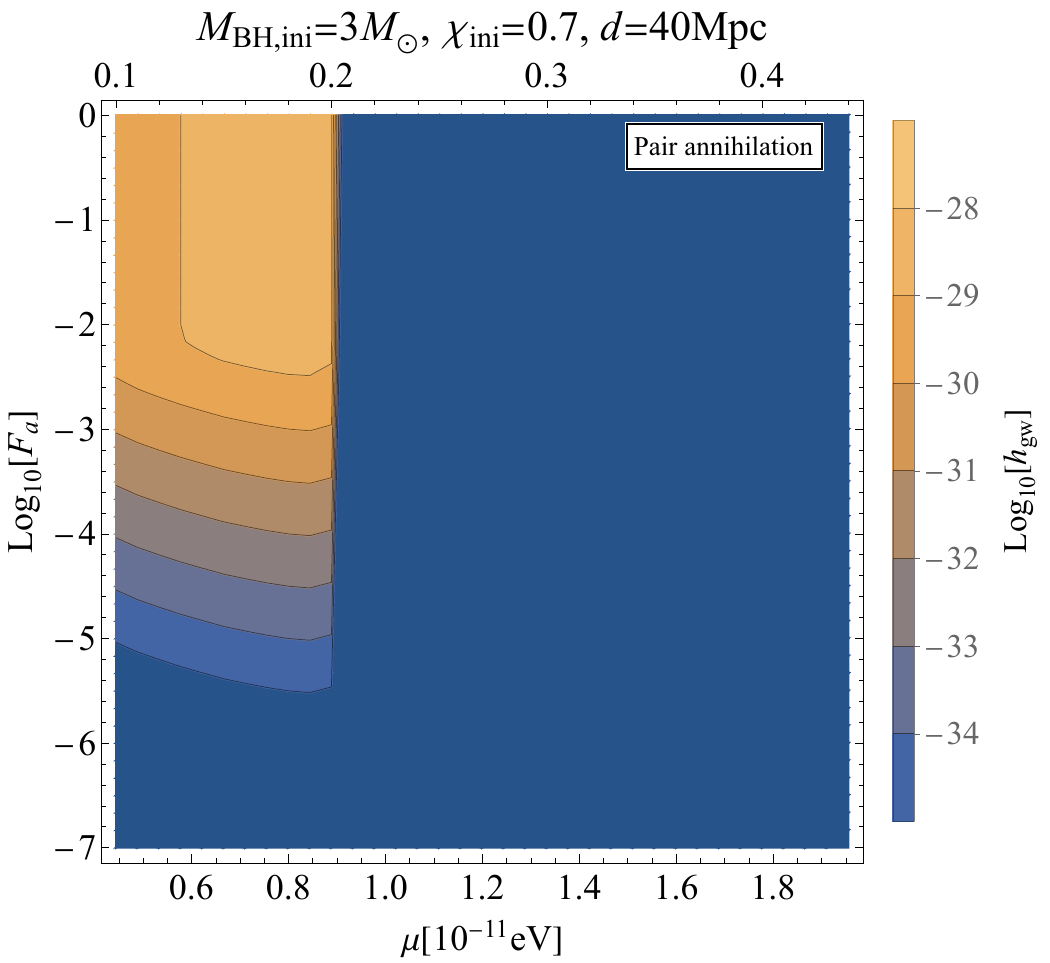}
\includegraphics[keepaspectratio, scale=0.42]{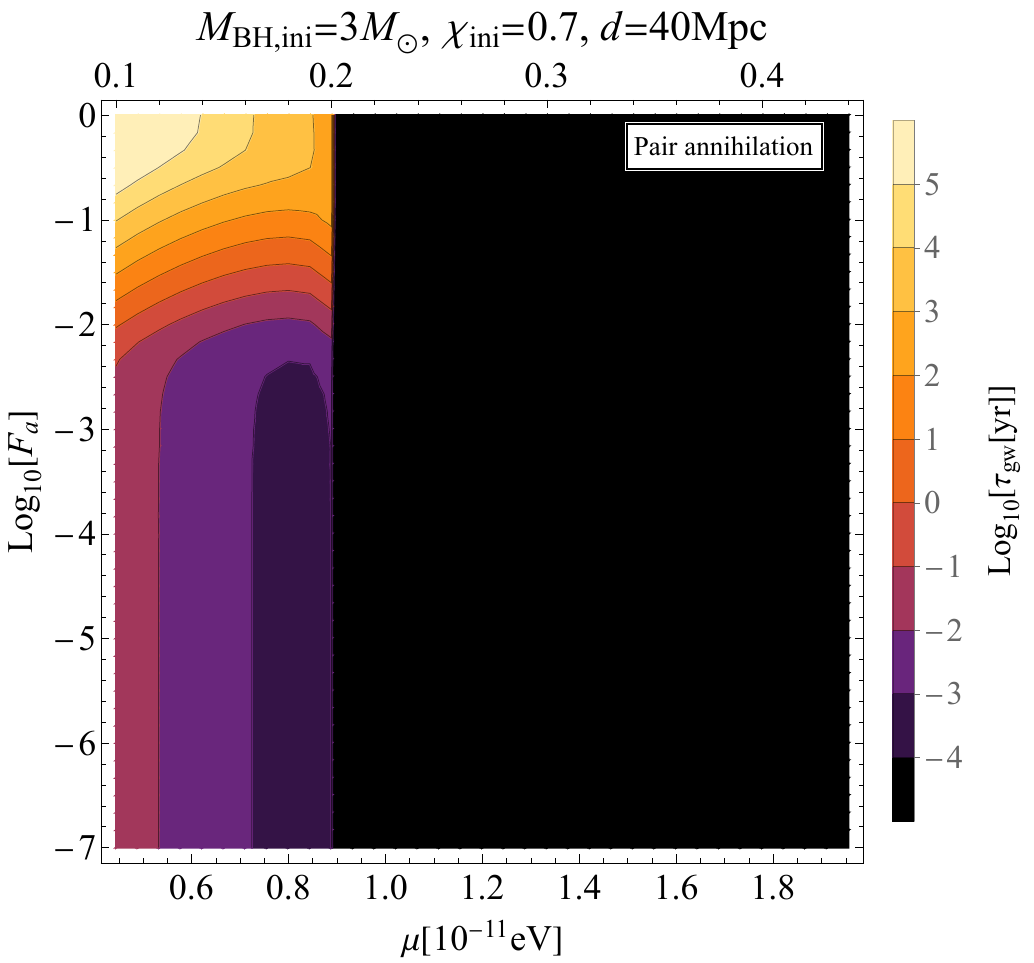}
 \caption{(Upper Left) The dependence of the peak amplitude of the gravitational wave from the pair annihilation signal of the $l=m=1$ cloud on axion mass $\mu$ and decay constant $F_a$. The parameters of the black hole are taken to be the ones similar to Cygnus X-1, $M_{\rm BH,ini} = 14.8 M_{\odot}, \chi_{\rm ini} = 0.99,$ and $d = 1.83{\rm kpc}$. We set a cutoff at $h_{\rm gw} = 10^{-28}$. The upper ticks correspond to the $\mu M_{\rm BH,ini}$. (Upper Right) The dependence of the duration of the pair annihilation signal, estimated by the FWHM, on $\mu$ and $F_a$. The parameters of the black hole are the same as the upper left panel.
 (Lower Left) The similar figure as the upper left figure but with the initial black hole parameters similar to the black hole of GW170817. Namely, $M_{\rm BH,ini} = 3 M_{\odot}, \chi_{\rm ini} = 0.7,$ and $d = 40 {\rm Mpc}$. We take the cutoff to be $h_{\rm gw} = 10^{-34}$. (Lower Right) The same figure as the upper right figure but with the same parameters as the lower left figure. }
 \label{fig:gwpairann}
\end{figure*}

\subsubsection{Dependence on the axion parameter}

In the previous section, we estimated the amplitudes of gravitational waves for the GUT scale decay constant. However, the result does not scale simply as one changes the parameters, $\mu$, $F_a$, the initial black hole mass $M_{\rm BH,ini}$ and the spin $\chi_{\rm BH, ini}$. 
Recall that the dependence of the evolution on these parameters is complicated  (see Sec.~\ref{sec:4A}). Here, we investigate how the peak amplitude $h_{\rm peak}$ and the duration of the signal $\tau_{\rm gw}$ depend on the parameters mentioned above by solving the evolution equation directly for various cases. We estimate the duration by the full width at the half-maximum (FWHM). 

As the initial parameters of the black hole, we choose the one similar to the Cygnus X-1~\cite{2011ApJ...742...85G, 2011ApJ...742...83R}, which has been proposed to be an interesting target for the axion search~\cite{Yoshino:2014wwa, Sun:2019mqb}, or to black holes produced from the compact binary coalescence such as GW170817~\cite{LIGOScientific:2017vwq}. The corresponding parameters are $M_{\rm BH,ini} = 14.8M_{\odot}, \chi_{\rm ini}=0.99,$ and $d = 1.82{\rm kpc}$ for Cygnus X-1-like black holes and  $M_{\rm BH,ini} = 3 M_{\odot}, \chi_{\rm ini}=0.7,$ and $d = 40{\rm Mpc}$ for the black holes from the compact binary coalescence, respectively. Here, $d$ is the distance to the black hole.

In Fig.~\ref{fig:gwpairann}, we show the peak amplitude and the duration of the gravitational wave from the pair annihilation of the $l=m=1$ cloud on the $(\mu,F_a)$ plane. The amplitude of the gravitational wave decreases as one decreases $\mu$ or $F_a$. The amplitude dependence on $\mu$ is evident from Fig.~\ref{fig:gwa099}. 
The dependence on $F_a$ is the consequence of Eq.~\eqref{eq:scaleamp}. The cut-off around $\mu M_{\rm BH,ini} \sim 0.2$ observed in the lower left panel of Fig.~\ref{fig:gwpairann} is reflecting the superradiant condition for $\chi = 0.7$ (see Eq.~\eqref{eq:SRcond}). Beyond this axion mass, the $l=m=1$ cloud will never be excited; thus, there is no emission of the pair annihilation gravitational wave.

\begin{figure*}[tb]
 \centering
  \includegraphics[keepaspectratio, scale=0.4]{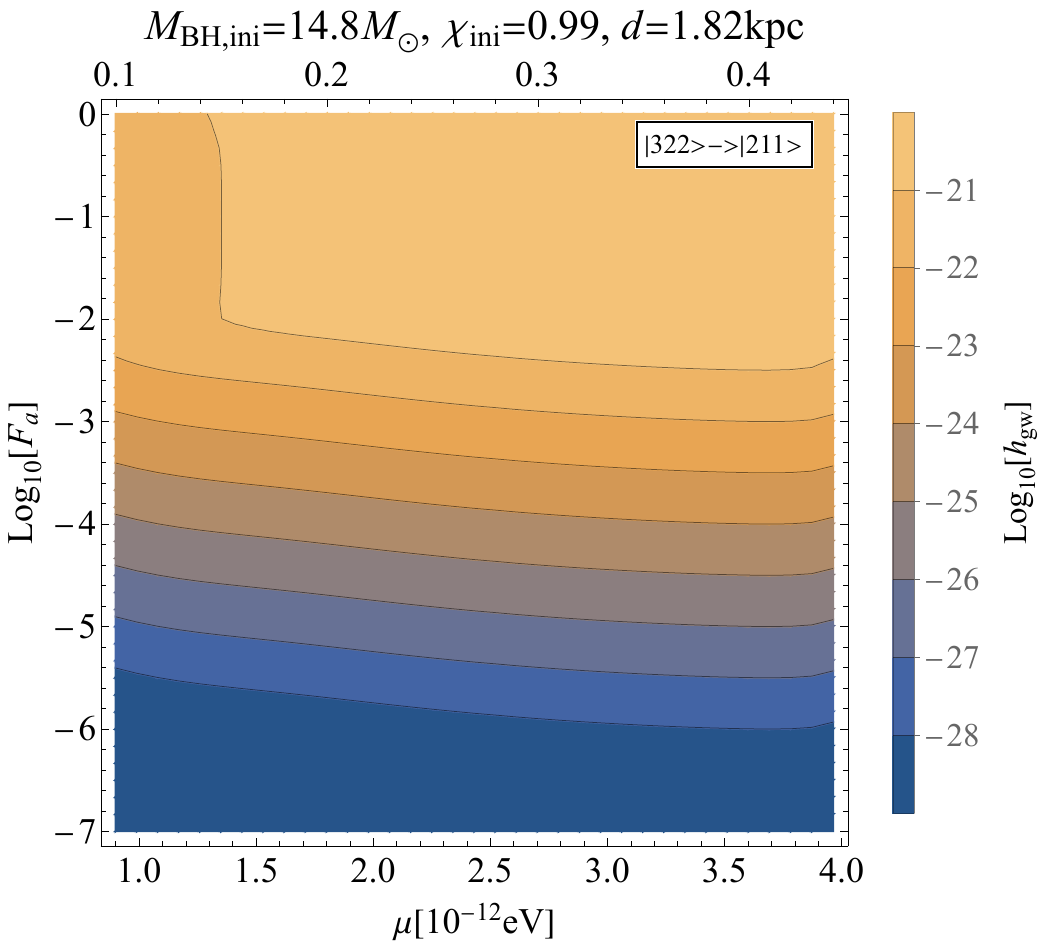}
 \includegraphics[keepaspectratio, scale=0.4]{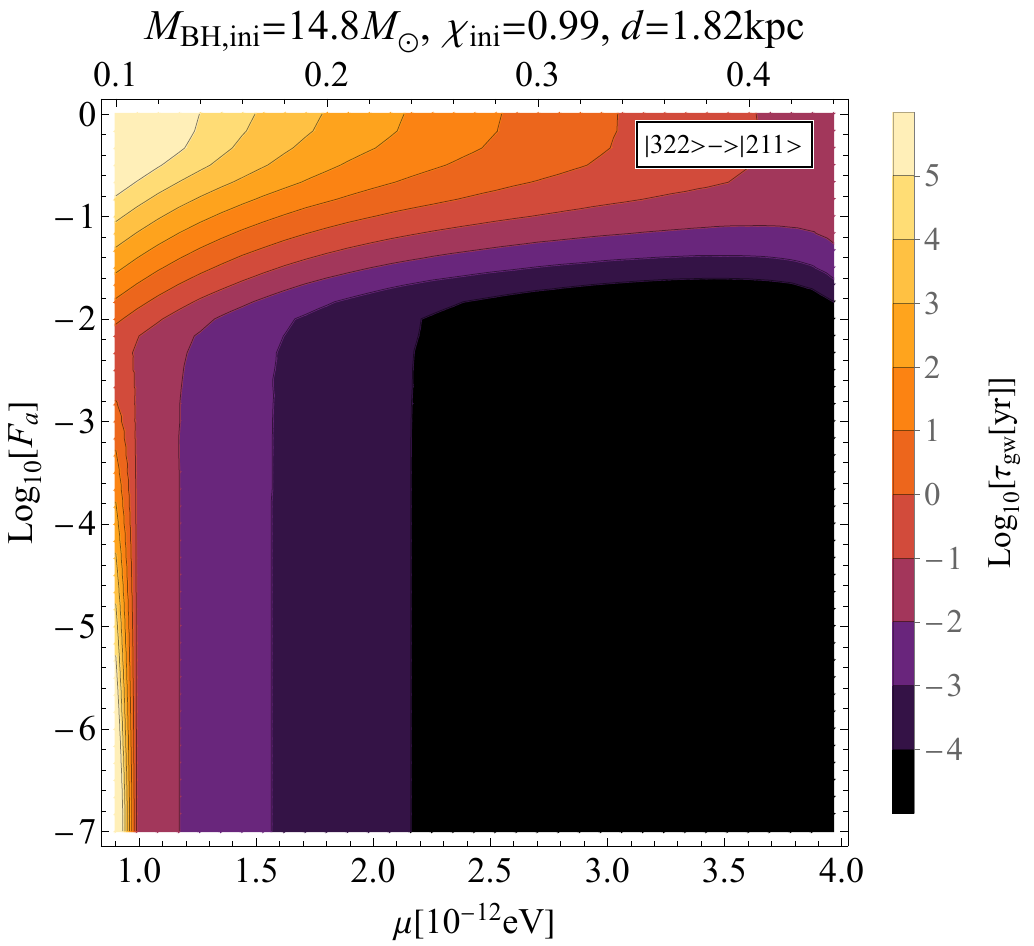}
     \includegraphics[keepaspectratio, scale=0.4]{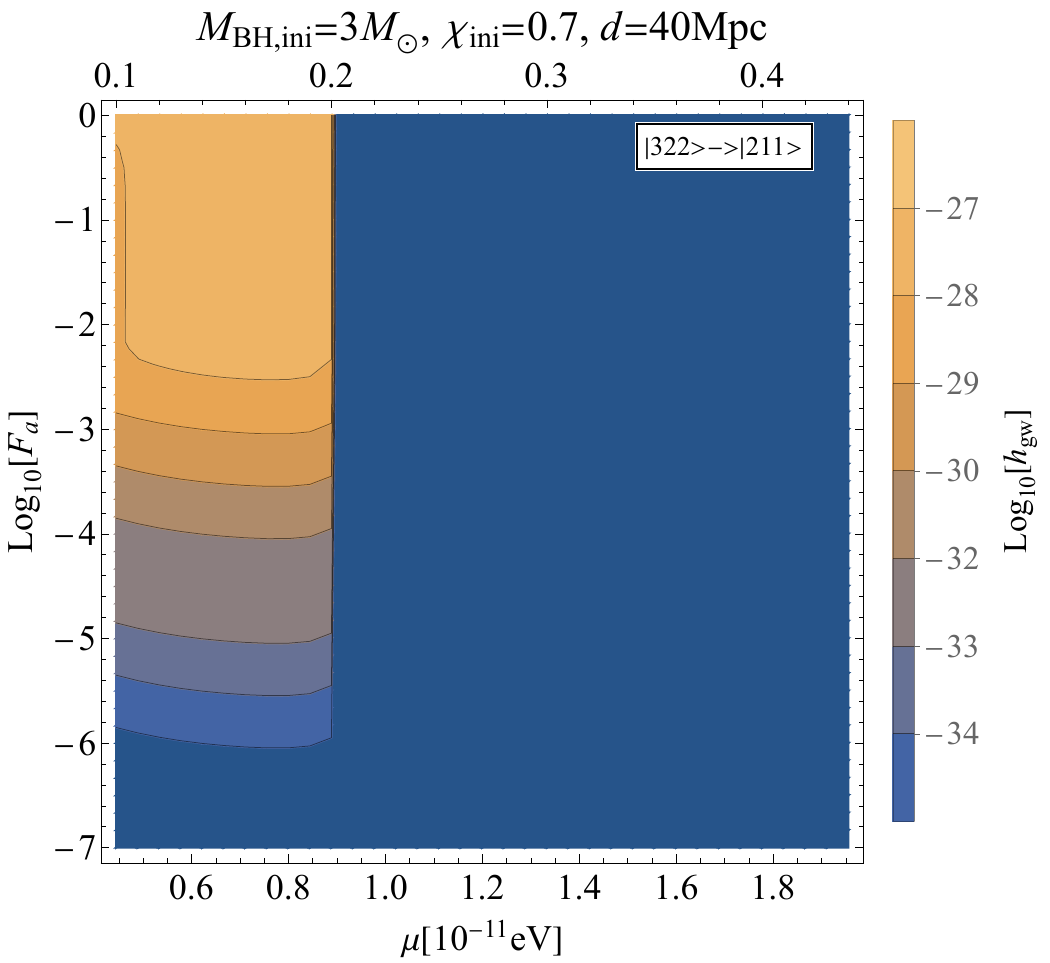}
  \includegraphics[keepaspectratio, scale=0.4]{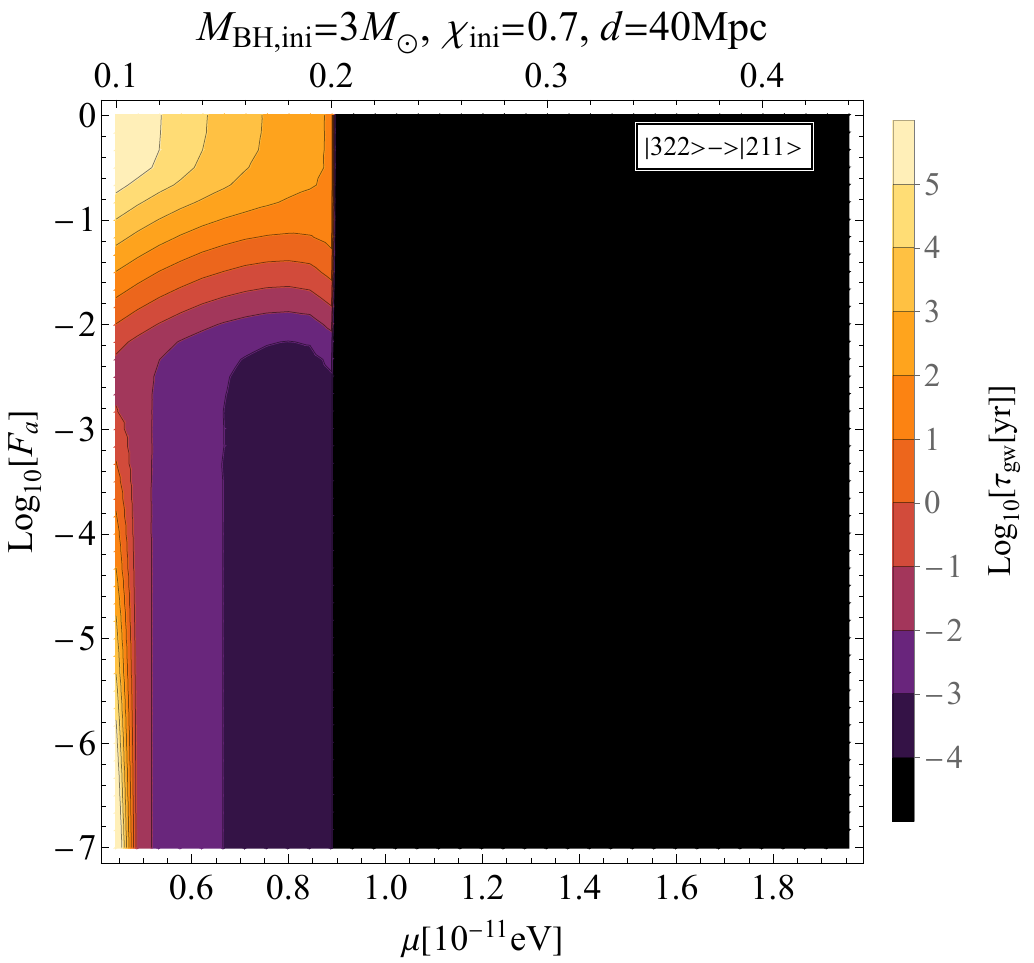}
 \caption{The same figures as Fig.~\ref{fig:gwpairann}, but with the gravitational waves from the level transition between the $l=m=2$ and the $l=m=1$ clouds.}
 \label{fig:gwtrans21}
\end{figure*}

Next, we consider the dependence of the duration on the parameters $\mu$ and $F_a$. For decay constant in the range $F_a \gtrsim 10^{-2}$, the duration becomes longer as one increases $F_a$ or decreases $\mu$. In this range, the duration is determined by the time when the $l=m=1$ cloud decays, which is determined by the spin down of the black hole due to the superradiance assisted by the presence of the $l=m=2$ cloud. Increasing the decay constant leads to relative suppression of the excitation of the $l=m=2$ cloud, which means that the spin down becomes slower. For a smaller axion mass, the superradiant instability rate is reduced, which leads to a slower spin down rate. The duration depends primarily on the axion mass for $F_a \lesssim 10^{-2}$. In this region, the excitation of the $l=m=2$ cloud happens before the spin down to the threshold for the superradiant instability of the $l=m=1$ cloud completes. Duration determined by the FWHM is short since a sharp peak occurs for the amplitude of gravitational waves from the pair annihilation in the middle of the transition process (see Fig.~\ref{fig:gwampevol}).

\begin{figure*}[tb]
 \centering
  \includegraphics[keepaspectratio, scale=0.4]{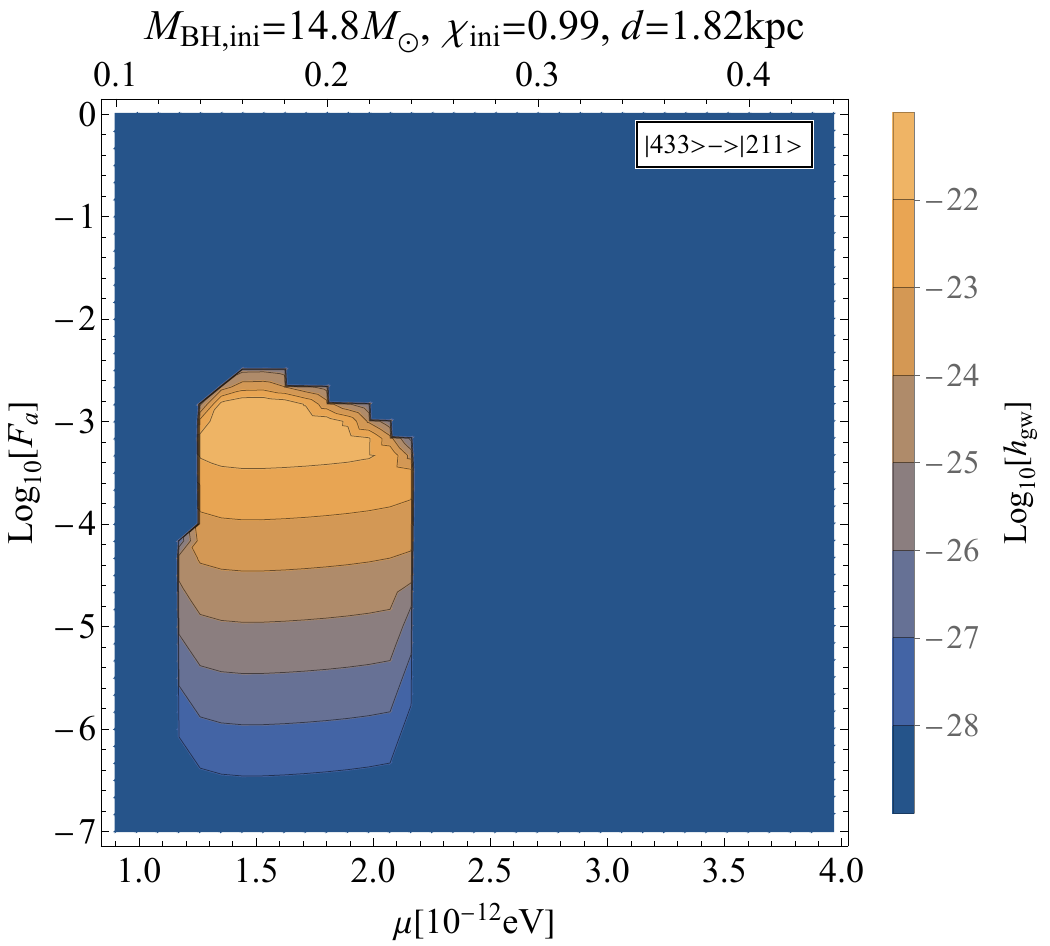}
 \includegraphics[keepaspectratio, scale=0.4]{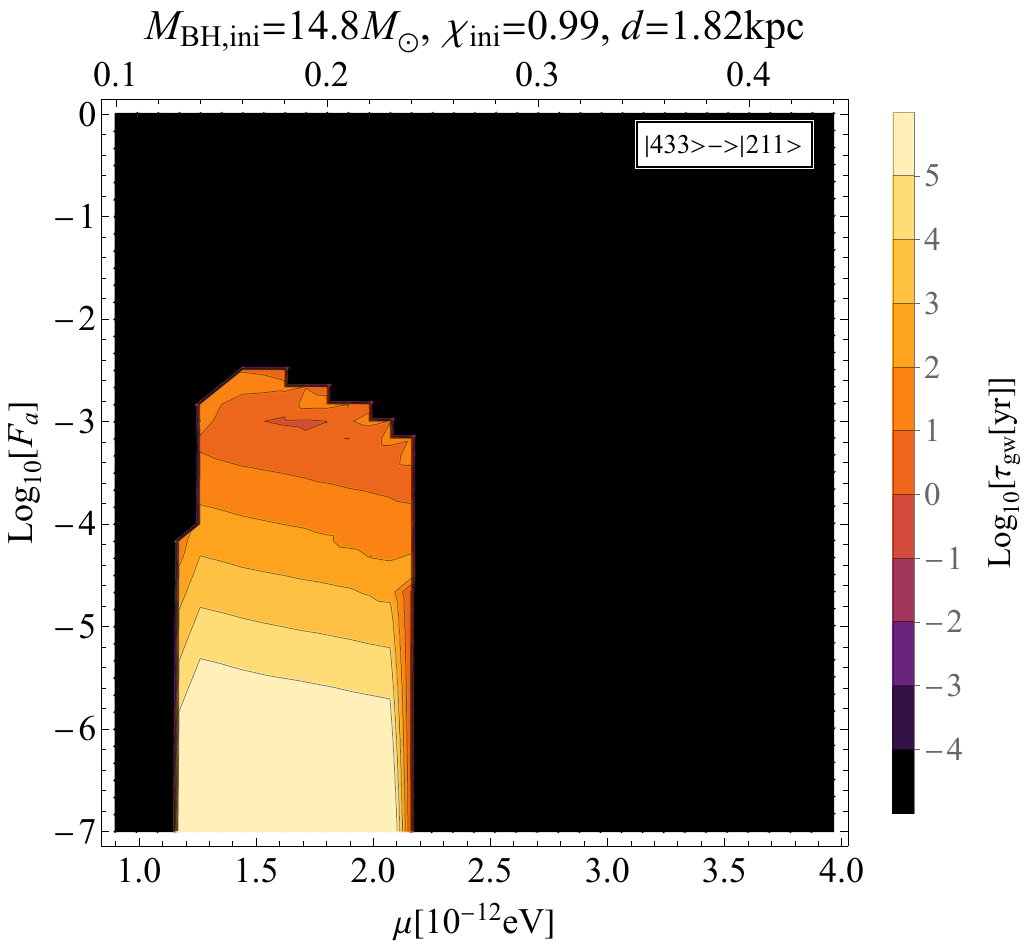}
     \includegraphics[keepaspectratio, scale=0.4]{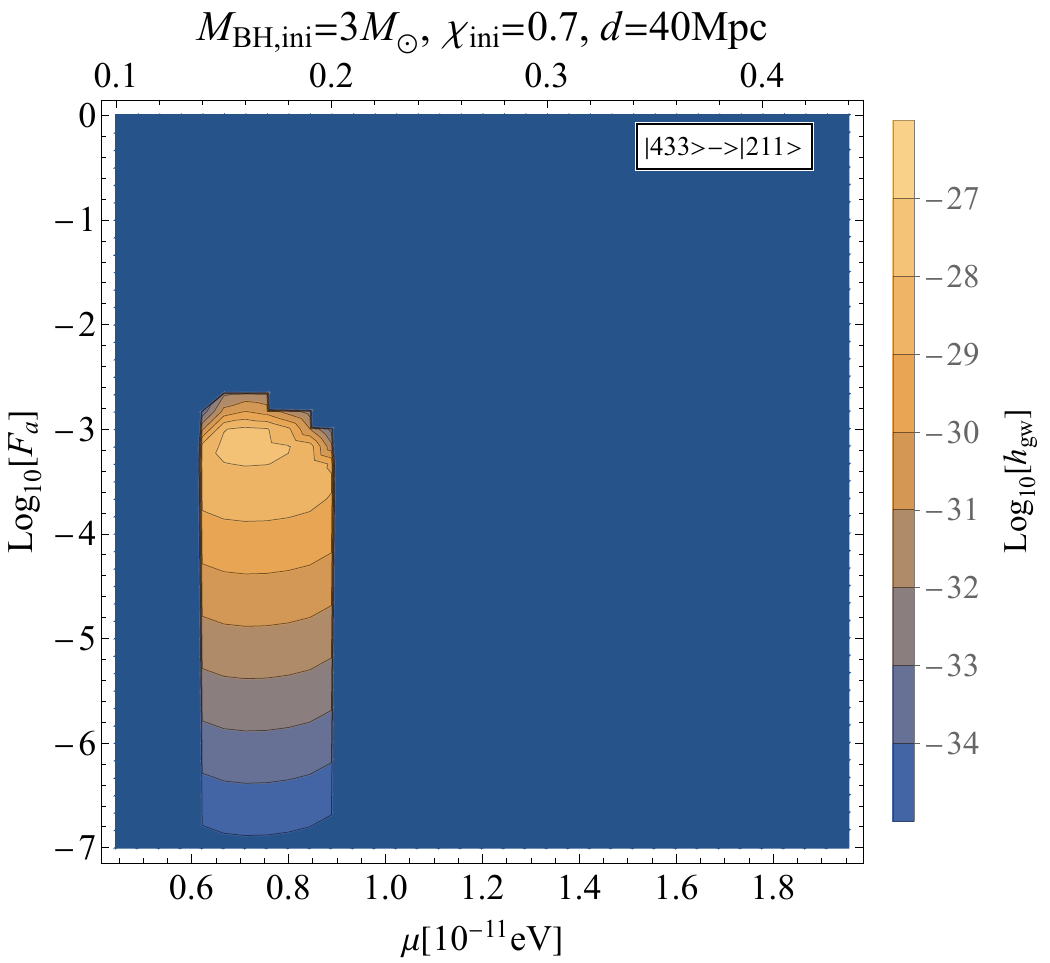}
  \includegraphics[keepaspectratio, scale=0.4]{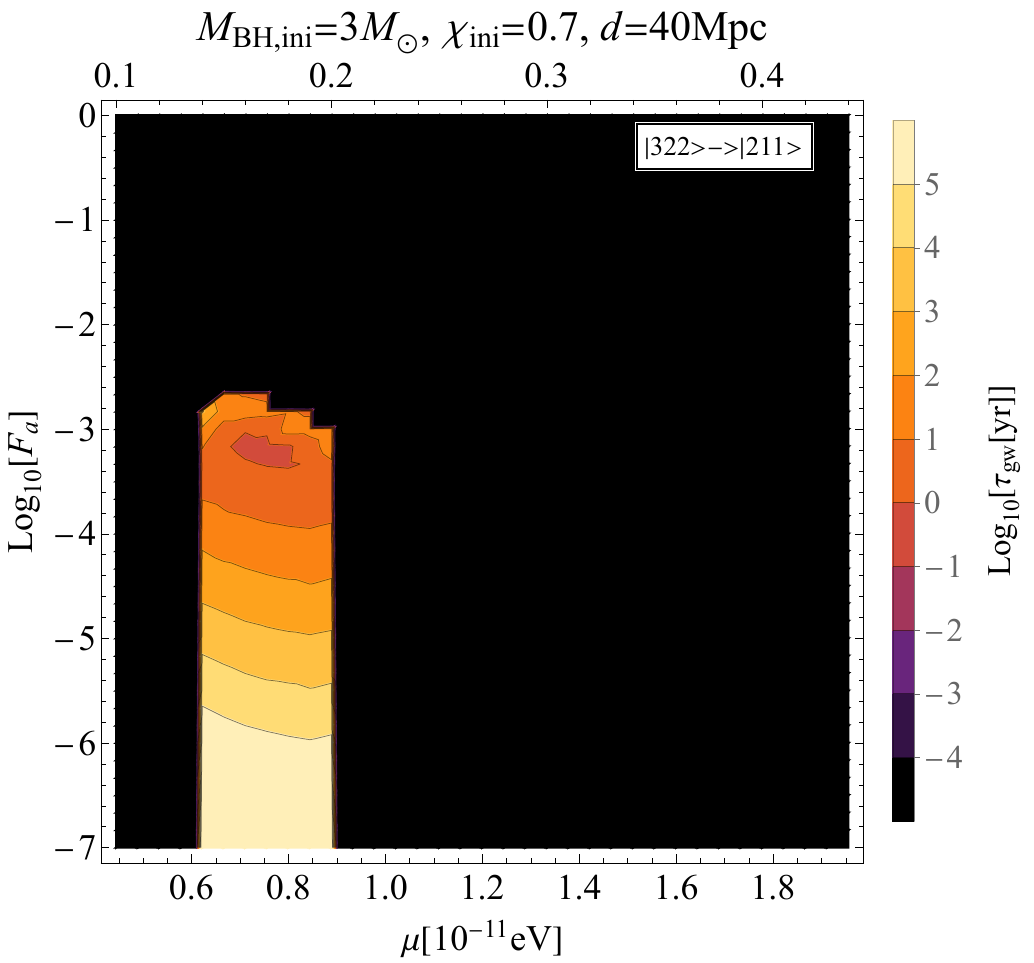}
 \caption{The same figures as Fig.~\ref{fig:gwpairann}, but with the gravitational waves from the level transition between the $l=m=3$ and the $l=m=1$ clouds.}
 \label{fig:gwtrans31}
\end{figure*}

\begin{figure*}[tb]
 \centering
  \includegraphics[keepaspectratio, scale=0.4]{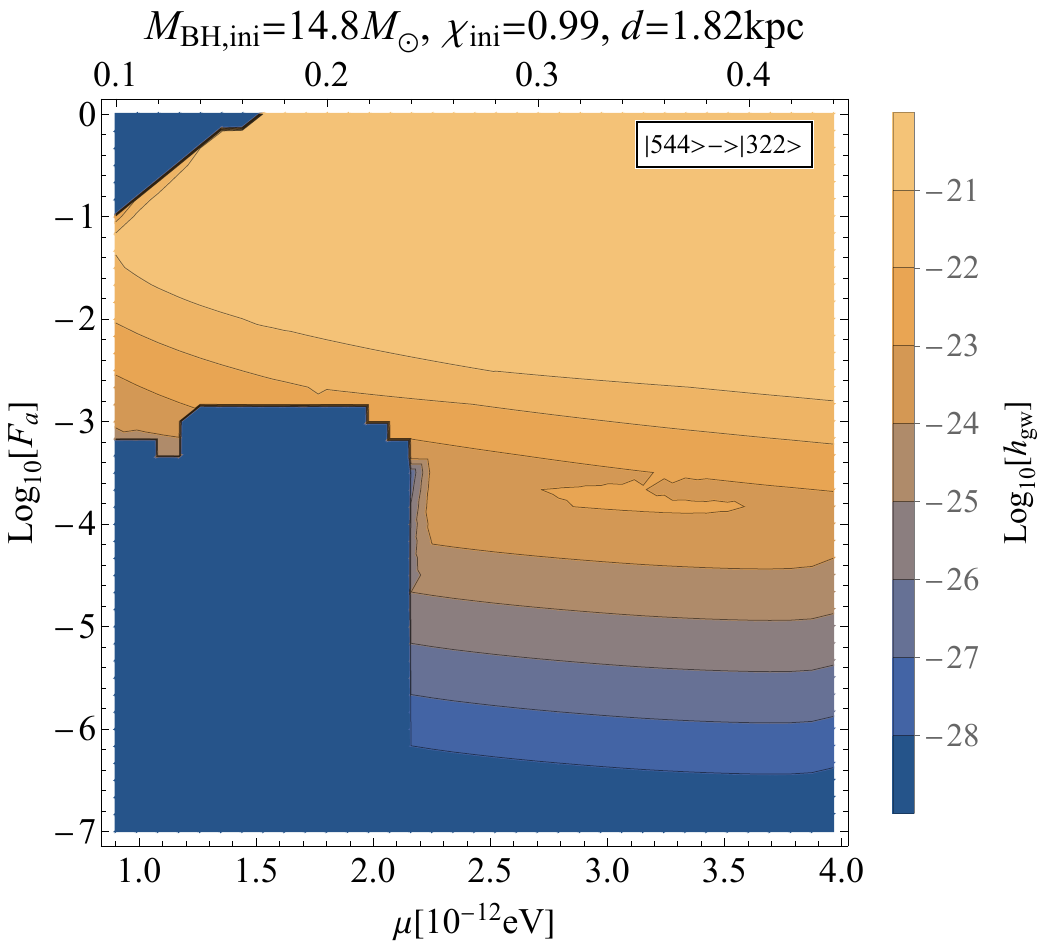}
 \includegraphics[keepaspectratio, scale=0.4]{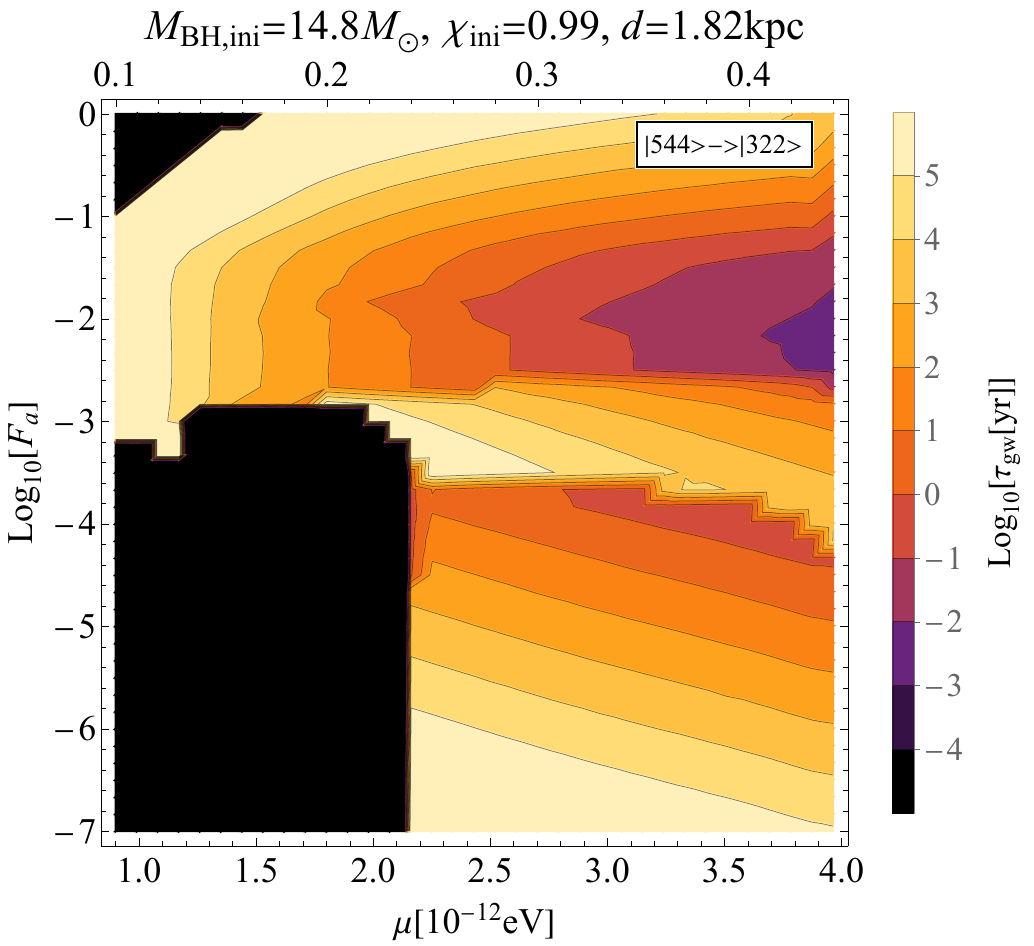}
     \includegraphics[keepaspectratio, scale=0.4]{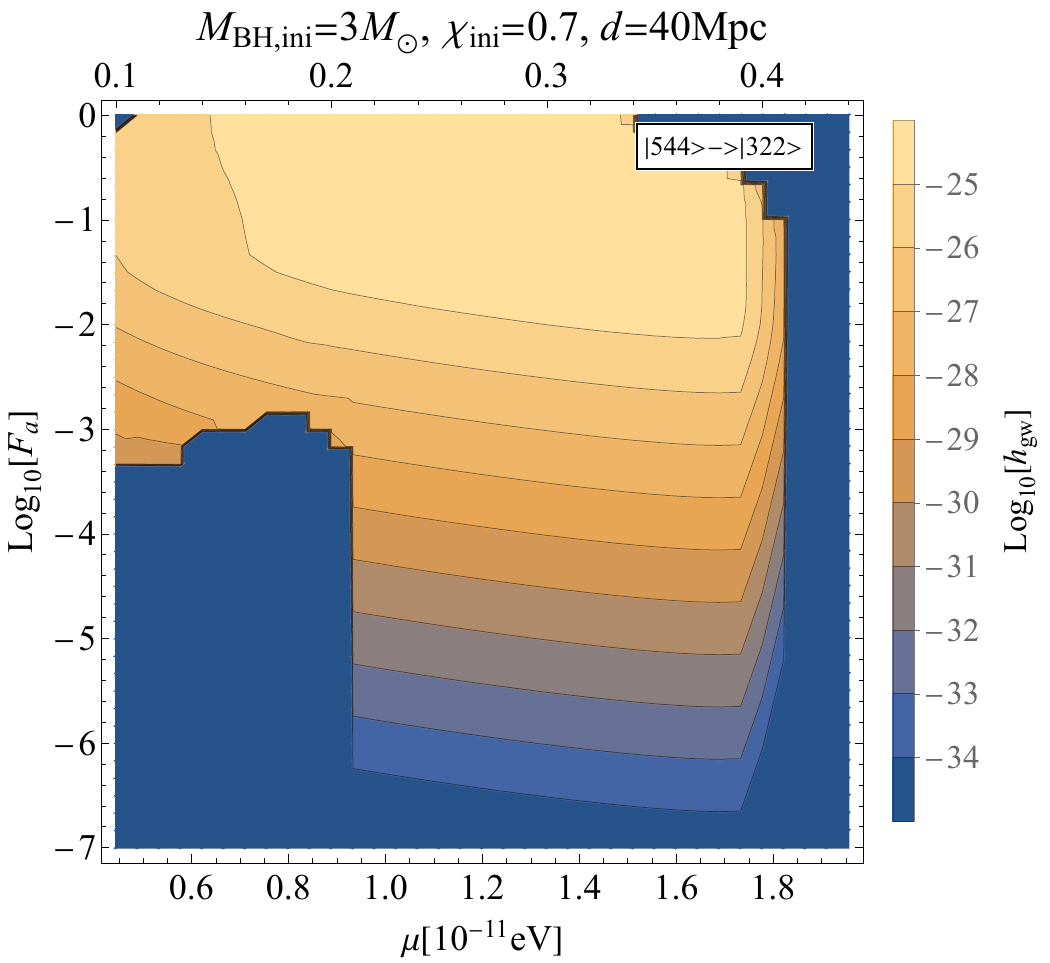}
  \includegraphics[keepaspectratio, scale=0.4]{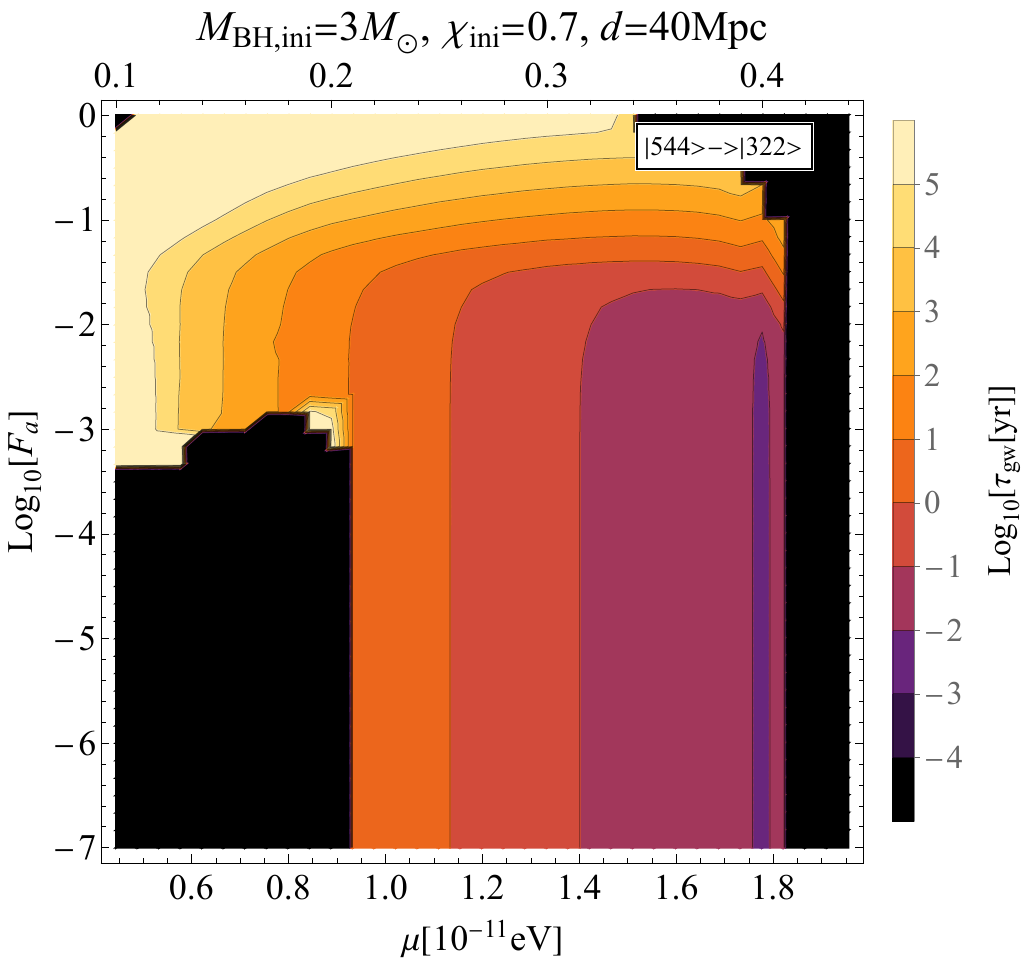}
 \caption{The same figures as Fig.~\ref{fig:gwpairann}, but with the gravitational waves from the level transition between the $l=m=4$ and the $l=m=2$ clouds.}
 \label{fig:gwtrans42}
\end{figure*}

Figure~\ref{fig:gwtrans21} shows the case for the level transition between the $l=m=2$ and the $l=m=1$ clouds. The basic feature is similar to the pair annihilation signal from the $l=m=1$ cloud. The main difference is in the dependence of the amplitude on $\mu$, which is much weaker than the pair annihilation signal. Another difference is in the size of the amplitude, which is about one order of magnitude larger for fixed $(\mu, F_a)$. The signal duration has the same order of magnitude as the $l=m=1$ pair annihilation process.

The case with the level transition process between the $l=m=3$ and the $l=m=1$ clouds is shown in Fig.~\ref{fig:gwtrans31}. In contrast to the pair annihilation or the level transition between the $l=m=2$ and the $l=m=1$ clouds, the emission occurs for the narrow parameter range, {\it i.e.}, $F_a \lesssim 10^{-3}$ and $0.12 \lesssim \mu M \lesssim 0.24\ (0.2)$ for $\chi = 0.99\ (0.7)$. The region is determined by the condition that the $l=m=3$ mode is excited before the $l=m=1$ mode decays. The $l=m=3$ cloud is not simultaneously excited in the large parameter range of $F_a$ since the spin down of the black hole is too fast and the $l=m=1$ cloud decays before the $l=m=3$ cloud appears (see Fig.~\ref{fig:examplecloudFa}). The region for the $\mu M$ agrees with the blue shaded region of Fig.~\ref{fig:wefflm34}.

Finally, we consider the case with the level transition process between the $l=m=4$ and the $l=m=2$ clouds (Fig.~\ref{fig:gwtrans42}). We cannot expect the signal from this process for some parameter regions: small $\mu$ and $F_a$ region, and the large $\mu$ region (only for $\chi_{\rm ini} = 0.7$, see the region $\mu \gtrsim 1.8\times 10^{-11}{\rm eV}$). The small $\mu$ and $F_a$ region corresponds to the region where the $l=m=3$ cloud is excited. As pointed out in Sec.~\ref{sec:4A}, when the $l=m=3$ cloud is excited, the $l=m=4$ cloud decays due to the mode coupling. The right edges in the lower two panels of Fig.~\ref{fig:gwtrans42} correspond to the threshold of the superradiance condition of the $l=m=2$ mode.

The duration is determined in a complicated way (see right panels of Fig.~\ref{fig:gwtrans42}). In the case of a relatively small initial spin (lower right panel of Fig.~\ref{fig:gwtrans42}), the behavior is similar to the level transition between the $l=m=2$ and the $l=m=1$ clouds in the sense that the duration becomes longer as $F_a$ is increased and $\mu$ is decreased (see right panels of Fig.~\ref{fig:gwtrans21}).

Behavior is complicated for a larger initial spin (upper right panel of Fig.~\ref{fig:gwtrans42}). We observe the transition around $F_a \sim 10^{-3}$ and $F_a \sim 10^{-4}$. This transition is determined by the competition between the decay of the $l=m=1$ cloud and the excitation of the $l=m=4$ cloud. In the large $F_a$ regime, the black hole spin down is so fast that the $l=m=1$ cloud disappears before the excitation of the $l=m=4$ cloud. Therefore, in this case, the transition signal is emitted after the complete decay of the $l=m=1$ cloud and the black hole spin down.

As $F_a$ is decreased, the spin down time scale due to the superradiance of the $l=m=1$ cloud becomes longer, and the transition signal between the $l=m=2$ and the $l=m=4$ clouds is emitted before the depletion of the $l=m=1$ cloud until $F_a \sim 10^{-3}$. But still, the configuration is almost determined by the stationary conditions for the $l=m=2$ and the $l=m=4$ clouds, and the $l=m=1$ cloud is subdominant. The presence of the $l=m=1$ cloud reduces the peak amplitude by assisting the growth of the $l=m=4$ through the processes $F_{114^*}^{\mathcal H}$ and $F_{124^*}^{\mathcal H}$. The earlier growth of the $l=m=4$ cloud results in the smoothing of the sharp peak.  Then, the peak amplitude becomes smaller than twice the amplitude at the plateau. 
In this case, the signal duration is determined by the spin down due to the $l=m=2$ superradiance, which greatly enhances the signal duration defined by the FWHM. 
For $F_a \lesssim 10^{-4}$, the excitation of the $l=m=4$ cloud occurs before the depletion of the $l=m=1$ cloud. Then, the configuration is determined by the balance among the three modes, not between the two. Now, the signal duration is determined by the spin down due to the superradiance of the $l=m=1$ mode, which is much faster than the spin down due to the $l=m=2$ mode.
When we further decrease $F_a$, the spin down time scale becomes longer, and the duration increases.

\subsection{Axion wave}\label{sec:5b}

Besides gravitational waves, the self-interacting axion condensate will emit the axion to infinity. Since the mass of the axion is ultra-light and the frequency is close to the axion mass, the emitted axion has properties similar to the ultra-light dark matter. Here, we estimate the amplitude and the axion number flux from the saturated configuration. 

From the definition of $F^{\mathcal{I}}_{ijk^*}$, the energy flux of the axion is given by
\begin{align}~\label{eq:fluxinfinity}
	\frac{dE_a}{dt} = F_a^2 F^{\mathcal{I}}_{ijk^*} M_{{\rm cl},i}M_{{\rm cl},j}M_{{\rm cl},k}~,
\end{align}
for each process. The axion field responsible for the energy flux~\eqref{eq:fluxinfinity} behaves near infinity as 
\begin{align}
	\phi \sim  A_{\infty} e^{-i (\omega t -m \varphi)} S_{mm\omega}(\theta) \frac{e^{ i k r}}{r}  + {\rm c.c.} ~,
\end{align}
with
\begin{align}
	A_{\infty} &= \sqrt{\frac{F^{\mathcal{I}}_{ijk^*}M_{cl,i}M_{cl,j}M_{cl,k}}{4 \pi \omega k}}~,\\
	k&=\sqrt{\omega^2 - \mu^2}~.
\end{align}
 Here, the frequency and the magnetic quantum number of the axion are given by $\omega = \omega_i + \omega_j - \omega_k$ and $m = m_i + m_j - m_k$, respectively, and used the fact that the contribution from the $l=m$ modes dominates the condensate in the index of the spheroidal harmonics. The amplitude is converted to the axion number flux per unit area as
\begin{align}
	\Phi_a  \sim \frac{1}{4 \pi r^2 \omega}\frac{dE_a}{dt}  = \frac{F_a^2 F^{\mathcal{I}}_{ijk^*} M_{{\rm cl},i}M_{{\rm cl},j}M_{{\rm cl},k}}{4 \pi r^2 \omega}~.
\end{align}

\begin{figure*}[t]
 \centering
   \includegraphics[keepaspectratio, scale=0.55]{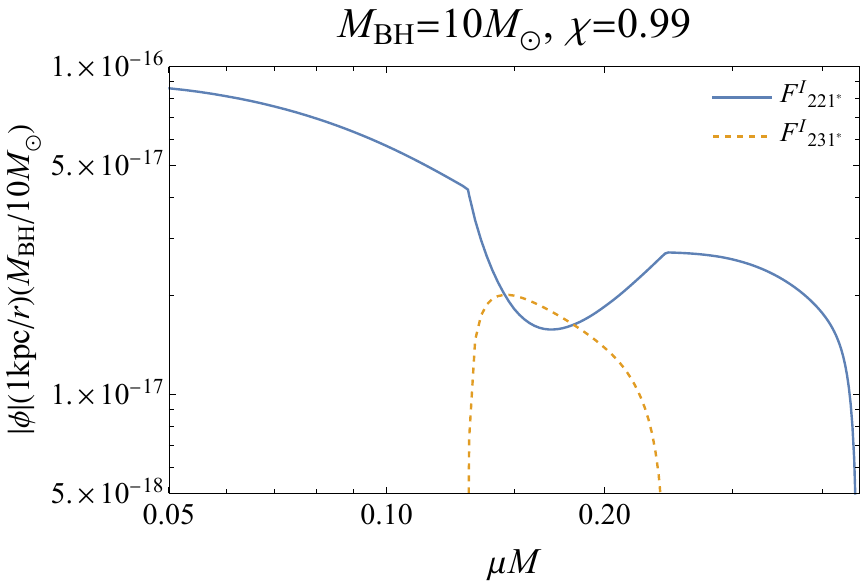}
 \includegraphics[keepaspectratio, scale=0.55]{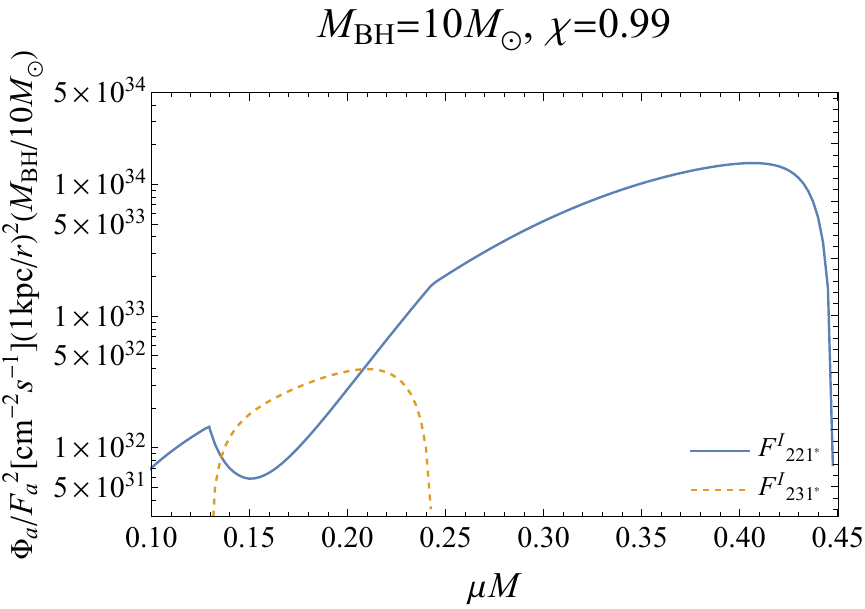}
 \caption{(Left) The blue solid and orange dotted curves show the amplitude of the axion wave normalized by the decay consent $F_a$ from the processes $F^{\mathcal{I}}_{221^*}$ and $F^{\mathcal{I}}_{231^*}$, respectively. We set the black hole mass to $M_{\rm BH} = 10M_{\odot}$ and the spin to $\chi = 0.99$. We put the black hole to be 1kpc away from us. (Right) Same as the left panel but with number density flux.}
 \label{fig:awamp}
\end{figure*}

When the $l=m=1$ cloud is present, the energy flux is dominated by the two processes, {\it i.e.}, $2\times \ket{322} \to \ket{211}+(\mbox{mode dissipating to inifinity})$ and $\ket{322}+ \ket{433} \to \ket{211} + (\mbox{ mode dissipating to inifinity})$. In Fig.~\ref{fig:awamp}, we show the amplitude and the axion number flux for these processes, assuming the black hole with $M_{\rm BH} = 10M_\odot$ and $\chi=0.99$ at 1kpc away from us. Note that if we assume the axion to be a dominant component of the dark matter, the amplitude (normalized by $F_a$) and the number flux in the solar system are estimated as
\begin{align}
	|\phi_{\rm DM}| \sim \sqrt{\frac{\rho_{\rm DM}}{F_a^2 \mu^2}} &\sim 2.6 \times 10^{-15}\left(\frac{\rho_{\rm DM}}{0.4 {\rm GeV/cm^3}}\right)^{1/2}\cr
 &\qquad \times \left(\frac{10^{-3}}{F_a}\right)\left(\frac{10^{-12}{\rm eV}}{\mu}\right)~,\\
	\Phi_{\rm DM} \sim \frac{v_{\rm DM}\rho_{\rm DM}}{\mu}&\sim 1.2\times 10^{28}[{\rm cm^2/s}]\left(\frac{\rho_{\rm DM}}{0.4 {\rm GeV/cm^3}}\right)\cr
 &\qquad \times \left(\frac{v_{\rm DM}}{10^{-3}}\right)\left(\frac{10^{-12}{\rm eV}}{\mu}\right)~.
\end{align}
Therefore, the amplitude of the axion from the black hole that we consider here becomes smaller than that of the ultra-light dark matter, but the number flux can be comparable. This is because the velocity of the emitted axion ($\sim \mu M_{\rm BH} $) is one order of magnitude larger than that of the dark matter particle ($v_{\rm DM} \sim 10^{-3}$). Since the amplitude is not so small as the dark matter particle, we might have a chance to observe the axion emitted from the black hole~\cite{Baryakhtar:2020gao}. We leave further investigation of the detectability of the axion wave from the black hole as a future work.

\section{Conclusion}\label{sec:6}

In this paper, we have numerically investigated the evolution of the self-interacting axion condensate and the observational signals of emitted gravitational waves in the relativistic regime $(\mu M \gtrsim 0.1)$, including the interaction between higher multipole modes up to $l=m=4$. We have found that the excitation of the higher multipole modes heavily depends on the axion mass $\mu$ and the axion decay constant $F_a$. These parameters control the relative size between the spin down rate and the energy transfer rate of clouds due to the self-interaction. As a result, the $l=m=3$ cloud can be excited only for the narrow parameter range: $0.12 \lesssim \mu M \lesssim 0.25$ and $F_a \lesssim 10^{-3}$. Except this parameter range, the $l=m=4$ cloud can be excited.

Owing to the presence of the higher multipole modes, we would expect gravitational wave signals in several frequency bands. The pair annihilation gives the signal in the high-frequency regime (kHz for a black hole with the mass $10M_\odot$), and the level transition process gives the low-frequency signal (Hz for a black hole with the mass $10M_\odot$). In this paper, we have shown that the gravitational waves from the level transition between the modes whose magnetic quantum numbers are different by two (i.e. $\Delta m = 2$) can provide a larger amplitude than the pair annihilation signal or the level transition signal with $\Delta m = 1$ (see Eqs.~\eqref{eq:estimateamp31}--~\eqref{eq:estimateamp21}). We have also estimated the flux of axion waves emitted from the condensate and have shown that it can be comparable to the flux of the dark matter.

Let us briefly consider the detection prospect of these level transition signals from black holes made by a binary neutron star merger. The frequency of the gravitational wave would be around Hz, which is the target of future detectors such as DECIGO~\cite{Kawamura:2011zz}, atomic interferometers~\cite{Proceedings:2023mkp}, TOBA~\cite{Shimoda:2018uiv}, TianGo~\cite{2020PhRvD.102d3001K}, DO~\cite{2020CQGra..37u5011A}, and AMIGO~\cite{2020IJMPD..2940007N}. For simplicity, let us focus on the case of DECIGO. The current sensitivity target of DECIGO is $\sim 4\times 10^{-24}{\rm Hz}^{-1/2}$ at around 1 Hz~\cite{Kawamura:2020pcg}. If we could integrate the signal for 1 yr, we could observe the signal with an amplitude of $h \sim 7 \times 10^{-28}$. Therefore, the gravitational wave signal of the transition between the $l=m=3$ mode and the $l=m=1$ mode could be observed assuming the GUT scale decay constant and 40 Mpc as the distance to the source (see lower panels of Fig.~\ref{fig:gwtrans31}). 
We can observe signals from black holes which are located much farther if we consider the heavier black holes. For example, we can observe it up to 800 Mpc for a black hole with the mass $60M_{\odot}$ (such as the one produced in the GW150914~\cite{LIGOScientific:2016aoc}). For the larger decay constant ($F_a \gtrsim 10^{-2}$), the dominant signal will be the gravitational wave from the level transition between the $l=m=4$ and $l=m=2$, which can be reached by the detectors with target sensitivity $\sim 10^{-21}{\rm Hz}^{-1/2}$ around 1 Hz (see lower panels of Fig.~\ref{fig:gwtrans42}).

However, to be more realistic, this estimation requires further study. We would need to calculate the frequency shift and fix the search strategy to estimate the detectability. The calculation of the frequency shift in the relativistic regime is already formulated in the literature~\cite{Omiya:2020vji, Cannizzaro:2023jle} but requires some further computation. Hence, we leave the detailed investigation of the detectability of the level transition signals to future work.

In this paper, we only considered the stellar mass black hole and did not consider the supermassive black holes. A huge signal would be expected if we scale our result to the supermassive black holes due to their large mass. However, we cannot simply scale our result since the environmental effect must be addressed. We must especially consider the effect of the accretion. The accretion would spin up the black hole, which would affect the evolution of the cloud~\cite{Brito:2014wla}. Furthermore, the tidal effect could potentially deplete the cloud~\cite{Du:2022trq}. We leave the investigation on the supermassive black hole for the future.

\begin{acknowledgments}
H.O. thanks Jiro Soda and Hiroki Takeda for a valuable discussion.
H. O. is supported by JSPS KAKENHI Grant Numbers JP23H00110.  
H. Y is in part supported by JSPS KAKENHI Grant Numbers JP22H01220 and JP21H05189,
and is partly supported by Osaka Central Advanced Mathematical Institute
(MEXT Joint Usage/Research Center on Mathematics and Theoretical Physics JPMXP0619217849). T. Takahashi is supported by JSPS KAKENHI Grant Number JP23KJ1214. T.T. is supported by JSPS KAKENHI Grant Nos. JP24H00963, JP24H01809, JP23H00110 and JP20K03928.
\end{acknowledgments}

\bibliographystyle{apsrev4-1}
\bibliography{axionref}

\begin{thebibliography}{78}%
\makeatletter
\providecommand \@ifxundefined [1]{%
 \@ifx{#1\undefined}
}%
\providecommand \@ifnum [1]{%
 \ifnum #1\expandafter \@firstoftwo
 \else \expandafter \@secondoftwo
 \fi
}%
\providecommand \@ifx [1]{%
 \ifx #1\expandafter \@firstoftwo
 \else \expandafter \@secondoftwo
 \fi
}%
\providecommand \natexlab [1]{#1}%
\providecommand \enquote  [1]{``#1''}%
\providecommand \bibnamefont  [1]{#1}%
\providecommand \bibfnamefont [1]{#1}%
\providecommand \citenamefont [1]{#1}%
\providecommand \href@noop [0]{\@secondoftwo}%
\providecommand \href [0]{\begingroup \@sanitize@url \@href}%
\providecommand \@href[1]{\@@startlink{#1}\@@href}%
\providecommand \@@href[1]{\endgroup#1\@@endlink}%
\providecommand \@sanitize@url [0]{\catcode `\\12\catcode `\$12\catcode `\&12\catcode `\#12\catcode `\^12\catcode `\_12\catcode `\%12\relax}%
\providecommand \@@startlink[1]{}%
\providecommand \@@endlink[0]{}%
\providecommand \url  [0]{\begingroup\@sanitize@url \@url }%
\providecommand \@url [1]{\endgroup\@href {#1}{\urlprefix }}%
\providecommand \urlprefix  [0]{URL }%
\providecommand \Eprint [0]{\href }%
\providecommand \doibase [0]{http://dx.doi.org/}%
\providecommand \selectlanguage [0]{\@gobble}%
\providecommand \bibinfo  [0]{\@secondoftwo}%
\providecommand \bibfield  [0]{\@secondoftwo}%
\providecommand \translation [1]{[#1]}%
\providecommand \BibitemOpen [0]{}%
\providecommand \bibitemStop [0]{}%
\providecommand \bibitemNoStop [0]{.\EOS\space}%
\providecommand \EOS [0]{\spacefactor3000\relax}%
\providecommand \BibitemShut  [1]{\csname bibitem#1\endcsname}%
\let\auto@bib@innerbib\@empty
\bibitem [{\citenamefont {Arvanitaki}\ \emph {et~al.}(2010)\citenamefont {Arvanitaki}, \citenamefont {Dimopoulos}, \citenamefont {Dubovsky}, \citenamefont {Kaloper},\ and\ \citenamefont {March-Russell}}]{Arvanitaki:2009fg}%
  \BibitemOpen
  \bibfield  {author} {\bibinfo {author} {\bibfnamefont {A.}~\bibnamefont {Arvanitaki}}, \bibinfo {author} {\bibfnamefont {S.}~\bibnamefont {Dimopoulos}}, \bibinfo {author} {\bibfnamefont {S.}~\bibnamefont {Dubovsky}}, \bibinfo {author} {\bibfnamefont {N.}~\bibnamefont {Kaloper}}, \ and\ \bibinfo {author} {\bibfnamefont {J.}~\bibnamefont {March-Russell}},\ }\href {\doibase 10.1103/PhysRevD.81.123530} {\bibfield  {journal} {\bibinfo  {journal} {Phys. Rev.}\ }\textbf {\bibinfo {volume} {D81}},\ \bibinfo {pages} {123530} (\bibinfo {year} {2010})},\ \Eprint {http://arxiv.org/abs/0905.4720} {arXiv:0905.4720 [hep-th]} \BibitemShut {NoStop}%
\bibitem [{\citenamefont {Cicoli}\ \emph {et~al.}(2012)\citenamefont {Cicoli}, \citenamefont {Goodsell},\ and\ \citenamefont {Ringwald}}]{Cicoli:2012sz}%
  \BibitemOpen
  \bibfield  {author} {\bibinfo {author} {\bibfnamefont {M.}~\bibnamefont {Cicoli}}, \bibinfo {author} {\bibfnamefont {M.}~\bibnamefont {Goodsell}}, \ and\ \bibinfo {author} {\bibfnamefont {A.}~\bibnamefont {Ringwald}},\ }\href {\doibase 10.1007/JHEP10(2012)146} {\bibfield  {journal} {\bibinfo  {journal} {JHEP}\ }\textbf {\bibinfo {volume} {10}},\ \bibinfo {pages} {146} (\bibinfo {year} {2012})},\ \Eprint {http://arxiv.org/abs/1206.0819} {arXiv:1206.0819 [hep-th]} \BibitemShut {NoStop}%
\bibitem [{\citenamefont {Demirtas}\ \emph {et~al.}(2020)\citenamefont {Demirtas}, \citenamefont {Long}, \citenamefont {McAllister},\ and\ \citenamefont {Stillman}}]{Demirtas:2018akl}%
  \BibitemOpen
  \bibfield  {author} {\bibinfo {author} {\bibfnamefont {M.}~\bibnamefont {Demirtas}}, \bibinfo {author} {\bibfnamefont {C.}~\bibnamefont {Long}}, \bibinfo {author} {\bibfnamefont {L.}~\bibnamefont {McAllister}}, \ and\ \bibinfo {author} {\bibfnamefont {M.}~\bibnamefont {Stillman}},\ }\href {\doibase 10.1007/JHEP04(2020)138} {\bibfield  {journal} {\bibinfo  {journal} {JHEP}\ }\textbf {\bibinfo {volume} {04}},\ \bibinfo {pages} {138} (\bibinfo {year} {2020})},\ \Eprint {http://arxiv.org/abs/1808.01282} {arXiv:1808.01282 [hep-th]} \BibitemShut {NoStop}%
\bibitem [{\citenamefont {Mehta}\ \emph {et~al.}(2021)\citenamefont {Mehta}, \citenamefont {Demirtas}, \citenamefont {Long}, \citenamefont {Marsh}, \citenamefont {McAllister},\ and\ \citenamefont {Stott}}]{Mehta:2021pwf}%
  \BibitemOpen
  \bibfield  {author} {\bibinfo {author} {\bibfnamefont {V.~M.}\ \bibnamefont {Mehta}}, \bibinfo {author} {\bibfnamefont {M.}~\bibnamefont {Demirtas}}, \bibinfo {author} {\bibfnamefont {C.}~\bibnamefont {Long}}, \bibinfo {author} {\bibfnamefont {D.~J.~E.}\ \bibnamefont {Marsh}}, \bibinfo {author} {\bibfnamefont {L.}~\bibnamefont {McAllister}}, \ and\ \bibinfo {author} {\bibfnamefont {M.~J.}\ \bibnamefont {Stott}},\ }\href {\doibase 10.1088/1475-7516/2021/07/033} {\bibfield  {journal} {\bibinfo  {journal} {JCAP}\ }\textbf {\bibinfo {volume} {07}},\ \bibinfo {pages} {033} (\bibinfo {year} {2021})},\ \Eprint {http://arxiv.org/abs/2103.06812} {arXiv:2103.06812 [hep-th]} \BibitemShut {NoStop}%
\bibitem [{\citenamefont {Svrcek}\ and\ \citenamefont {Witten}(2006)}]{Svrcek:2006yi}%
  \BibitemOpen
  \bibfield  {author} {\bibinfo {author} {\bibfnamefont {P.}~\bibnamefont {Svrcek}}\ and\ \bibinfo {author} {\bibfnamefont {E.}~\bibnamefont {Witten}},\ }\href {\doibase 10.1088/1126-6708/2006/06/051} {\bibfield  {journal} {\bibinfo  {journal} {JHEP}\ }\textbf {\bibinfo {volume} {06}},\ \bibinfo {pages} {051} (\bibinfo {year} {2006})},\ \Eprint {http://arxiv.org/abs/hep-th/0605206} {arXiv:hep-th/0605206 [hep-th]} \BibitemShut {NoStop}%
\bibitem [{\citenamefont {Peccei}\ and\ \citenamefont {Quinn}(1977)}]{Peccei:1977hh}%
  \BibitemOpen
  \bibfield  {author} {\bibinfo {author} {\bibfnamefont {R.~D.}\ \bibnamefont {Peccei}}\ and\ \bibinfo {author} {\bibfnamefont {H.~R.}\ \bibnamefont {Quinn}},\ }\href {\doibase 10.1103/PhysRevLett.38.1440} {\bibfield  {journal} {\bibinfo  {journal} {Phys. Rev. Lett.}\ }\textbf {\bibinfo {volume} {38}},\ \bibinfo {pages} {1440} (\bibinfo {year} {1977})},\ \bibinfo {note} {[,328(1977)]}\BibitemShut {NoStop}%
\bibitem [{\citenamefont {Weinberg}(1978)}]{Weinberg:1977ma}%
  \BibitemOpen
  \bibfield  {author} {\bibinfo {author} {\bibfnamefont {S.}~\bibnamefont {Weinberg}},\ }\href {\doibase 10.1103/PhysRevLett.40.223} {\bibfield  {journal} {\bibinfo  {journal} {Phys. Rev. Lett.}\ }\textbf {\bibinfo {volume} {40}},\ \bibinfo {pages} {223} (\bibinfo {year} {1978})}\BibitemShut {NoStop}%
\bibitem [{\citenamefont {Wilczek}(1978)}]{Wilczek:1977pj}%
  \BibitemOpen
  \bibfield  {author} {\bibinfo {author} {\bibfnamefont {F.}~\bibnamefont {Wilczek}},\ }\href {\doibase 10.1103/PhysRevLett.40.279} {\bibfield  {journal} {\bibinfo  {journal} {Phys. Rev. Lett.}\ }\textbf {\bibinfo {volume} {40}},\ \bibinfo {pages} {279} (\bibinfo {year} {1978})}\BibitemShut {NoStop}%
\bibitem [{\citenamefont {Kim}(1979)}]{Kim:1979if}%
  \BibitemOpen
  \bibfield  {author} {\bibinfo {author} {\bibfnamefont {J.~E.}\ \bibnamefont {Kim}},\ }\href {\doibase 10.1103/PhysRevLett.43.103} {\bibfield  {journal} {\bibinfo  {journal} {Phys. Rev. Lett.}\ }\textbf {\bibinfo {volume} {43}},\ \bibinfo {pages} {103} (\bibinfo {year} {1979})}\BibitemShut {NoStop}%
\bibitem [{\citenamefont {Shifman}\ \emph {et~al.}(1980)\citenamefont {Shifman}, \citenamefont {Vainshtein},\ and\ \citenamefont {Zakharov}}]{Shifman:1979if}%
  \BibitemOpen
  \bibfield  {author} {\bibinfo {author} {\bibfnamefont {M.~A.}\ \bibnamefont {Shifman}}, \bibinfo {author} {\bibfnamefont {A.~I.}\ \bibnamefont {Vainshtein}}, \ and\ \bibinfo {author} {\bibfnamefont {V.~I.}\ \bibnamefont {Zakharov}},\ }\href {\doibase 10.1016/0550-3213(80)90209-6} {\bibfield  {journal} {\bibinfo  {journal} {Nucl. Phys. B}\ }\textbf {\bibinfo {volume} {166}},\ \bibinfo {pages} {493} (\bibinfo {year} {1980})}\BibitemShut {NoStop}%
\bibitem [{\citenamefont {Zhitnitsky}(1980)}]{Zhitnitsky:1980tq}%
  \BibitemOpen
  \bibfield  {author} {\bibinfo {author} {\bibfnamefont {A.~R.}\ \bibnamefont {Zhitnitsky}},\ }\href@noop {} {\bibfield  {journal} {\bibinfo  {journal} {Sov. J. Nucl. Phys.}\ }\textbf {\bibinfo {volume} {31}},\ \bibinfo {pages} {260} (\bibinfo {year} {1980})}\BibitemShut {NoStop}%
\bibitem [{\citenamefont {Dine}\ \emph {et~al.}(1981)\citenamefont {Dine}, \citenamefont {Fischler},\ and\ \citenamefont {Srednicki}}]{Dine:1981rt}%
  \BibitemOpen
  \bibfield  {author} {\bibinfo {author} {\bibfnamefont {M.}~\bibnamefont {Dine}}, \bibinfo {author} {\bibfnamefont {W.}~\bibnamefont {Fischler}}, \ and\ \bibinfo {author} {\bibfnamefont {M.}~\bibnamefont {Srednicki}},\ }\href {\doibase 10.1016/0370-2693(81)90590-6} {\bibfield  {journal} {\bibinfo  {journal} {Phys. Lett. B}\ }\textbf {\bibinfo {volume} {104}},\ \bibinfo {pages} {199} (\bibinfo {year} {1981})}\BibitemShut {NoStop}%
\bibitem [{\citenamefont {Abbott}\ and\ \citenamefont {Sikivie}(1983)}]{Abbott:1982af}%
  \BibitemOpen
  \bibfield  {author} {\bibinfo {author} {\bibfnamefont {L.~F.}\ \bibnamefont {Abbott}}\ and\ \bibinfo {author} {\bibfnamefont {P.}~\bibnamefont {Sikivie}},\ }\href {\doibase 10.1016/0370-2693(83)90638-X} {\bibfield  {journal} {\bibinfo  {journal} {Phys. Lett.}\ }\textbf {\bibinfo {volume} {120B}},\ \bibinfo {pages} {133} (\bibinfo {year} {1983})}\BibitemShut {NoStop}%
\bibitem [{\citenamefont {Preskill}\ \emph {et~al.}(1983)\citenamefont {Preskill}, \citenamefont {Wise},\ and\ \citenamefont {Wilczek}}]{Preskill:1982cy}%
  \BibitemOpen
  \bibfield  {author} {\bibinfo {author} {\bibfnamefont {J.}~\bibnamefont {Preskill}}, \bibinfo {author} {\bibfnamefont {M.~B.}\ \bibnamefont {Wise}}, \ and\ \bibinfo {author} {\bibfnamefont {F.}~\bibnamefont {Wilczek}},\ }\href {\doibase 10.1016/0370-2693(83)90637-8} {\bibfield  {journal} {\bibinfo  {journal} {Phys. Lett.}\ }\textbf {\bibinfo {volume} {120B}},\ \bibinfo {pages} {127} (\bibinfo {year} {1983})}\BibitemShut {NoStop}%
\bibitem [{\citenamefont {Dine}\ and\ \citenamefont {Fischler}(1983)}]{Dine:1982ah}%
  \BibitemOpen
  \bibfield  {author} {\bibinfo {author} {\bibfnamefont {M.}~\bibnamefont {Dine}}\ and\ \bibinfo {author} {\bibfnamefont {W.}~\bibnamefont {Fischler}},\ }\href {\doibase 10.1016/0370-2693(83)90639-1} {\bibfield  {journal} {\bibinfo  {journal} {Phys. Lett.}\ }\textbf {\bibinfo {volume} {120B}},\ \bibinfo {pages} {137} (\bibinfo {year} {1983})}\BibitemShut {NoStop}%
\bibitem [{\citenamefont {Hui}\ \emph {et~al.}(2017)\citenamefont {Hui}, \citenamefont {Ostriker}, \citenamefont {Tremaine},\ and\ \citenamefont {~}}]{Hui:2016ltb}%
  \BibitemOpen
  \bibfield  {author} {\bibinfo {author} {\bibfnamefont {L.}~\bibnamefont {Hui}}, \bibinfo {author} {\bibfnamefont {J.~P.}\ \bibnamefont {Ostriker}}, \bibinfo {author} {\bibfnamefont {S.}~\bibnamefont {Tremaine}}, \ and\ \bibinfo {author} {\bibfnamefont {E.}~\bibnamefont {~}},\ }\href {\doibase 10.1103/PhysRevD.95.043541} {\bibfield  {journal} {\bibinfo  {journal} {Phys. Rev. D}\ }\textbf {\bibinfo {volume} {95}},\ \bibinfo {pages} {043541} (\bibinfo {year} {2017})},\ \Eprint {http://arxiv.org/abs/1610.08297} {arXiv:1610.08297 [astro-ph.CO]} \BibitemShut {NoStop}%
\bibitem [{\citenamefont {Arvanitaki}\ and\ \citenamefont {Dubovsky}(2011)}]{Arvanitaki:2010sy}%
  \BibitemOpen
  \bibfield  {author} {\bibinfo {author} {\bibfnamefont {A.}~\bibnamefont {Arvanitaki}}\ and\ \bibinfo {author} {\bibfnamefont {S.}~\bibnamefont {Dubovsky}},\ }\href {\doibase 10.1103/PhysRevD.83.044026} {\bibfield  {journal} {\bibinfo  {journal} {Phys. Rev.}\ }\textbf {\bibinfo {volume} {D83}},\ \bibinfo {pages} {044026} (\bibinfo {year} {2011})},\ \Eprint {http://arxiv.org/abs/1004.3558} {arXiv:1004.3558 [hep-th]} \BibitemShut {NoStop}%
\bibitem [{\citenamefont {Zouros}\ and\ \citenamefont {Eardley}(1979)}]{Zouros:1979iw}%
  \BibitemOpen
  \bibfield  {author} {\bibinfo {author} {\bibfnamefont {T.~J.~M.}\ \bibnamefont {Zouros}}\ and\ \bibinfo {author} {\bibfnamefont {D.~M.}\ \bibnamefont {Eardley}},\ }\href {\doibase 10.1016/0003-4916(79)90237-9} {\bibfield  {journal} {\bibinfo  {journal} {Annals Phys.}\ }\textbf {\bibinfo {volume} {118}},\ \bibinfo {pages} {139} (\bibinfo {year} {1979})}\BibitemShut {NoStop}%
\bibitem [{\citenamefont {Detweiler}(1980)}]{Detweiler:1980uk}%
  \BibitemOpen
  \bibfield  {author} {\bibinfo {author} {\bibfnamefont {S.~L.}\ \bibnamefont {Detweiler}},\ }\href {\doibase 10.1103/PhysRevD.22.2323} {\bibfield  {journal} {\bibinfo  {journal} {Phys. Rev.}\ }\textbf {\bibinfo {volume} {D22}},\ \bibinfo {pages} {2323} (\bibinfo {year} {1980})}\BibitemShut {NoStop}%
\bibitem [{\citenamefont {Brito}\ \emph {et~al.}(2015{\natexlab{a}})\citenamefont {Brito}, \citenamefont {Cardoso},\ and\ \citenamefont {Pani}}]{Brito:2015oca}%
  \BibitemOpen
  \bibfield  {author} {\bibinfo {author} {\bibfnamefont {R.}~\bibnamefont {Brito}}, \bibinfo {author} {\bibfnamefont {V.}~\bibnamefont {Cardoso}}, \ and\ \bibinfo {author} {\bibfnamefont {P.}~\bibnamefont {Pani}},\ }\href {\doibase 10.1007/978-3-319-19000-6} {\bibfield  {journal} {\bibinfo  {journal} {Lect. Notes Phys.}\ }\textbf {\bibinfo {volume} {906}},\ \bibinfo {pages} {pp.1} (\bibinfo {year} {2015}{\natexlab{a}})},\ \Eprint {http://arxiv.org/abs/1501.06570} {arXiv:1501.06570 [gr-qc]} \BibitemShut {NoStop}%
\bibitem [{\citenamefont {Arvanitaki}\ \emph {et~al.}(2015)\citenamefont {Arvanitaki}, \citenamefont {Baryakhtar},\ and\ \citenamefont {Huang}}]{Arvanitaki:2014wva}%
  \BibitemOpen
  \bibfield  {author} {\bibinfo {author} {\bibfnamefont {A.}~\bibnamefont {Arvanitaki}}, \bibinfo {author} {\bibfnamefont {M.}~\bibnamefont {Baryakhtar}}, \ and\ \bibinfo {author} {\bibfnamefont {X.}~\bibnamefont {Huang}},\ }\href {\doibase 10.1103/PhysRevD.91.084011} {\bibfield  {journal} {\bibinfo  {journal} {Phys. Rev.}\ }\textbf {\bibinfo {volume} {D91}},\ \bibinfo {pages} {084011} (\bibinfo {year} {2015})},\ \Eprint {http://arxiv.org/abs/1411.2263} {arXiv:1411.2263 [hep-ph]} \BibitemShut {NoStop}%
\bibitem [{\citenamefont {Yoshino}\ and\ \citenamefont {Kodama}(2015{\natexlab{a}})}]{Yoshino:2014wwa}%
  \BibitemOpen
  \bibfield  {author} {\bibinfo {author} {\bibfnamefont {H.}~\bibnamefont {Yoshino}}\ and\ \bibinfo {author} {\bibfnamefont {H.}~\bibnamefont {Kodama}},\ }\href {\doibase 10.1093/ptep/ptv067} {\bibfield  {journal} {\bibinfo  {journal} {PTEP}\ }\textbf {\bibinfo {volume} {2015}},\ \bibinfo {pages} {061E01} (\bibinfo {year} {2015}{\natexlab{a}})},\ \Eprint {http://arxiv.org/abs/1407.2030} {arXiv:1407.2030 [gr-qc]} \BibitemShut {NoStop}%
\bibitem [{\citenamefont {Brito}\ \emph {et~al.}(2017)\citenamefont {Brito}, \citenamefont {Ghosh}, \citenamefont {Barausse}, \citenamefont {Berti}, \citenamefont {Cardoso}, \citenamefont {Dvorkin}, \citenamefont {Klein},\ and\ \citenamefont {Pani}}]{Brito:2017zvb}%
  \BibitemOpen
  \bibfield  {author} {\bibinfo {author} {\bibfnamefont {R.}~\bibnamefont {Brito}}, \bibinfo {author} {\bibfnamefont {S.}~\bibnamefont {Ghosh}}, \bibinfo {author} {\bibfnamefont {E.}~\bibnamefont {Barausse}}, \bibinfo {author} {\bibfnamefont {E.}~\bibnamefont {Berti}}, \bibinfo {author} {\bibfnamefont {V.}~\bibnamefont {Cardoso}}, \bibinfo {author} {\bibfnamefont {I.}~\bibnamefont {Dvorkin}}, \bibinfo {author} {\bibfnamefont {A.}~\bibnamefont {Klein}}, \ and\ \bibinfo {author} {\bibfnamefont {P.}~\bibnamefont {Pani}},\ }\href {\doibase 10.1103/PhysRevD.96.064050} {\bibfield  {journal} {\bibinfo  {journal} {Phys. Rev.}\ }\textbf {\bibinfo {volume} {D96}},\ \bibinfo {pages} {064050} (\bibinfo {year} {2017})},\ \Eprint {http://arxiv.org/abs/1706.06311} {arXiv:1706.06311 [gr-qc]} \BibitemShut {NoStop}%
\bibitem [{\citenamefont {Abbott}\ \emph {et~al.}(2021)\citenamefont {Abbott} \emph {et~al.}}]{LIGOScientific:2021jlr}%
  \BibitemOpen
  \bibfield  {author} {\bibinfo {author} {\bibfnamefont {R.}~\bibnamefont {Abbott}} \emph {et~al.} (\bibinfo {collaboration} {LIGO Scientific, VIRGO, KAGRA}),\ }\href@noop {} {\  (\bibinfo {year} {2021})},\ \Eprint {http://arxiv.org/abs/2111.15507} {arXiv:2111.15507 [astro-ph.HE]} \BibitemShut {NoStop}%
\bibitem [{\citenamefont {Zhu}\ \emph {et~al.}(2020)\citenamefont {Zhu}, \citenamefont {Baryakhtar}, \citenamefont {Papa}, \citenamefont {Tsuna}, \citenamefont {Kawanaka},\ and\ \citenamefont {Eggenstein}}]{Zhu:2020tht}%
  \BibitemOpen
  \bibfield  {author} {\bibinfo {author} {\bibfnamefont {S.~J.}\ \bibnamefont {Zhu}}, \bibinfo {author} {\bibfnamefont {M.}~\bibnamefont {Baryakhtar}}, \bibinfo {author} {\bibfnamefont {M.~A.}\ \bibnamefont {Papa}}, \bibinfo {author} {\bibfnamefont {D.}~\bibnamefont {Tsuna}}, \bibinfo {author} {\bibfnamefont {N.}~\bibnamefont {Kawanaka}}, \ and\ \bibinfo {author} {\bibfnamefont {H.-B.}\ \bibnamefont {Eggenstein}},\ }\href {\doibase 10.1103/PhysRevD.102.063020} {\bibfield  {journal} {\bibinfo  {journal} {Phys. Rev. D}\ }\textbf {\bibinfo {volume} {102}},\ \bibinfo {pages} {063020} (\bibinfo {year} {2020})},\ \Eprint {http://arxiv.org/abs/2003.03359} {arXiv:2003.03359 [gr-qc]} \BibitemShut {NoStop}%
\bibitem [{\citenamefont {Tsukada}\ \emph {et~al.}(2019)\citenamefont {Tsukada}, \citenamefont {Callister}, \citenamefont {Matas},\ and\ \citenamefont {Meyers}}]{Tsukada:2018mbp}%
  \BibitemOpen
  \bibfield  {author} {\bibinfo {author} {\bibfnamefont {L.}~\bibnamefont {Tsukada}}, \bibinfo {author} {\bibfnamefont {T.}~\bibnamefont {Callister}}, \bibinfo {author} {\bibfnamefont {A.}~\bibnamefont {Matas}}, \ and\ \bibinfo {author} {\bibfnamefont {P.}~\bibnamefont {Meyers}},\ }\href {\doibase 10.1103/PhysRevD.99.103015} {\bibfield  {journal} {\bibinfo  {journal} {Phys. Rev. D}\ }\textbf {\bibinfo {volume} {99}},\ \bibinfo {pages} {103015} (\bibinfo {year} {2019})},\ \Eprint {http://arxiv.org/abs/1812.09622} {arXiv:1812.09622 [astro-ph.HE]} \BibitemShut {NoStop}%
\bibitem [{\citenamefont {Isi}\ \emph {et~al.}(2019{\natexlab{a}})\citenamefont {Isi}, \citenamefont {Sun}, \citenamefont {Brito},\ and\ \citenamefont {Melatos}}]{Isi:2018pzk}%
  \BibitemOpen
  \bibfield  {author} {\bibinfo {author} {\bibfnamefont {M.}~\bibnamefont {Isi}}, \bibinfo {author} {\bibfnamefont {L.}~\bibnamefont {Sun}}, \bibinfo {author} {\bibfnamefont {R.}~\bibnamefont {Brito}}, \ and\ \bibinfo {author} {\bibfnamefont {A.}~\bibnamefont {Melatos}},\ }\href {\doibase 10.1103/PhysRevD.99.084042} {\bibfield  {journal} {\bibinfo  {journal} {Phys. Rev. D}\ }\textbf {\bibinfo {volume} {99}},\ \bibinfo {pages} {084042} (\bibinfo {year} {2019}{\natexlab{a}})},\ \bibinfo {note} {[Erratum: Phys.Rev.D 102, 049901 (2020)]},\ \Eprint {http://arxiv.org/abs/1810.03812} {arXiv:1810.03812 [gr-qc]} \BibitemShut {NoStop}%
\bibitem [{\citenamefont {Siemonsen}\ \emph {et~al.}(2022)\citenamefont {Siemonsen}, \citenamefont {May},\ and\ \citenamefont {East}}]{Siemonsen:2022yyf}%
  \BibitemOpen
  \bibfield  {author} {\bibinfo {author} {\bibfnamefont {N.}~\bibnamefont {Siemonsen}}, \bibinfo {author} {\bibfnamefont {T.}~\bibnamefont {May}}, \ and\ \bibinfo {author} {\bibfnamefont {W.~E.}\ \bibnamefont {East}},\ }\href@noop {} {\  (\bibinfo {year} {2022})},\ \Eprint {http://arxiv.org/abs/2211.03845} {arXiv:2211.03845 [gr-qc]} \BibitemShut {NoStop}%
\bibitem [{\citenamefont {Baumann}\ \emph {et~al.}(2019{\natexlab{a}})\citenamefont {Baumann}, \citenamefont {Chia},\ and\ \citenamefont {Porto}}]{Baumann:2018vus}%
  \BibitemOpen
  \bibfield  {author} {\bibinfo {author} {\bibfnamefont {D.}~\bibnamefont {Baumann}}, \bibinfo {author} {\bibfnamefont {H.~S.}\ \bibnamefont {Chia}}, \ and\ \bibinfo {author} {\bibfnamefont {R.~A.}\ \bibnamefont {Porto}},\ }\href {\doibase 10.1103/PhysRevD.99.044001} {\bibfield  {journal} {\bibinfo  {journal} {Phys. Rev. D}\ }\textbf {\bibinfo {volume} {99}},\ \bibinfo {pages} {044001} (\bibinfo {year} {2019}{\natexlab{a}})},\ \Eprint {http://arxiv.org/abs/1804.03208} {arXiv:1804.03208 [gr-qc]} \BibitemShut {NoStop}%
\bibitem [{\citenamefont {Chen}\ \emph {et~al.}(2020)\citenamefont {Chen}, \citenamefont {Shu}, \citenamefont {Xue}, \citenamefont {Yuan},\ and\ \citenamefont {Zhao}}]{Chen:2019fsq}%
  \BibitemOpen
  \bibfield  {author} {\bibinfo {author} {\bibfnamefont {Y.}~\bibnamefont {Chen}}, \bibinfo {author} {\bibfnamefont {J.}~\bibnamefont {Shu}}, \bibinfo {author} {\bibfnamefont {X.}~\bibnamefont {Xue}}, \bibinfo {author} {\bibfnamefont {Q.}~\bibnamefont {Yuan}}, \ and\ \bibinfo {author} {\bibfnamefont {Y.}~\bibnamefont {Zhao}},\ }\href {\doibase 10.1103/PhysRevLett.124.061102} {\bibfield  {journal} {\bibinfo  {journal} {Phys. Rev. Lett.}\ }\textbf {\bibinfo {volume} {124}},\ \bibinfo {pages} {061102} (\bibinfo {year} {2020})},\ \Eprint {http://arxiv.org/abs/1905.02213} {arXiv:1905.02213 [hep-ph]} \BibitemShut {NoStop}%
\bibitem [{\citenamefont {Ikeda}\ \emph {et~al.}(2019)\citenamefont {Ikeda}, \citenamefont {Brito},\ and\ \citenamefont {Cardoso}}]{Ikeda:2018nhb}%
  \BibitemOpen
  \bibfield  {author} {\bibinfo {author} {\bibfnamefont {T.}~\bibnamefont {Ikeda}}, \bibinfo {author} {\bibfnamefont {R.}~\bibnamefont {Brito}}, \ and\ \bibinfo {author} {\bibfnamefont {V.}~\bibnamefont {Cardoso}},\ }\href {\doibase 10.1103/PhysRevLett.122.081101} {\bibfield  {journal} {\bibinfo  {journal} {Phys. Rev. Lett.}\ }\textbf {\bibinfo {volume} {122}},\ \bibinfo {pages} {081101} (\bibinfo {year} {2019})},\ \Eprint {http://arxiv.org/abs/1811.04950} {arXiv:1811.04950 [gr-qc]} \BibitemShut {NoStop}%
\bibitem [{\citenamefont {Ding}\ \emph {et~al.}(2021)\citenamefont {Ding}, \citenamefont {Tong},\ and\ \citenamefont {Wang}}]{Ding:2020bnl}%
  \BibitemOpen
  \bibfield  {author} {\bibinfo {author} {\bibfnamefont {Q.}~\bibnamefont {Ding}}, \bibinfo {author} {\bibfnamefont {X.}~\bibnamefont {Tong}}, \ and\ \bibinfo {author} {\bibfnamefont {Y.}~\bibnamefont {Wang}},\ }\href {\doibase 10.3847/1538-4357/abd803} {\bibfield  {journal} {\bibinfo  {journal} {Astrophys. J.}\ }\textbf {\bibinfo {volume} {908}},\ \bibinfo {pages} {78} (\bibinfo {year} {2021})},\ \Eprint {http://arxiv.org/abs/2009.11106} {arXiv:2009.11106 [astro-ph.HE]} \BibitemShut {NoStop}%
\bibitem [{\citenamefont {Tong}\ \emph {et~al.}(2022{\natexlab{a}})\citenamefont {Tong}, \citenamefont {Wang},\ and\ \citenamefont {Zhu}}]{Tong:2021whq}%
  \BibitemOpen
  \bibfield  {author} {\bibinfo {author} {\bibfnamefont {X.}~\bibnamefont {Tong}}, \bibinfo {author} {\bibfnamefont {Y.}~\bibnamefont {Wang}}, \ and\ \bibinfo {author} {\bibfnamefont {H.-Y.}\ \bibnamefont {Zhu}},\ }\href {\doibase 10.3847/1538-4357/ac36db} {\bibfield  {journal} {\bibinfo  {journal} {Astrophys. J.}\ }\textbf {\bibinfo {volume} {924}},\ \bibinfo {pages} {99} (\bibinfo {year} {2022}{\natexlab{a}})},\ \Eprint {http://arxiv.org/abs/2106.13484} {arXiv:2106.13484 [astro-ph.HE]} \BibitemShut {NoStop}%
\bibitem [{\citenamefont {Choudhary}\ \emph {et~al.}(2021)\citenamefont {Choudhary}, \citenamefont {Sanchis-Gual}, \citenamefont {Gupta}, \citenamefont {Degollado}, \citenamefont {Bose},\ and\ \citenamefont {Font}}]{Choudhary:2020pxy}%
  \BibitemOpen
  \bibfield  {author} {\bibinfo {author} {\bibfnamefont {S.}~\bibnamefont {Choudhary}}, \bibinfo {author} {\bibfnamefont {N.}~\bibnamefont {Sanchis-Gual}}, \bibinfo {author} {\bibfnamefont {A.}~\bibnamefont {Gupta}}, \bibinfo {author} {\bibfnamefont {J.~C.}\ \bibnamefont {Degollado}}, \bibinfo {author} {\bibfnamefont {S.}~\bibnamefont {Bose}}, \ and\ \bibinfo {author} {\bibfnamefont {J.~A.}\ \bibnamefont {Font}},\ }\href {\doibase 10.1103/PhysRevD.103.044032} {\bibfield  {journal} {\bibinfo  {journal} {Phys. Rev. D}\ }\textbf {\bibinfo {volume} {103}},\ \bibinfo {pages} {044032} (\bibinfo {year} {2021})},\ \Eprint {http://arxiv.org/abs/2010.00935} {arXiv:2010.00935 [gr-qc]} \BibitemShut {NoStop}%
\bibitem [{\citenamefont {Yoo}\ \emph {et~al.}(2022)\citenamefont {Yoo}, \citenamefont {Naruko}, \citenamefont {Sakurai}, \citenamefont {Takahashi}, \citenamefont {Takamori},\ and\ \citenamefont {Yamauchi}}]{Yoo:2021kyv}%
  \BibitemOpen
  \bibfield  {author} {\bibinfo {author} {\bibfnamefont {C.-M.}\ \bibnamefont {Yoo}}, \bibinfo {author} {\bibfnamefont {A.}~\bibnamefont {Naruko}}, \bibinfo {author} {\bibfnamefont {Y.}~\bibnamefont {Sakurai}}, \bibinfo {author} {\bibfnamefont {K.}~\bibnamefont {Takahashi}}, \bibinfo {author} {\bibfnamefont {Y.}~\bibnamefont {Takamori}}, \ and\ \bibinfo {author} {\bibfnamefont {D.}~\bibnamefont {Yamauchi}},\ }\href {\doibase 10.1093/pasj/psab110} {\bibfield  {journal} {\bibinfo  {journal} {Publ. Astron. Soc. Jap.}\ }\textbf {\bibinfo {volume} {74}},\ \bibinfo {pages} {64} (\bibinfo {year} {2022})},\ \Eprint {http://arxiv.org/abs/2103.13227} {arXiv:2103.13227 [hep-ph]} \BibitemShut {NoStop}%
\bibitem [{\citenamefont {Tong}\ \emph {et~al.}(2022{\natexlab{b}})\citenamefont {Tong}, \citenamefont {Wang},\ and\ \citenamefont {Zhu}}]{Tong:2022bbl}%
  \BibitemOpen
  \bibfield  {author} {\bibinfo {author} {\bibfnamefont {X.}~\bibnamefont {Tong}}, \bibinfo {author} {\bibfnamefont {Y.}~\bibnamefont {Wang}}, \ and\ \bibinfo {author} {\bibfnamefont {H.-Y.}\ \bibnamefont {Zhu}},\ }\href {\doibase 10.1103/PhysRevD.106.043002} {\bibfield  {journal} {\bibinfo  {journal} {Phys. Rev. D}\ }\textbf {\bibinfo {volume} {106}},\ \bibinfo {pages} {043002} (\bibinfo {year} {2022}{\natexlab{b}})},\ \Eprint {http://arxiv.org/abs/2205.10527} {arXiv:2205.10527 [gr-qc]} \BibitemShut {NoStop}%
\bibitem [{\citenamefont {Takahashi}\ and\ \citenamefont {Tanaka}(2021)}]{Takahashi:2021eso}%
  \BibitemOpen
  \bibfield  {author} {\bibinfo {author} {\bibfnamefont {T.}~\bibnamefont {Takahashi}}\ and\ \bibinfo {author} {\bibfnamefont {T.}~\bibnamefont {Tanaka}},\ }\href {\doibase 10.1088/1475-7516/2021/10/031} {\bibfield  {journal} {\bibinfo  {journal} {JCAP}\ }\textbf {\bibinfo {volume} {10}},\ \bibinfo {pages} {031} (\bibinfo {year} {2021})},\ \Eprint {http://arxiv.org/abs/2106.08836} {arXiv:2106.08836 [gr-qc]} \BibitemShut {NoStop}%
\bibitem [{\citenamefont {Takahashi}\ \emph {et~al.}(2022)\citenamefont {Takahashi}, \citenamefont {Omiya},\ and\ \citenamefont {Tanaka}}]{Takahashi:2021yhy}%
  \BibitemOpen
  \bibfield  {author} {\bibinfo {author} {\bibfnamefont {T.}~\bibnamefont {Takahashi}}, \bibinfo {author} {\bibfnamefont {H.}~\bibnamefont {Omiya}}, \ and\ \bibinfo {author} {\bibfnamefont {T.}~\bibnamefont {Tanaka}},\ }\href {\doibase 10.1093/ptep/ptac044} {\bibfield  {journal} {\bibinfo  {journal} {PTEP}\ }\textbf {\bibinfo {volume} {2022}},\ \bibinfo {pages} {043E01} (\bibinfo {year} {2022})},\ \Eprint {http://arxiv.org/abs/2112.05774} {arXiv:2112.05774 [gr-qc]} \BibitemShut {NoStop}%
\bibitem [{\citenamefont {Takahashi}\ \emph {et~al.}(2023)\citenamefont {Takahashi}, \citenamefont {Omiya},\ and\ \citenamefont {Tanaka}}]{Takahashi:2023flk}%
  \BibitemOpen
  \bibfield  {author} {\bibinfo {author} {\bibfnamefont {T.}~\bibnamefont {Takahashi}}, \bibinfo {author} {\bibfnamefont {H.}~\bibnamefont {Omiya}}, \ and\ \bibinfo {author} {\bibfnamefont {T.}~\bibnamefont {Tanaka}},\ }\href {\doibase 10.1103/PhysRevD.107.103020} {\bibfield  {journal} {\bibinfo  {journal} {Phys. Rev. D}\ }\textbf {\bibinfo {volume} {107}},\ \bibinfo {pages} {103020} (\bibinfo {year} {2023})},\ \Eprint {http://arxiv.org/abs/2301.13213} {arXiv:2301.13213 [gr-qc]} \BibitemShut {NoStop}%
\bibitem [{\citenamefont {Sakurai}\ \emph {et~al.}(2023)\citenamefont {Sakurai}, \citenamefont {Yoo}, \citenamefont {Naruko},\ and\ \citenamefont {Yamauchi}}]{Sakurai:2023hkg}%
  \BibitemOpen
  \bibfield  {author} {\bibinfo {author} {\bibfnamefont {Y.}~\bibnamefont {Sakurai}}, \bibinfo {author} {\bibfnamefont {C.-M.}\ \bibnamefont {Yoo}}, \bibinfo {author} {\bibfnamefont {A.}~\bibnamefont {Naruko}}, \ and\ \bibinfo {author} {\bibfnamefont {D.}~\bibnamefont {Yamauchi}},\ }\href@noop {} {\  (\bibinfo {year} {2023})},\ \Eprint {http://arxiv.org/abs/2312.07058} {arXiv:2312.07058 [hep-ph]} \BibitemShut {NoStop}%
\bibitem [{\citenamefont {Spieksma}\ \emph {et~al.}(2023)\citenamefont {Spieksma}, \citenamefont {Cannizzaro}, \citenamefont {Ikeda}, \citenamefont {Cardoso},\ and\ \citenamefont {Chen}}]{Spieksma:2023vwl}%
  \BibitemOpen
  \bibfield  {author} {\bibinfo {author} {\bibfnamefont {T.~F.~M.}\ \bibnamefont {Spieksma}}, \bibinfo {author} {\bibfnamefont {E.}~\bibnamefont {Cannizzaro}}, \bibinfo {author} {\bibfnamefont {T.}~\bibnamefont {Ikeda}}, \bibinfo {author} {\bibfnamefont {V.}~\bibnamefont {Cardoso}}, \ and\ \bibinfo {author} {\bibfnamefont {Y.}~\bibnamefont {Chen}},\ }\href {\doibase 10.1103/PhysRevD.108.063013} {\bibfield  {journal} {\bibinfo  {journal} {Phys. Rev. D}\ }\textbf {\bibinfo {volume} {108}},\ \bibinfo {pages} {063013} (\bibinfo {year} {2023})},\ \Eprint {http://arxiv.org/abs/2306.16447} {arXiv:2306.16447 [gr-qc]} \BibitemShut {NoStop}%
\bibitem [{\citenamefont {Cannizzaro}\ \emph {et~al.}(2024)\citenamefont {Cannizzaro}, \citenamefont {Sberna}, \citenamefont {Green},\ and\ \citenamefont {Hollands}}]{Cannizzaro:2023jle}%
  \BibitemOpen
  \bibfield  {author} {\bibinfo {author} {\bibfnamefont {E.}~\bibnamefont {Cannizzaro}}, \bibinfo {author} {\bibfnamefont {L.}~\bibnamefont {Sberna}}, \bibinfo {author} {\bibfnamefont {S.~R.}\ \bibnamefont {Green}}, \ and\ \bibinfo {author} {\bibfnamefont {S.}~\bibnamefont {Hollands}},\ }\href {\doibase 10.1103/PhysRevLett.132.051401} {\bibfield  {journal} {\bibinfo  {journal} {Phys. Rev. Lett.}\ }\textbf {\bibinfo {volume} {132}},\ \bibinfo {pages} {051401} (\bibinfo {year} {2024})},\ \Eprint {http://arxiv.org/abs/2309.10021} {arXiv:2309.10021 [gr-qc]} \BibitemShut {NoStop}%
\bibitem [{\citenamefont {Sarmah}\ \emph {et~al.}(2024)\citenamefont {Sarmah}, \citenamefont {Verma}, \citenamefont {Cheung},\ and\ \citenamefont {Silk}}]{Sarmah:2024nst}%
  \BibitemOpen
  \bibfield  {author} {\bibinfo {author} {\bibfnamefont {P.}~\bibnamefont {Sarmah}}, \bibinfo {author} {\bibfnamefont {H.}~\bibnamefont {Verma}}, \bibinfo {author} {\bibfnamefont {K.}~\bibnamefont {Cheung}}, \ and\ \bibinfo {author} {\bibfnamefont {J.}~\bibnamefont {Silk}},\ }\href@noop {} {\  (\bibinfo {year} {2024})},\ \Eprint {http://arxiv.org/abs/2404.09955} {arXiv:2404.09955 [astro-ph.HE]} \BibitemShut {NoStop}%
\bibitem [{\citenamefont {Yoshino}\ and\ \citenamefont {Kodama}(2015{\natexlab{b}})}]{Yoshino:2015nsa}%
  \BibitemOpen
  \bibfield  {author} {\bibinfo {author} {\bibfnamefont {H.}~\bibnamefont {Yoshino}}\ and\ \bibinfo {author} {\bibfnamefont {H.}~\bibnamefont {Kodama}},\ }\href {\doibase 10.1088/0264-9381/32/21/214001} {\bibfield  {journal} {\bibinfo  {journal} {Class. Quant. Grav.}\ }\textbf {\bibinfo {volume} {32}},\ \bibinfo {pages} {214001} (\bibinfo {year} {2015}{\natexlab{b}})},\ \Eprint {http://arxiv.org/abs/1505.00714} {arXiv:1505.00714 [gr-qc]} \BibitemShut {NoStop}%
\bibitem [{\citenamefont {Fukuda}\ and\ \citenamefont {Nakayama}(2020)}]{Fukuda:2019ewf}%
  \BibitemOpen
  \bibfield  {author} {\bibinfo {author} {\bibfnamefont {H.}~\bibnamefont {Fukuda}}\ and\ \bibinfo {author} {\bibfnamefont {K.}~\bibnamefont {Nakayama}},\ }\href {\doibase 10.1007/JHEP01(2020)128} {\bibfield  {journal} {\bibinfo  {journal} {JHEP}\ }\textbf {\bibinfo {volume} {01}},\ \bibinfo {pages} {128} (\bibinfo {year} {2020})},\ \Eprint {http://arxiv.org/abs/1910.06308} {arXiv:1910.06308 [hep-ph]} \BibitemShut {NoStop}%
\bibitem [{\citenamefont {Omiya}\ \emph {et~al.}(2021)\citenamefont {Omiya}, \citenamefont {Takahashi},\ and\ \citenamefont {Tanaka}}]{Omiya:2020vji}%
  \BibitemOpen
  \bibfield  {author} {\bibinfo {author} {\bibfnamefont {H.}~\bibnamefont {Omiya}}, \bibinfo {author} {\bibfnamefont {T.}~\bibnamefont {Takahashi}}, \ and\ \bibinfo {author} {\bibfnamefont {T.}~\bibnamefont {Tanaka}},\ }\href {\doibase 10.1093/ptep/ptab032} {\bibfield  {journal} {\bibinfo  {journal} {PTEP}\ }\textbf {\bibinfo {volume} {2021}},\ \bibinfo {pages} {043E02} (\bibinfo {year} {2021})},\ \Eprint {http://arxiv.org/abs/2012.03473} {arXiv:2012.03473 [gr-qc]} \BibitemShut {NoStop}%
\bibitem [{\citenamefont {Omiya}\ \emph {et~al.}(2022)\citenamefont {Omiya}, \citenamefont {Takahashi},\ and\ \citenamefont {Tanaka}}]{Omiya:2022mwv}%
  \BibitemOpen
  \bibfield  {author} {\bibinfo {author} {\bibfnamefont {H.}~\bibnamefont {Omiya}}, \bibinfo {author} {\bibfnamefont {T.}~\bibnamefont {Takahashi}}, \ and\ \bibinfo {author} {\bibfnamefont {T.}~\bibnamefont {Tanaka}},\ }\href {\doibase 10.1093/ptep/ptac058} {\bibfield  {journal} {\bibinfo  {journal} {PTEP}\ }\textbf {\bibinfo {volume} {2022}},\ \bibinfo {pages} {043E03} (\bibinfo {year} {2022})},\ \Eprint {http://arxiv.org/abs/2201.04382} {arXiv:2201.04382 [gr-qc]} \BibitemShut {NoStop}%
\bibitem [{\citenamefont {Gruzinov}(2016)}]{Gruzinov:2016hcq}%
  \BibitemOpen
  \bibfield  {author} {\bibinfo {author} {\bibfnamefont {A.}~\bibnamefont {Gruzinov}},\ }\href@noop {} {\  (\bibinfo {year} {2016})},\ \Eprint {http://arxiv.org/abs/1604.06422} {arXiv:1604.06422 [astro-ph.HE]} \BibitemShut {NoStop}%
\bibitem [{\citenamefont {Baryakhtar}\ \emph {et~al.}(2021)\citenamefont {Baryakhtar}, \citenamefont {Galanis}, \citenamefont {Lasenby},\ and\ \citenamefont {Simon}}]{Baryakhtar:2020gao}%
  \BibitemOpen
  \bibfield  {author} {\bibinfo {author} {\bibfnamefont {M.}~\bibnamefont {Baryakhtar}}, \bibinfo {author} {\bibfnamefont {M.}~\bibnamefont {Galanis}}, \bibinfo {author} {\bibfnamefont {R.}~\bibnamefont {Lasenby}}, \ and\ \bibinfo {author} {\bibfnamefont {O.}~\bibnamefont {Simon}},\ }\href {\doibase 10.1103/PhysRevD.103.095019} {\bibfield  {journal} {\bibinfo  {journal} {Phys. Rev. D}\ }\textbf {\bibinfo {volume} {103}},\ \bibinfo {pages} {095019} (\bibinfo {year} {2021})},\ \Eprint {http://arxiv.org/abs/2011.11646} {arXiv:2011.11646 [hep-ph]} \BibitemShut {NoStop}%
\bibitem [{\citenamefont {Omiya}\ \emph {et~al.}(2023)\citenamefont {Omiya}, \citenamefont {Takahashi}, \citenamefont {Tanaka},\ and\ \citenamefont {Yoshino}}]{Omiya:2022gwu}%
  \BibitemOpen
  \bibfield  {author} {\bibinfo {author} {\bibfnamefont {H.}~\bibnamefont {Omiya}}, \bibinfo {author} {\bibfnamefont {T.}~\bibnamefont {Takahashi}}, \bibinfo {author} {\bibfnamefont {T.}~\bibnamefont {Tanaka}}, \ and\ \bibinfo {author} {\bibfnamefont {H.}~\bibnamefont {Yoshino}},\ }\href {\doibase 10.1088/1475-7516/2023/06/016} {\bibfield  {journal} {\bibinfo  {journal} {JCAP}\ }\textbf {\bibinfo {volume} {06}},\ \bibinfo {pages} {016} (\bibinfo {year} {2023})},\ \Eprint {http://arxiv.org/abs/2211.01949} {arXiv:2211.01949 [gr-qc]} \BibitemShut {NoStop}%
\bibitem [{\citenamefont {Calz\`a}\ \emph {et~al.}(2023)\citenamefont {Calz\`a}, \citenamefont {Rosa},\ and\ \citenamefont {Serrano}}]{Calza:2023rjt}%
  \BibitemOpen
  \bibfield  {author} {\bibinfo {author} {\bibfnamefont {M.}~\bibnamefont {Calz\`a}}, \bibinfo {author} {\bibfnamefont {J.~a.~G.}\ \bibnamefont {Rosa}}, \ and\ \bibinfo {author} {\bibfnamefont {F.}~\bibnamefont {Serrano}},\ }\href@noop {} {\  (\bibinfo {year} {2023})},\ \Eprint {http://arxiv.org/abs/2306.09430} {arXiv:2306.09430 [hep-ph]} \BibitemShut {NoStop}%
\bibitem [{\citenamefont {Brito}\ \emph {et~al.}(2015{\natexlab{b}})\citenamefont {Brito}, \citenamefont {Cardoso},\ and\ \citenamefont {Pani}}]{Brito:2014wla}%
  \BibitemOpen
  \bibfield  {author} {\bibinfo {author} {\bibfnamefont {R.}~\bibnamefont {Brito}}, \bibinfo {author} {\bibfnamefont {V.}~\bibnamefont {Cardoso}}, \ and\ \bibinfo {author} {\bibfnamefont {P.}~\bibnamefont {Pani}},\ }\href {\doibase 10.1088/0264-9381/32/13/134001} {\bibfield  {journal} {\bibinfo  {journal} {Class. Quant. Grav.}\ }\textbf {\bibinfo {volume} {32}},\ \bibinfo {pages} {134001} (\bibinfo {year} {2015}{\natexlab{b}})},\ \Eprint {http://arxiv.org/abs/1411.0686} {arXiv:1411.0686 [gr-qc]} \BibitemShut {NoStop}%
\bibitem [{\citenamefont {Brill}\ \emph {et~al.}(1972)\citenamefont {Brill}, \citenamefont {Chrzanowski}, \citenamefont {Martin~Pereira}, \citenamefont {Fackerell},\ and\ \citenamefont {Ipser}}]{Brill:1972xj}%
  \BibitemOpen
  \bibfield  {author} {\bibinfo {author} {\bibfnamefont {D.}~\bibnamefont {Brill}}, \bibinfo {author} {\bibfnamefont {P.}~\bibnamefont {Chrzanowski}}, \bibinfo {author} {\bibfnamefont {C.}~\bibnamefont {Martin~Pereira}}, \bibinfo {author} {\bibfnamefont {E.}~\bibnamefont {Fackerell}}, \ and\ \bibinfo {author} {\bibfnamefont {J.}~\bibnamefont {Ipser}},\ }\href {\doibase 10.1103/PhysRevD.5.1913} {\bibfield  {journal} {\bibinfo  {journal} {Phys. Rev. D}\ }\textbf {\bibinfo {volume} {5}},\ \bibinfo {pages} {1913} (\bibinfo {year} {1972})}\BibitemShut {NoStop}%
\bibitem [{\citenamefont {Dolan}(2007)}]{Dolan:2007mj}%
  \BibitemOpen
  \bibfield  {author} {\bibinfo {author} {\bibfnamefont {S.~R.}\ \bibnamefont {Dolan}},\ }\href {\doibase 10.1103/PhysRevD.76.084001} {\bibfield  {journal} {\bibinfo  {journal} {Phys. Rev.}\ }\textbf {\bibinfo {volume} {D76}},\ \bibinfo {pages} {084001} (\bibinfo {year} {2007})},\ \Eprint {http://arxiv.org/abs/0705.2880} {arXiv:0705.2880 [gr-qc]} \BibitemShut {NoStop}%
\bibitem [{\citenamefont {Baumann}\ \emph {et~al.}(2019{\natexlab{b}})\citenamefont {Baumann}, \citenamefont {Chia}, \citenamefont {Stout},\ and\ \citenamefont {ter Haar}}]{Baumann:2019eav}%
  \BibitemOpen
  \bibfield  {author} {\bibinfo {author} {\bibfnamefont {D.}~\bibnamefont {Baumann}}, \bibinfo {author} {\bibfnamefont {H.~S.}\ \bibnamefont {Chia}}, \bibinfo {author} {\bibfnamefont {J.}~\bibnamefont {Stout}}, \ and\ \bibinfo {author} {\bibfnamefont {L.}~\bibnamefont {ter Haar}},\ }\href {\doibase 10.1088/1475-7516/2019/12/006} {\bibfield  {journal} {\bibinfo  {journal} {JCAP}\ }\textbf {\bibinfo {volume} {12}},\ \bibinfo {pages} {006} (\bibinfo {year} {2019}{\natexlab{b}})},\ \Eprint {http://arxiv.org/abs/1908.10370} {arXiv:1908.10370 [gr-qc]} \BibitemShut {NoStop}%
\bibitem [{\citenamefont {Bao}\ \emph {et~al.}(2022)\citenamefont {Bao}, \citenamefont {Xu},\ and\ \citenamefont {Zhang}}]{Bao:2022hew}%
  \BibitemOpen
  \bibfield  {author} {\bibinfo {author} {\bibfnamefont {S.}~\bibnamefont {Bao}}, \bibinfo {author} {\bibfnamefont {Q.}~\bibnamefont {Xu}}, \ and\ \bibinfo {author} {\bibfnamefont {H.}~\bibnamefont {Zhang}},\ }\href {\doibase 10.1103/PhysRevD.106.064016} {\bibfield  {journal} {\bibinfo  {journal} {Phys. Rev. D}\ }\textbf {\bibinfo {volume} {106}},\ \bibinfo {pages} {064016} (\bibinfo {year} {2022})},\ \Eprint {http://arxiv.org/abs/2201.10941} {arXiv:2201.10941 [gr-qc]} \BibitemShut {NoStop}%
\bibitem [{\citenamefont {Mocanu}\ and\ \citenamefont {Grumiller}(2012)}]{Mocanu:2012fd}%
  \BibitemOpen
  \bibfield  {author} {\bibinfo {author} {\bibfnamefont {G.}~\bibnamefont {Mocanu}}\ and\ \bibinfo {author} {\bibfnamefont {D.}~\bibnamefont {Grumiller}},\ }\href {\doibase 10.1103/PhysRevD.85.105022} {\bibfield  {journal} {\bibinfo  {journal} {Phys. Rev. D}\ }\textbf {\bibinfo {volume} {85}},\ \bibinfo {pages} {105022} (\bibinfo {year} {2012})},\ \Eprint {http://arxiv.org/abs/1203.4681} {arXiv:1203.4681 [astro-ph.CO]} \BibitemShut {NoStop}%
\bibitem [{\citenamefont {Yoshino}\ and\ \citenamefont {Kodama}(2012)}]{Yoshino:2012kn}%
  \BibitemOpen
  \bibfield  {author} {\bibinfo {author} {\bibfnamefont {H.}~\bibnamefont {Yoshino}}\ and\ \bibinfo {author} {\bibfnamefont {H.}~\bibnamefont {Kodama}},\ }\href {\doibase 10.1143/PTP.128.153} {\bibfield  {journal} {\bibinfo  {journal} {Prog. Theor. Phys.}\ }\textbf {\bibinfo {volume} {128}},\ \bibinfo {pages} {153} (\bibinfo {year} {2012})},\ \Eprint {http://arxiv.org/abs/1203.5070} {arXiv:1203.5070 [gr-qc]} \BibitemShut {NoStop}%
\bibitem [{\citenamefont {Isi}\ \emph {et~al.}(2019{\natexlab{b}})\citenamefont {Isi}, \citenamefont {Sun}, \citenamefont {Brito},\ and\ \citenamefont {Melatos}}]{PhysRevD.99.084042}%
  \BibitemOpen
  \bibfield  {author} {\bibinfo {author} {\bibfnamefont {M.}~\bibnamefont {Isi}}, \bibinfo {author} {\bibfnamefont {L.}~\bibnamefont {Sun}}, \bibinfo {author} {\bibfnamefont {R.}~\bibnamefont {Brito}}, \ and\ \bibinfo {author} {\bibfnamefont {A.}~\bibnamefont {Melatos}},\ }\href {\doibase 10.1103/PhysRevD.99.084042} {\bibfield  {journal} {\bibinfo  {journal} {Phys. Rev. D}\ }\textbf {\bibinfo {volume} {99}},\ \bibinfo {pages} {084042} (\bibinfo {year} {2019}{\natexlab{b}})}\BibitemShut {NoStop}%
\bibitem [{\citenamefont {Yoshino}\ and\ \citenamefont {Kodama}(2014)}]{Yoshino:2013ofa}%
  \BibitemOpen
  \bibfield  {author} {\bibinfo {author} {\bibfnamefont {H.}~\bibnamefont {Yoshino}}\ and\ \bibinfo {author} {\bibfnamefont {H.}~\bibnamefont {Kodama}},\ }\href {\doibase 10.1093/ptep/ptu029} {\bibfield  {journal} {\bibinfo  {journal} {PTEP}\ }\textbf {\bibinfo {volume} {2014}},\ \bibinfo {pages} {043E02} (\bibinfo {year} {2014})},\ \Eprint {http://arxiv.org/abs/1312.2326} {arXiv:1312.2326 [gr-qc]} \BibitemShut {NoStop}%
\bibitem [{\citenamefont {Chia}\ \emph {et~al.}(2023)\citenamefont {Chia}, \citenamefont {Doorman}, \citenamefont {Wernersson}, \citenamefont {Hinderer},\ and\ \citenamefont {Nissanke}}]{Chia:2022udn}%
  \BibitemOpen
  \bibfield  {author} {\bibinfo {author} {\bibfnamefont {H.~S.}\ \bibnamefont {Chia}}, \bibinfo {author} {\bibfnamefont {C.}~\bibnamefont {Doorman}}, \bibinfo {author} {\bibfnamefont {A.}~\bibnamefont {Wernersson}}, \bibinfo {author} {\bibfnamefont {T.}~\bibnamefont {Hinderer}}, \ and\ \bibinfo {author} {\bibfnamefont {S.}~\bibnamefont {Nissanke}},\ }\href {\doibase 10.1088/1475-7516/2023/04/018} {\bibfield  {journal} {\bibinfo  {journal} {JCAP}\ }\textbf {\bibinfo {volume} {04}},\ \bibinfo {pages} {018} (\bibinfo {year} {2023})},\ \Eprint {http://arxiv.org/abs/2212.11948} {arXiv:2212.11948 [gr-qc]} \BibitemShut {NoStop}%
\bibitem [{Note1()}]{Note1}%
  \BibitemOpen
  \bibinfo {note} {Note that we normalized the axion field by $F_a$, so $\phi $ is non-dimension. In addition, $M_{{\protect \rm cl},i}$ is normalized by $F_a^2 M_{\protect \rm BH}$, so also nondimension.}\BibitemShut {Stop}%
\bibitem [{\citenamefont {Guo}\ \emph {et~al.}(2023)\citenamefont {Guo}, \citenamefont {Bao},\ and\ \citenamefont {Zhang}}]{Guo:2022mpr}%
  \BibitemOpen
  \bibfield  {author} {\bibinfo {author} {\bibfnamefont {Y.-d.}\ \bibnamefont {Guo}}, \bibinfo {author} {\bibfnamefont {S.-s.}\ \bibnamefont {Bao}}, \ and\ \bibinfo {author} {\bibfnamefont {H.}~\bibnamefont {Zhang}},\ }\href {\doibase 10.1103/PhysRevD.107.075009} {\bibfield  {journal} {\bibinfo  {journal} {Phys. Rev. D}\ }\textbf {\bibinfo {volume} {107}},\ \bibinfo {pages} {075009} (\bibinfo {year} {2023})},\ \Eprint {http://arxiv.org/abs/2212.07186} {arXiv:2212.07186 [gr-qc]} \BibitemShut {NoStop}%
\bibitem [{\citenamefont {Teukolsky}(1973)}]{Teukolsky:1973ha}%
  \BibitemOpen
  \bibfield  {author} {\bibinfo {author} {\bibfnamefont {S.~A.}\ \bibnamefont {Teukolsky}},\ }\href {\doibase 10.1086/152444} {\bibfield  {journal} {\bibinfo  {journal} {Astrophys. J.}\ }\textbf {\bibinfo {volume} {185}},\ \bibinfo {pages} {635} (\bibinfo {year} {1973})}\BibitemShut {NoStop}%
\bibitem [{\citenamefont {Kawamura}\ \emph {et~al.}(2011)\citenamefont {Kawamura} \emph {et~al.}}]{Kawamura:2011zz}%
  \BibitemOpen
  \bibfield  {author} {\bibinfo {author} {\bibfnamefont {S.}~\bibnamefont {Kawamura}} \emph {et~al.},\ }\href {\doibase 10.1088/0264-9381/28/9/094011} {\bibfield  {journal} {\bibinfo  {journal} {Class. Quant. Grav.}\ }\textbf {\bibinfo {volume} {28}},\ \bibinfo {pages} {094011} (\bibinfo {year} {2011})}\BibitemShut {NoStop}%
\bibitem [{Pro(2023)}]{Proceedings:2023mkp}%
  \BibitemOpen
  \href@noop {} {\emph {\bibinfo {title} {{Terrestrial Very-Long-Baseline Atom Interferometry: Workshop Summary}}}}\ (\bibinfo {year} {2023})\ \Eprint {http://arxiv.org/abs/2310.08183} {arXiv:2310.08183 [hep-ex]} \BibitemShut {NoStop}%
\bibitem [{\citenamefont {Shimoda}\ \emph {et~al.}(2020)\citenamefont {Shimoda}, \citenamefont {Takano}, \citenamefont {Ooi}, \citenamefont {Aritomi}, \citenamefont {Michimura}, \citenamefont {Ando},\ and\ \citenamefont {Shoda}}]{Shimoda:2018uiv}%
  \BibitemOpen
  \bibfield  {author} {\bibinfo {author} {\bibfnamefont {T.}~\bibnamefont {Shimoda}}, \bibinfo {author} {\bibfnamefont {S.}~\bibnamefont {Takano}}, \bibinfo {author} {\bibfnamefont {C.~P.}\ \bibnamefont {Ooi}}, \bibinfo {author} {\bibfnamefont {N.}~\bibnamefont {Aritomi}}, \bibinfo {author} {\bibfnamefont {Y.}~\bibnamefont {Michimura}}, \bibinfo {author} {\bibfnamefont {M.}~\bibnamefont {Ando}}, \ and\ \bibinfo {author} {\bibfnamefont {A.}~\bibnamefont {Shoda}},\ }\href {\doibase 10.1142/S0218271819400030} {\bibfield  {journal} {\bibinfo  {journal} {Int. J. Mod. Phys. D}\ }\textbf {\bibinfo {volume} {29}},\ \bibinfo {pages} {1940003} (\bibinfo {year} {2020})},\ \Eprint {http://arxiv.org/abs/1812.01835} {arXiv:1812.01835 [physics.ins-det]} \BibitemShut {NoStop}%
\bibitem [{\citenamefont {{Kuns}}\ \emph {et~al.}(2020)\citenamefont {{Kuns}}, \citenamefont {{Yu}}, \citenamefont {{Chen}},\ and\ \citenamefont {{Adhikari}}}]{2020PhRvD.102d3001K}%
  \BibitemOpen
  \bibfield  {author} {\bibinfo {author} {\bibfnamefont {K.~A.}\ \bibnamefont {{Kuns}}}, \bibinfo {author} {\bibfnamefont {H.}~\bibnamefont {{Yu}}}, \bibinfo {author} {\bibfnamefont {Y.}~\bibnamefont {{Chen}}}, \ and\ \bibinfo {author} {\bibfnamefont {R.~X.}\ \bibnamefont {{Adhikari}}},\ }\href {\doibase 10.1103/PhysRevD.102.043001} {\bibfield  {journal} {\bibinfo  {journal} {\prd}\ }\textbf {\bibinfo {volume} {102}},\ \bibinfo {eid} {043001} (\bibinfo {year} {2020})},\ \Eprint {http://arxiv.org/abs/1908.06004} {arXiv:1908.06004 [gr-qc]} \BibitemShut {NoStop}%
\bibitem [{\citenamefont {{Arca Sedda}}\ \emph {et~al.}(2020)\citenamefont {{Arca Sedda}}, \citenamefont {{Berry}}, \citenamefont {{Jani}}, \citenamefont {{Amaro-Seoane}}, \citenamefont {{Auclair}}, \citenamefont {{Baird}}, \citenamefont {{Baker}}, \citenamefont {{Berti}}, \citenamefont {{Breivik}}, \citenamefont {{Burrows}}, \citenamefont {{Caprini}}, \citenamefont {{Chen}}, \citenamefont {{Doneva}}, \citenamefont {{Ezquiaga}}, \citenamefont {{Saavik Ford}}, \citenamefont {{Katz}}, \citenamefont {{Kolkowitz}}, \citenamefont {{McKernan}}, \citenamefont {{Mueller}}, \citenamefont {{Nardini}}, \citenamefont {{Pikovski}}, \citenamefont {{Rajendran}}, \citenamefont {{Sesana}}, \citenamefont {{Shao}}, \citenamefont {{Tamanini}}, \citenamefont {{Vartanyan}}, \citenamefont {{Warburton}}, \citenamefont {{Witek}}, \citenamefont {{Wong}},\ and\ \citenamefont {{Zevin}}}]{2020CQGra..37u5011A}%
  \BibitemOpen
  \bibfield  {author} {\bibinfo {author} {\bibfnamefont {M.}~\bibnamefont {{Arca Sedda}}}, \bibinfo {author} {\bibfnamefont {C.~P.~L.}\ \bibnamefont {{Berry}}}, \bibinfo {author} {\bibfnamefont {K.}~\bibnamefont {{Jani}}}, \bibinfo {author} {\bibfnamefont {P.}~\bibnamefont {{Amaro-Seoane}}}, \bibinfo {author} {\bibfnamefont {P.}~\bibnamefont {{Auclair}}}, \bibinfo {author} {\bibfnamefont {J.}~\bibnamefont {{Baird}}}, \bibinfo {author} {\bibfnamefont {T.}~\bibnamefont {{Baker}}}, \bibinfo {author} {\bibfnamefont {E.}~\bibnamefont {{Berti}}}, \bibinfo {author} {\bibfnamefont {K.}~\bibnamefont {{Breivik}}}, \bibinfo {author} {\bibfnamefont {A.}~\bibnamefont {{Burrows}}}, \bibinfo {author} {\bibfnamefont {C.}~\bibnamefont {{Caprini}}}, \bibinfo {author} {\bibfnamefont {X.}~\bibnamefont {{Chen}}}, \bibinfo {author} {\bibfnamefont {D.}~\bibnamefont {{Doneva}}}, \bibinfo {author} {\bibfnamefont {J.~M.}\ \bibnamefont {{Ezquiaga}}}, \bibinfo {author} {\bibfnamefont {K.~E.}\ \bibnamefont {{Saavik Ford}}}, \bibinfo
  {author} {\bibfnamefont {M.~L.}\ \bibnamefont {{Katz}}}, \bibinfo {author} {\bibfnamefont {S.}~\bibnamefont {{Kolkowitz}}}, \bibinfo {author} {\bibfnamefont {B.}~\bibnamefont {{McKernan}}}, \bibinfo {author} {\bibfnamefont {G.}~\bibnamefont {{Mueller}}}, \bibinfo {author} {\bibfnamefont {G.}~\bibnamefont {{Nardini}}}, \bibinfo {author} {\bibfnamefont {I.}~\bibnamefont {{Pikovski}}}, \bibinfo {author} {\bibfnamefont {S.}~\bibnamefont {{Rajendran}}}, \bibinfo {author} {\bibfnamefont {A.}~\bibnamefont {{Sesana}}}, \bibinfo {author} {\bibfnamefont {L.}~\bibnamefont {{Shao}}}, \bibinfo {author} {\bibfnamefont {N.}~\bibnamefont {{Tamanini}}}, \bibinfo {author} {\bibfnamefont {D.}~\bibnamefont {{Vartanyan}}}, \bibinfo {author} {\bibfnamefont {N.}~\bibnamefont {{Warburton}}}, \bibinfo {author} {\bibfnamefont {H.}~\bibnamefont {{Witek}}}, \bibinfo {author} {\bibfnamefont {K.}~\bibnamefont {{Wong}}}, \ and\ \bibinfo {author} {\bibfnamefont {M.}~\bibnamefont {{Zevin}}},\ }\href {\doibase 10.1088/1361-6382/abb5c1}
  {\bibfield  {journal} {\bibinfo  {journal} {Classical and Quantum Gravity}\ }\textbf {\bibinfo {volume} {37}},\ \bibinfo {eid} {215011} (\bibinfo {year} {2020})},\ \Eprint {http://arxiv.org/abs/1908.11375} {arXiv:1908.11375 [gr-qc]} \BibitemShut {NoStop}%
\bibitem [{\citenamefont {{Ni}}\ \emph {et~al.}(2020)\citenamefont {{Ni}}, \citenamefont {{Wang}},\ and\ \citenamefont {{Wu}}}]{2020IJMPD..2940007N}%
  \BibitemOpen
  \bibfield  {author} {\bibinfo {author} {\bibfnamefont {W.-T.}\ \bibnamefont {{Ni}}}, \bibinfo {author} {\bibfnamefont {G.}~\bibnamefont {{Wang}}}, \ and\ \bibinfo {author} {\bibfnamefont {A.-M.}\ \bibnamefont {{Wu}}},\ }\href {\doibase 10.1142/S0218271819400078} {\bibfield  {journal} {\bibinfo  {journal} {International Journal of Modern Physics D}\ }\textbf {\bibinfo {volume} {29}},\ \bibinfo {eid} {1940007-129} (\bibinfo {year} {2020})},\ \Eprint {http://arxiv.org/abs/1909.04995} {arXiv:1909.04995 [gr-qc]} \BibitemShut {NoStop}%
\bibitem [{\citenamefont {Ajith}\ \emph {et~al.}(2024)\citenamefont {Ajith} \emph {et~al.}}]{Ajith:2024mie}%
  \BibitemOpen
  \bibfield  {author} {\bibinfo {author} {\bibfnamefont {P.}~\bibnamefont {Ajith}} \emph {et~al.},\ }\href@noop {} {\  (\bibinfo {year} {2024})},\ \Eprint {http://arxiv.org/abs/2404.09181} {arXiv:2404.09181 [gr-qc]} \BibitemShut {NoStop}%
\bibitem [{\citenamefont {{Gou}}\ \emph {et~al.}(2011)\citenamefont {{Gou}}, \citenamefont {{McClintock}}, \citenamefont {{Reid}}, \citenamefont {{Orosz}}, \citenamefont {{Steiner}}, \citenamefont {{Narayan}}, \citenamefont {{Xiang}}, \citenamefont {{Remillard}}, \citenamefont {{Arnaud}},\ and\ \citenamefont {{Davis}}}]{2011ApJ...742...85G}%
  \BibitemOpen
  \bibfield  {author} {\bibinfo {author} {\bibfnamefont {L.}~\bibnamefont {{Gou}}}, \bibinfo {author} {\bibfnamefont {J.~E.}\ \bibnamefont {{McClintock}}}, \bibinfo {author} {\bibfnamefont {M.~J.}\ \bibnamefont {{Reid}}}, \bibinfo {author} {\bibfnamefont {J.~A.}\ \bibnamefont {{Orosz}}}, \bibinfo {author} {\bibfnamefont {J.~F.}\ \bibnamefont {{Steiner}}}, \bibinfo {author} {\bibfnamefont {R.}~\bibnamefont {{Narayan}}}, \bibinfo {author} {\bibfnamefont {J.}~\bibnamefont {{Xiang}}}, \bibinfo {author} {\bibfnamefont {R.~A.}\ \bibnamefont {{Remillard}}}, \bibinfo {author} {\bibfnamefont {K.~A.}\ \bibnamefont {{Arnaud}}}, \ and\ \bibinfo {author} {\bibfnamefont {S.~W.}\ \bibnamefont {{Davis}}},\ }\href {\doibase 10.1088/0004-637X/742/2/85} {\bibfield  {journal} {\bibinfo  {journal} {\apj}\ }\textbf {\bibinfo {volume} {742}},\ \bibinfo {eid} {85} (\bibinfo {year} {2011})},\ \Eprint {http://arxiv.org/abs/1106.3690} {arXiv:1106.3690 [astro-ph.HE]} \BibitemShut {NoStop}%
\bibitem [{\citenamefont {{Reid}}\ \emph {et~al.}(2011)\citenamefont {{Reid}}, \citenamefont {{McClintock}}, \citenamefont {{Narayan}}, \citenamefont {{Gou}}, \citenamefont {{Remillard}},\ and\ \citenamefont {{Orosz}}}]{2011ApJ...742...83R}%
  \BibitemOpen
  \bibfield  {author} {\bibinfo {author} {\bibfnamefont {M.~J.}\ \bibnamefont {{Reid}}}, \bibinfo {author} {\bibfnamefont {J.~E.}\ \bibnamefont {{McClintock}}}, \bibinfo {author} {\bibfnamefont {R.}~\bibnamefont {{Narayan}}}, \bibinfo {author} {\bibfnamefont {L.}~\bibnamefont {{Gou}}}, \bibinfo {author} {\bibfnamefont {R.~A.}\ \bibnamefont {{Remillard}}}, \ and\ \bibinfo {author} {\bibfnamefont {J.~A.}\ \bibnamefont {{Orosz}}},\ }\href {\doibase 10.1088/0004-637X/742/2/83} {\bibfield  {journal} {\bibinfo  {journal} {\apj}\ }\textbf {\bibinfo {volume} {742}},\ \bibinfo {eid} {83} (\bibinfo {year} {2011})},\ \Eprint {http://arxiv.org/abs/1106.3688} {arXiv:1106.3688 [astro-ph.HE]} \BibitemShut {NoStop}%
\bibitem [{\citenamefont {Sun}\ \emph {et~al.}(2020)\citenamefont {Sun}, \citenamefont {Brito},\ and\ \citenamefont {Isi}}]{Sun:2019mqb}%
  \BibitemOpen
  \bibfield  {author} {\bibinfo {author} {\bibfnamefont {L.}~\bibnamefont {Sun}}, \bibinfo {author} {\bibfnamefont {R.}~\bibnamefont {Brito}}, \ and\ \bibinfo {author} {\bibfnamefont {M.}~\bibnamefont {Isi}},\ }\href {\doibase 10.1103/PhysRevD.101.063020} {\bibfield  {journal} {\bibinfo  {journal} {Phys. Rev. D}\ }\textbf {\bibinfo {volume} {101}},\ \bibinfo {pages} {063020} (\bibinfo {year} {2020})},\ \bibinfo {note} {[Erratum: Phys.Rev.D 102, 089902 (2020)]},\ \Eprint {http://arxiv.org/abs/1909.11267} {arXiv:1909.11267 [gr-qc]} \BibitemShut {NoStop}%
\bibitem [{\citenamefont {Abbott}\ \emph {et~al.}(2017)\citenamefont {Abbott} \emph {et~al.}}]{LIGOScientific:2017vwq}%
  \BibitemOpen
  \bibfield  {author} {\bibinfo {author} {\bibfnamefont {B.~P.}\ \bibnamefont {Abbott}} \emph {et~al.} (\bibinfo {collaboration} {LIGO Scientific, Virgo}),\ }\href {\doibase 10.1103/PhysRevLett.119.161101} {\bibfield  {journal} {\bibinfo  {journal} {Phys. Rev. Lett.}\ }\textbf {\bibinfo {volume} {119}},\ \bibinfo {pages} {161101} (\bibinfo {year} {2017})},\ \Eprint {http://arxiv.org/abs/1710.05832} {arXiv:1710.05832 [gr-qc]} \BibitemShut {NoStop}%
\bibitem [{\citenamefont {Kawamura}\ \emph {et~al.}(2021)\citenamefont {Kawamura} \emph {et~al.}}]{Kawamura:2020pcg}%
  \BibitemOpen
  \bibfield  {author} {\bibinfo {author} {\bibfnamefont {S.}~\bibnamefont {Kawamura}} \emph {et~al.},\ }\href {\doibase 10.1093/ptep/ptab019} {\bibfield  {journal} {\bibinfo  {journal} {PTEP}\ }\textbf {\bibinfo {volume} {2021}},\ \bibinfo {pages} {05A105} (\bibinfo {year} {2021})},\ \Eprint {http://arxiv.org/abs/2006.13545} {arXiv:2006.13545 [gr-qc]} \BibitemShut {NoStop}%
\bibitem [{\citenamefont {Abbott}\ \emph {et~al.}(2016)\citenamefont {Abbott} \emph {et~al.}}]{LIGOScientific:2016aoc}%
  \BibitemOpen
  \bibfield  {author} {\bibinfo {author} {\bibfnamefont {B.~P.}\ \bibnamefont {Abbott}} \emph {et~al.} (\bibinfo {collaboration} {LIGO Scientific, Virgo}),\ }\href {\doibase 10.1103/PhysRevLett.116.061102} {\bibfield  {journal} {\bibinfo  {journal} {Phys. Rev. Lett.}\ }\textbf {\bibinfo {volume} {116}},\ \bibinfo {pages} {061102} (\bibinfo {year} {2016})},\ \Eprint {http://arxiv.org/abs/1602.03837} {arXiv:1602.03837 [gr-qc]} \BibitemShut {NoStop}%
\bibitem [{\citenamefont {Du}\ \emph {et~al.}(2022)\citenamefont {Du}, \citenamefont {Egana-Ugrinovic}, \citenamefont {Essig}, \citenamefont {Fragione},\ and\ \citenamefont {Perna}}]{Du:2022trq}%
  \BibitemOpen
  \bibfield  {author} {\bibinfo {author} {\bibfnamefont {P.}~\bibnamefont {Du}}, \bibinfo {author} {\bibfnamefont {D.}~\bibnamefont {Egana-Ugrinovic}}, \bibinfo {author} {\bibfnamefont {R.}~\bibnamefont {Essig}}, \bibinfo {author} {\bibfnamefont {G.}~\bibnamefont {Fragione}}, \ and\ \bibinfo {author} {\bibfnamefont {R.}~\bibnamefont {Perna}},\ }\href {\doibase 10.1038/s41467-022-32301-4} {\bibfield  {journal} {\bibinfo  {journal} {Nature Commun.}\ }\textbf {\bibinfo {volume} {13}},\ \bibinfo {pages} {4626} (\bibinfo {year} {2022})},\ \Eprint {http://arxiv.org/abs/2202.01215} {arXiv:2202.01215 [hep-ph]} \BibitemShut {NoStop}%
\end{thebibliography}%

\end{document}